\documentclass[a4paper,fleqn,usenatbib]{mnras}
\usepackage{amsmath,amssymb,natbib,latexsym,times}
\usepackage{lineno}
\DeclareMathOperator\arctanh{arctanh}

\usepackage[T1]{fontenc}
\usepackage{aecompl}

\usepackage{verbatim}
\usepackage{hhline}
\usepackage[usenames,dvipsnames]{xcolor}
\usepackage{tabulary}
\usepackage{tabularx}
\usepackage{todonotes}
\usepackage{hyperref}
\usepackage{xspace}
\usepackage{eso-pic}

\AddToShipoutPictureBG*{%
  \AtPageUpperLeft{%
    \hspace{0.75\paperwidth}%
    \raisebox{-3.5\baselineskip}{%
      \makebox[0pt][l]{\textnormal{DES 2016-0208}}
}}}%

\AddToShipoutPictureBG*{%
  \AtPageUpperLeft{%
    \hspace{0.75\paperwidth}%
    \raisebox{-4.5\baselineskip}{%
      \makebox[0pt][l]{\textnormal{Fermilab PUB-17-290-A-AE}}
}}}%

\numberwithin{equation}{section}

\newcommand{\changedtext}[1]{#1}

\newcommand{\meds}{MEDS}

\newcommand{\uberseg}{{\"u}berseg}

\newcommand{\photoz}{photo-$z$}
\newcommand\degree{\ensuremath{\,^\circ}}
\newcommand{\snr}{\ensuremath{S/N}}
\newcommand{\snrw}{\ensuremath{S/N}}

\newcommand{\rgp}{\ensuremath{R_{gp}/R_p}}
\newcommand{\epsf}{\ensuremath{e_\textsc{psf}}}
\newcommand{\Tpsf}{\ensuremath{T_\textsc{psf}}}
\newcommand{\Tgal}{\ensuremath{T_\mathrm{gal}}}
\newcommand{\bfx}{\ensuremath{\mathbf{x}}}
\newcommand{\bfxpt}{\ensuremath{\mathbf{x} + \boldsymbol{\theta}}}

\newcommand*\justify{%
  \fontdimen2\font=0.4em
  \fontdimen3\font=0.2em
  \fontdimen4\font=0.1em
  \fontdimen7\font=0.1em
  \hyphenchar\font=`\-
}

\newcommand\code[1]{\texttt{\small\justify #1}}

\newcommand{\sex}{\textsc{SEx\-tractor}}
\newcommand{\psfex}{\textsc{PSF\-Ex}}

\newcommand{\magauto}{\code{MAG\_AUTO}}

\newcommand{\spreadmodel}{\code{SPREAD\_MODEL}}

\newcommand{\imshape}{{\textsc{im3shape}}}
\newcommand{\ngmix}{\textsc{ngmix}}
\newcommand{\metacal}{\textsc{metacalibration}}
\newcommand{\galsim}{\textsc{GalSim}}

\newcommand{\scamp}{\textsc{SCamp}}

\newcommand{\hoopoe}{\textsc{Hoopoe}}

\newcommand{\mcalR}{\mbox{\boldmath $R$}}
\newcommand{\mcalRmean}{\mbox{\boldmath $\langle R \rangle$}}
\newcommand{\mcalRgmean}{\mbox{\boldmath $\langle R_\gamma \rangle$}}
\newcommand{\mcalRSmean}{\mbox{\boldmath $\langle R_S \rangle$}}
\newcommand{\mcalRg}{\mbox{\boldmath $R_\gamma$}}
\newcommand{\mcalRS}{\mbox{\boldmath $R_S$}}

\newcommand{\est}{e}
\newcommand{\vest}{\mbox{\boldmath $e$}}
\newcommand{\vecg}{\mbox{\boldmath $\gamma$}}

\newcommand{\devauc}{de Vaucouleurs}
\newcommand{\sersic}{S{\'e}rsic}

\newcommand{\levmar}{Levenberg-Marquardt}

\newcommand{\blockfont}[1]{{\textsc{#1}}\xspace}

\defcitealias{jarvis16}{J16}
\defcitealias{SheldonHuff2017}{SH17}

\newcommand\eqn[1]{equation~\ref{#1}}

\newcommand\be{\begin{equation}}
\newcommand\ee{\end{equation}}

\allowdisplaybreaks

\title[DES Year 1 Results: Weak Lensing Shape Catalogues]{Dark Energy Survey Year 1 Results:\\Weak Lensing Shape Catalogues}

\author[J. Zuntz et al.]{
\parbox{\textwidth}{
\Large
J.~Zuntz$^{1}$\thanks{joe.zuntz@ed.ac.uk},
E.~Sheldon$^{2}$\thanks{erin.sheldon@gmail.com},
S.~Samuroff$^{3}$,
M.~A.~Troxel$^{4,5}$,
M.~Jarvis$^{6}$,
N.~MacCrann$^{4,5}$,
D.~Gruen$^{7,8,\ddagger}$,
J.~Prat$^{9}$,
C.~S{\'a}nchez$^{9}$,
A.~Choi$^{4}$
% ...
\begin{center} (DES Collaboration) \end{center}
}
\vspace{0.4cm}
\\
\parbox{\textwidth}{
The authors' affiliations are shown in Appendix \ref{sec:affiliations}.
}
}

\date{Accepted XXX. Received YYY; in original form ZZZ}

\pubyear{2015}

\begin{document}

\label{firstpage}
\pagerange{\pageref{firstpage}--\pageref{lastpage}}
\maketitle
\begin{abstract} We present two galaxy shape catalogues from the Dark Energy Survey Year 1 data set, covering 1500
square degrees with a median redshift of $0.59$. The catalogues cover two main fields: Stripe 82, and an area overlapping the South Pole Telescope survey region.  We describe our data analysis process and in particular our shape
measurement using two independent shear measurement pipelines, \metacal\ and \imshape. 
The \metacal\ catalogue uses a Gaussian model with an innovative internal calibration scheme, and was applied to $riz$ bands, yielding 34.8M objects.
The \imshape\ catalogue uses a
maximum-likelihood bulge/disc model calibrated using simulations, and
was applied to $r$-band data, yielding 21.9M objects.
Both catalogues pass a suite of null tests that demonstrate their
fitness for use in weak lensing science.  We estimate the $1\sigma$
uncertainties in multiplicative shear calibration to be 0.013 and 0.025
for the \metacal\ and \imshape\ catalogues, respectively.
\end{abstract}

\begin{keywords}
gravitational lensing: weak -- cosmology: observations --
surveys -- catalogues --
methods: data analysis -- techniques: image processing
\end{keywords}

\tableofcontents

\section{Introduction}

%
%

Weak lensing, the gravitational bending of light paths by wide-field matter distributions, presents
a powerful probe of cosmological physics and the laws of gravity.  The angle by which light is bent
by any lens depends on two factors: the geometry of the source-lens-observer system, and the
inherent strength of the lens.  In the cosmic case, the former depends on the expansion history of
the Universe via the relationship between redshift and distance.  The latter depends on laws of
gravity and the amount of structure in the Universe - the variance of the cosmic density field.
Through both these dependencies we can put limits on the history of the Universe, the cosmological
parameters, and most interestingly the behaviour of dark matter and the equation of state of dark
energy.

The most direct way to measure weak lensing is to measure the ellipticity of distant galaxies. The
effect of the intermediate gravitational fields on the light from a source is to shear it,
coherently stretching the galaxies in a region in the same direction.  The magnitude of this
effect on a single galaxy is only a few percent, which is much smaller than either the intrinsic scatter in galaxy
shapes or the atmospheric and optical image distortion. The intrinsic scatter means we require
large surveys, to obtain as much statistical power as possible, and the 
atmospheric and optical effects mean we 
require careful optical design and precision modelling of the induced distortions (the point-spread function, PSF).


%
%
The Dark Energy Survey (DES) is the largest ongoing lensing survey designed to meet these requirements, and is part of the current ``Stage III'' group of lensing surveys \citep{DETF}.  The earliest Stage I surveys, including VIRMOS-Descart \citep{vanwaerbeke2005}, CTIO \citep{jarvis2006}, SDSS \citep{hirata2004} and COSMOS \citep{schrabback2007}, mostly measured tens of square degrees, and made some of the first detections of cosmic shear.  Stage II surveys included DLS \citep{jee2013}, SDSS \citep{Lin11,Huff14}, RCSLenS \citep{hildebrandt2016}, CFHTLenS \citep{heymans2012} as well as early science verification (SV) DES results in \citet{jarvis16}. They included both deep and wide surveys, up to hundreds of square degrees, and obtained significant cosmological constraints.  The current Stage III generation includes DES, KiDS \citep{hildebrandt2017,amon17},  and HSC \citep{aihara2017}, which are each surveying at least $1000$ square degrees and will obtain cosmological constraints comparable in power to all other cosmological data.  Upcoming Stage IV surveys, including Euclid, LSST, WFIRST, and SKA, will measure the dark energy equation of state with $1\%$ precision when combined with  data from the cosmic microwave background (CMB).
DES will eventually survey $5000$ square degrees.  It has currently completed four out of its five planned full seasons of observations.  The catalogues described in this paper use observations from the first of those four years, and cover $1500$ square degrees. Processing and analysis of the entirety of existing DES data is underway.

%
%

\begin{figure}
\centerline{\includegraphics[width=7cm]{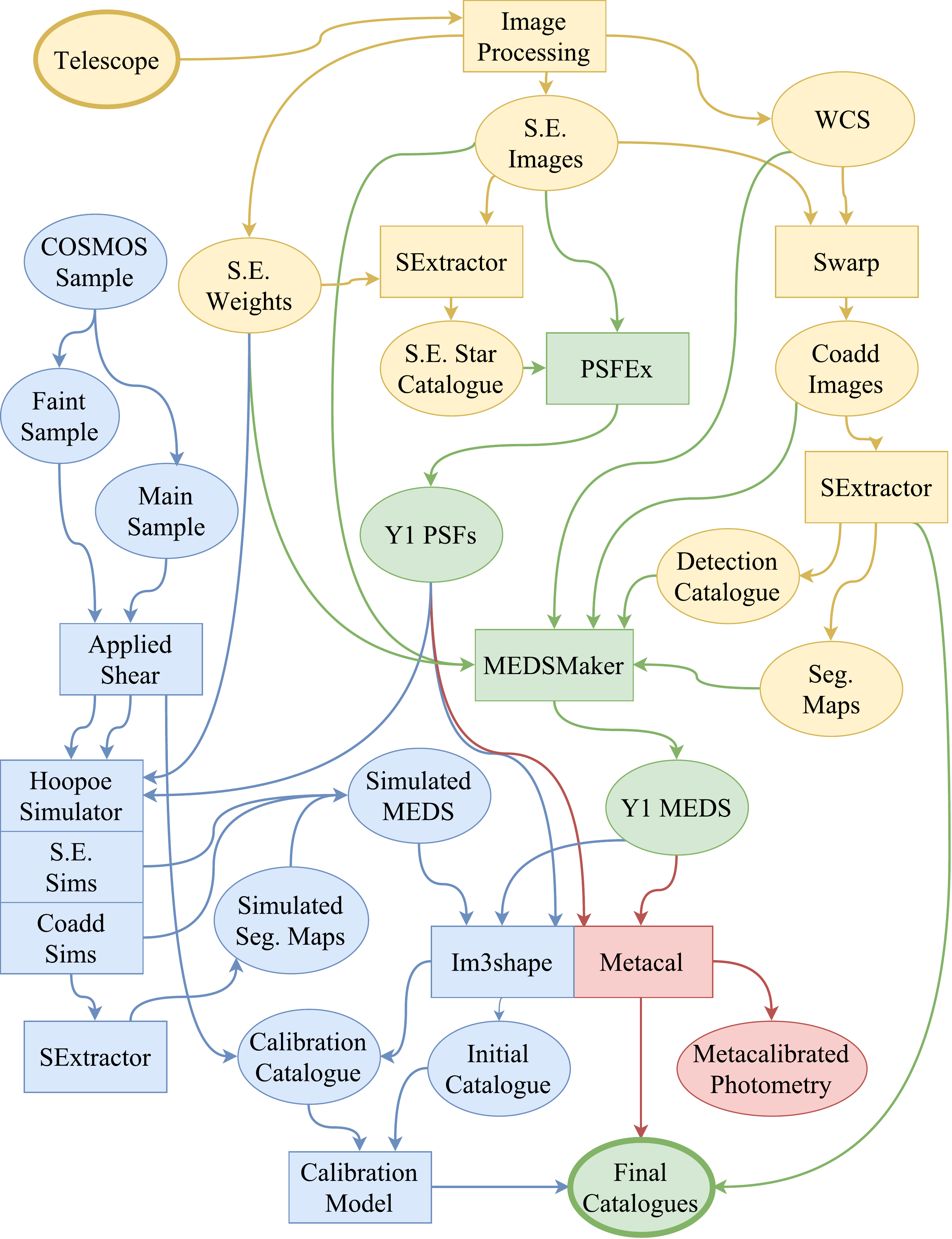}}
\caption{A flow-chart showing the steps in the DES Year 1 shape analysis, starting from low-level calibrated data products made by DES Data Management (DES-DM) and ending with final output catalogues.  Yellow stages are performed in the DES-DM software process.  Green stages are performed in the Weak Lensing analysis process. Blue stages are part of the \imshape\ process, mostly simulation and calibration, and red stages part of the \metacal\ analysis.  ``S.E.'' stands for ``single epoch''.
}
\label{fig:flowchart}
\end{figure}

Building a catalogue of galaxy ellipticities (a shape catalogue) from image data is a long process with many steps, each of which must be performed with careful attention to potential induced biases.  The DES implementation of these steps is shown visually in Figure~\ref{fig:flowchart}. The first stage is low-level calibration to detect artifacts, measure noise, and regularize images.  We build coadded images and detect and classify stars and galaxies in them. We measure the astrometry and PSF in each single-epoch image.  We collect together single-epoch ``postage-stamp'' images for each source into a single multi-epoch
data structure (MEDS). Finally we come to the shape measurement process itself, which forms the bulk of this paper. 
We measure galaxy ellipticities with two quantities $e_1$ and $e_2$, and the ensemble shear in terms of either $\gamma_1$ and $\gamma_2$ or the reduced shears $g_1$ and $g_2$ \citep{BartelmannSchneider99}.

The difficulty of accurately recovering ellipticities and shears from noisy, pixelized
data, as well as the value of exploring multiple approaches to it, was quickly recognized.
In response, a series of shape measurement challenges have sought to compare
and test the various codes available. 
The past decade has seen several such exercises, most notably the Shear Testing Programme (STEP) and GRavitational lEnsing Accuracy Testing (GREAT) 
challenges \citep{step1,step2,great08,great10,great3},
which have illuminated many of the issues that the field must solve.

Galaxy shape measurement methods can be split into two broad categories. Each must correct for the imaging processes, such as PSF convolution, which alter the apparent shapes of galaxies. 
The first is forward-modelling methods, in which parametric models of galaxy images 
are generated, propagated through the observing processes, and compared to the data in order to obtain a likelihood 
or other goodness-of-fit metric for the galaxy parameters.
The second class, inverse methods, measure second-order moments or other values 
on the image data, then apply corrections to compensate for the effects of the observing process.
Early methods like KSB \citep{ksb} and Shapelets \citep{shapelets} largely fall
into the latter category, but recent work has mostly focused on model-based methods.

Within each of these categories there are a great many methodologies
and specific codes, each with different assumptions and designs,
which lead to advantages and drawbacks in different domains.  
One advantage of model fitting-methods is that it is easier to enumerate the biases that can afflict
them\footnote{Problems analogous to these issues affect model-independent methods too, but it is typically harder to
interpret their impact.}. 

We can characterize these biases with a Taylor expansion as \citep{step1}:
\begin{equation}\label{eq:mc_bias}
g_i = (1+m_i) g^\mathrm{tr}_i + c_i,
\end{equation}
where $g_i$ is a shear estimate for the $i = (1,2)$ component of shear and $g^\mathrm{tr}_i$ is the true value.  
The dominant contribution to the $c_i$ term usually arises from the PSF ellipticity, so we sometimes re-write this as:
\begin{equation}
g_i = (1+m_i) g^\mathrm{tr}_i + \alpha e^\mathrm{PSF}_i + c_i \; .
\label{eqn:m-c-alpha}
\end{equation}
for some $\alpha$ and the PSF ellipticity $e^\mathrm{PSF}$.  The three largest biases that generate various combinations of $m$, $c$, and $\alpha$ are usually model bias, noise bias, and selection bias.

Model bias, the mismatch between an assumed galaxy image model and the true one, was shown in the GREAT3 challenge to
cause an error of up to $\sim 1\%$, which is comparable to the target errors in the current generation of surveys
\citep{great3}.

Noise bias is often the dominant shear measurement bias, and is more properly understood as an estimator bias.
It affects methods that use the maximum point in the likelihood of model parameters or similar
quantities as a point-wise estimator of the ellipticity, since these quantities are inherently biased if the
probability distributions are asymmetric \citep{hirata2003,kacprzak2012,bernsteinjarvis02}, as is almost always 
the case for shear estimation. 
It typically causes a $\sim 10\%$ bias if untreated.  One solution is to account for the shape of the posterior surface; methods for doing this
have been developed in \citet{miller2007} and \citet{bernsteinarmstrong} 
and was used by the DES-SV analysis in the \ngmix\ code \citep{ngmix}.

Selection bias is the result of objects being included or excluded from the catalogue in a way that depends on their intrinsic shapes or the shear to which they are subject. Every catalogue has some selection function, and nearly all will
result in biased shear estimates. 
Even if the measurements on individual galaxies are completely accurate (i.e., the histogram of their shapes can be recovered
perfectly), if we preferentially select, for example, the roundest galaxies, we will systematically underestimate the cosmological
shear. 
If noise bias is an estimator bias, then selection bias can be thought of as a \emph{representativeness} bias. 
These effects have been found to be more pervasive than previously believed, and were found to cause
$5\%$ biases in \citet{jarvis16}.  They make comparison between shear samples particularly difficult, and can arise from
the detection process itself or from cuts or binning applied to measured results---the latter was found to be much more
significant in \citet{fc16}.

%
%
%
%

There are multiple practical paths to the elimination of these various shear estimation biases. The simplest is to accept their existence and estimate the shear errors by processing simulated data with known input shear through the same pipeline as the real data.
Early calibration methods using simulations used a single global calibration factor \citep{schrabback2007,jee2013}.  More
recent methods have derived a calibration value per-object as a function of measured galaxy properties, e.g.
\citet{jarvis16}, \citet{hildebrandt2017}.  This is the approach taken by the \imshape\ code in this paper.

These calibration methods require simulations that are very carefully matched to the properties of the given data;
otherwise the calibration factors used can be incorrect.  Methods which do not depend on simulations can reduce 
the scale of this challenge, or avoid it completely.
There has been a flurry of interest in recent years in the various ways one could do this.
In \citet{fc16}, the KiDS collaboration used \emph{self-calibration}, in which a simulated version of
each object is generated from the best-fit model parameters and re-measured---this removes about half of the noise bias
and reduces required simulation volumes.  \citet{HuffMandelbaum2017} and \citet{SheldonHuff2017} describe the 
\emph{metacalibration} method used by the \metacal\ pipeline in this paper, which calibrates the estimator biases by applying 
an added shear to the \emph{real} galaxy images and gauging its effect on galaxy measurement and selection.  
This proves highly effective in tests on simulations.
Another recent approach, the Bayesian Fourier Domain method  (BFD; \citealt{bernsteinarmstrong}), uses deeper data to provide an implicit
model, avoiding model bias, and prescribes a selection process for which biases are calculable from
a full probabilistic treatment.  BFD estimates will be investigated in future DES shear catalogues.

The DES shape measurement methodology in the DES-SV period was exhaustively detailed in
\citet{jarvis16}, hereafter \citetalias{jarvis16}.  Many aspects of our methodology are the same as in SV, so this paper
builds on that work---unchanged aspects of the process that are not explained here are detailed there.

%
%

This paper is organized as follows: in \S\ref{sec:data} we describe the observations analyzed in this work.  In
\S\ref{sec:psf} we describe the measurement of the PSF.  \S\ref{sec:metacal} and \S\ref{sec:im3shape}  describe the construction of our two catalogues, \metacal~ and \imshape~ respectively, including
the calibration simulations used in the latter.  \S\ref{sec:tests} describes a series of tests validating that the
catalogues have sufficient accuracy for cosmic shear, cross correlations, and other measurements of the lensing signal.  \S\ref{sec:cats} discusses procedures for use
of the catalogues, including appropriate systematic error priors and the correct use of the calibration systems.
We conclude in \S\ref{sec:conclusions}.

\section{Data}
\label{sec:data}

\subsection{Observing Period and Conditions}
\label{sec:data:conditions}
The Dark Energy Survey (DES) Year One (Y1) catalogues described here are 
based on observations taken
using the Dark Energy Camera \citep[DECam,][]{Flaugher15} on the Blanco telescope
at the Cerro Tololo Inter-American Observatory
during the first full season of DES operations. Y1 images were acquired
between 31 Aug 2013 and 9 Feb 2014 \citep{diehl2014}.  The nominal plan for the DES Wide
Survey is to image the entire 5000~deg$^2$ footprint 10 times in each
of the $g, r, i, z,$ and $Y$ filters over 5 seasons of operation.
\changedtext{DECam images have an average pixel scale of 0.263 arcsec.}
In Y1 we opted to target only the regions overlapping the South Pole
Telescope (SPT) survey footprint at $-60^\circ \lesssim \delta \lesssim
-40^\circ$ and the equatorial SDSS ``Stripe 82'' region covering $-1.26\degree < \delta < +1.26\degree$ 
and $20:00h < \mathrm{RA} < 04:00h$, comprising about 30\% of the full
footprint.  The goal was to obtain 4 ``tilings'' per filter over this
region in Y1, rather than cover the full footprint with 2 tilings,
because 4-tiling coverage is much more robust to cosmic rays
and per-exposure systematic errors, especially after considering the
gaps in the functional imaging area of DECam. Given these factors, a 2-tiling coverage
would not have led to a viable shape catalogue. The vagaries of the
weather led to non-uniform coverage of the Y1 target area.
Figure~\ref{fig:sptemap} shows the footprint of the Y1 \metacal\ shape catalogue
after the cuts described below for minimum depth in each filter.  

In comparison to the SV catalogues described by \citetalias{jarvis16},
the main areas of the Y1 shape catalogues cover a much larger area (1500 vs
140~deg$^2$) but with a lower integrated exposure time (up to $4\times90$~s exposures
per filter in $griz$ vs. $10\times90$~s nominal in SV).  The quality of the Y1 imaging
is superior to that taken in SV in several respects:
\begin{itemize}
\item The telescope tracking servos exhibited oscillations in right
  ascension during most of the SV period, leading to more elliptical
  and less stable
  PSFs.  This was fully remedied for Y1.
\item More rigorous assessment of image quality was in place for Y1,
  and exposures failing to meet certain thresholds for seeing, cloud
  extinction, and sky brightness were rejected after each night's
  observing and placed back onto the observing queue \citep{neilsen2016}.
\item The feedback system using out-of-focus stellar images to
  maintain focus and alignment of the camera \citep{roodman} was
  improved substantially by the start of Y1, further stabilizing the PSF
  quality. 
\item Thermal control of the Blanco mirror and dome was improved between
  the SV and Y1 periods.
\item Improved baffling of the filters reduced the incidence of
  stray-light contamination, and improvements in software
  identification of image artifacts also reduced the number of
  spurious features in the images.
\item The SV observing sequences concentrated most of the observations
  of a given part of the sky into a small number of nights.  By Y1 we
  had adopted a wide-survey scheduler which penalizes repeat coverage
  in a given filter on a given night.  This decorrelates weather
  variation from the sky coordinates and leads to more uniform survey quality.
\item The shutter-closed time between exposures was reduced, increasing the observing efficiency $\sim$ 2.5\%. 

\end{itemize}
One degradation in camera performance during Y1 is that one of the 62 CCDs in
the DECam science array failed on 30 Nov 2013. Most of the
Y1 data therefore has one less usable CCD's worth of data per exposure.

\subsection{Object Catalogue}
\label{sec:gold}

\begin{figure}
\includegraphics[width=\columnwidth]{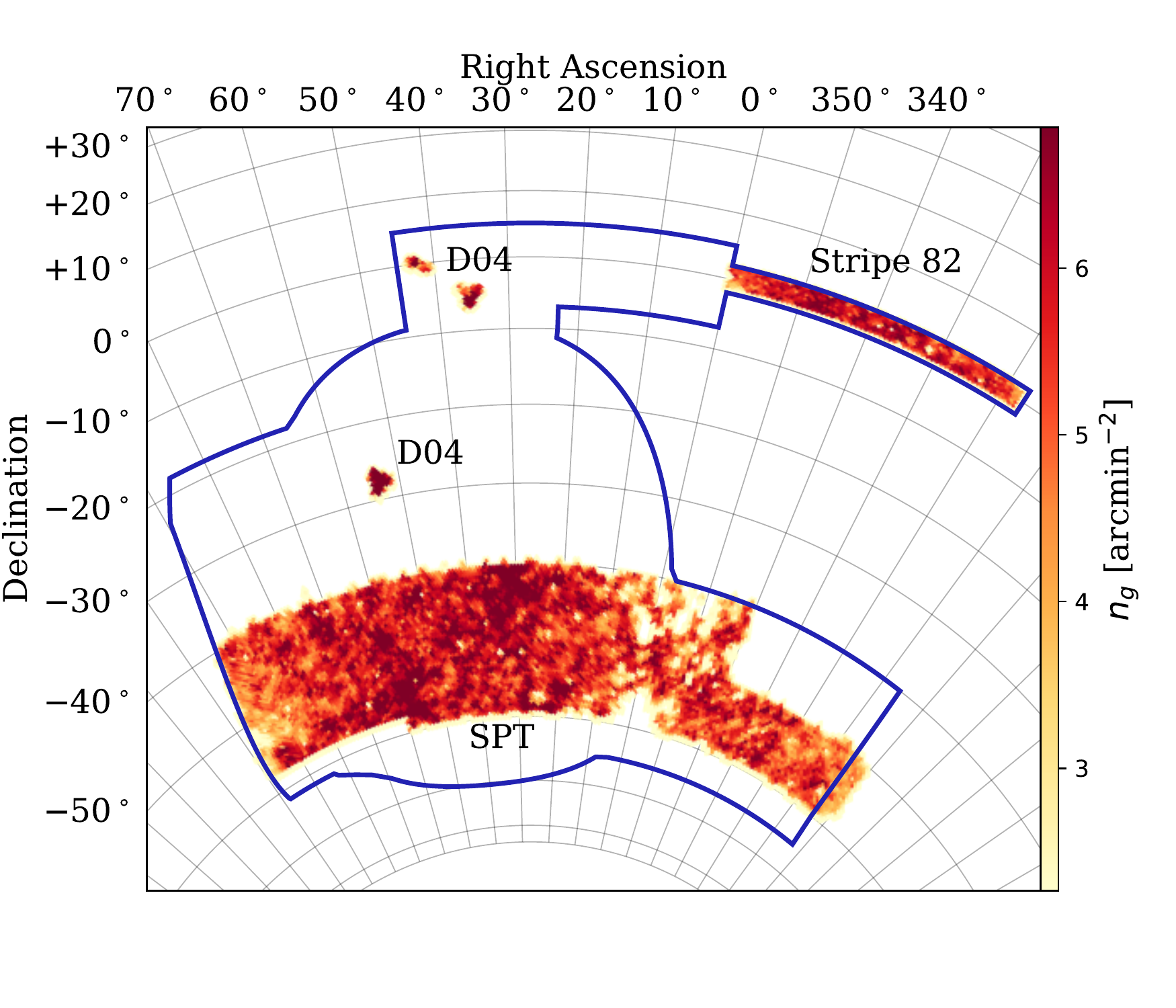}
\caption{The DES Y1 shear catalogue footprint with galaxy density of the \metacal\ catalogue shown with the nominal 5-year DES footprint outline overlayed. \imshape~is qualitatively similar, but slightly shallower. We define three fields: 1) The large, southern field overlapping with SPT, which has been selected for DES Y1 science applications due to contiguity. 2) The long equatorial strip overlapping with SDSS Stripe 82. 3) The disjoint supernovae and spectroscopic-overlap fields, which have been selected from the 4 exposure depth (D04) {\sc Gold} catalogue. Additional D04 fields far from the SPT region are not shown. The densities are not corrected for the detection fraction within each pixel. The Albers equal-area projection is used.
\label{fig:sptemap}
}
\end{figure}

The initial selection of galaxies on which shape measurement was performed is detailed in 
\citet{y1gold}, and the selection described therein is denoted the {\sc Gold} catalogue.  The image
reduction, photometric calibration, and detection from coadded images
to the catalogues are described in that paper, and the star-galaxy separation described therein is applied to the \imshape\ catalogue.  The full region the catalogue covers is shown in 
Figure~\ref{fig:sptemap}, though our cosmological analyses will use only the southern region that overlaps with the SPT survey.

\subsection{Galaxy Selection}
\label{sec:galaxy selection}

\begin{figure}
\includegraphics[width=\columnwidth]{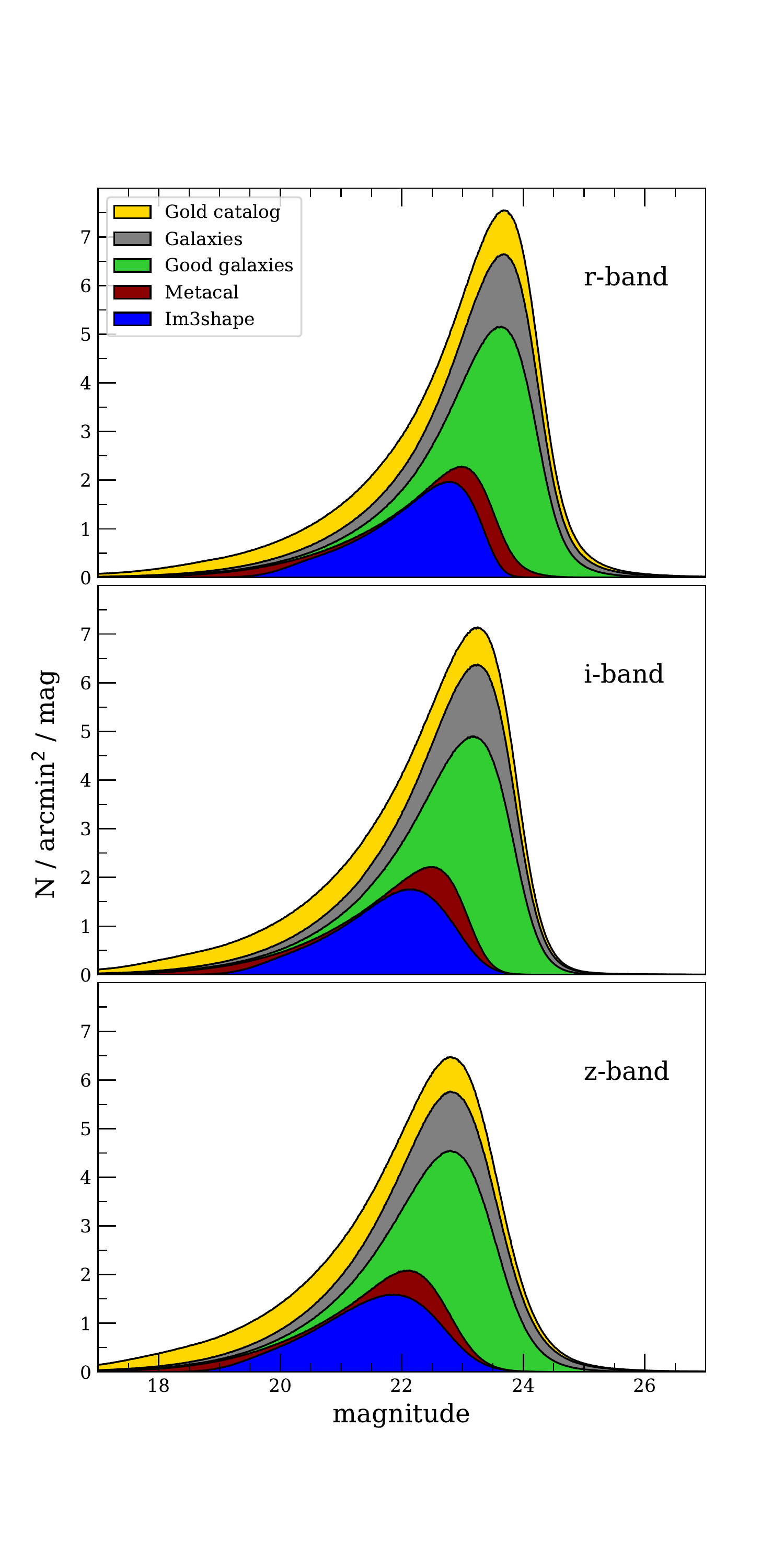}
\caption{Magnitude histograms showing different selections of the DES Y1 catalogues.  Values are measured with the multi-object fitting (MOF) method described in \citet{y1gold}. The {\sc Gold} catalogue is the input detection catalogue described in \S \ref{sec:gold}. ``Galaxies'' are those identified as galaxies by the process described in \S \ref{sec:galaxy selection}. ``Good'' galaxies are those with no indication of blending or extreme colours from \sex. The \metacal\ and \imshape\ histograms show objects in the final shape catalogues, after method-specific cuts.
\label{fig:maghist}
}
\end{figure}

Galaxies are distinguished from stars in {\sc Gold} using a classifier called {\sc Modest}, which is
based on the \sex~ \spreadmodel~variable \citep{BertinSExtractor1996,soumagnac15}, which discriminates between objects best fit as a 
point source vs. an extended object.  In this paper the \imshape~ selection cuts made use of
{\sc Modest}, in the high-purity variant described in \citet{y1gold}.  The \metacal~ catalogue and the PSF star selection used alternative criteria. The overall
magnitude distributions of the selections, and of the final shape catalogues, are shown in Figure~\ref{fig:maghist}.

Images within 30 pixels of the edge of a CCD are removed from the selection because of a ``glowing-edge''  effect which gives
pixels there a different effective size \citep{plazas}.

\subsection{Astrometry}
\label{sec:astrometry}

The DES Y1 single-epoch pipeline derives an astrometric solution for each exposure by comparing
object positions across the focal plane to the reference catalogue UCAC-4 \citep{ucac4} using the AstrOmatic utility
\scamp\ \citep{scamp1,scamp2}.  These solutions typically have 200-300 milliarcsecond RMS in their residuals with respect to the reference catalogue.  In order
to produce high quality co-added images in the multi-epoch pipeline, an astrometric refinement step
is used prior to combining the images.  That step considers catalogued objects with S/N > 10 from all
exposures (at all bands) that overlap the coadd tile.  A simultaneous astrometric fit is made, again
using \scamp\ but now using the 2MASS Point Source Catalog as an astrometric reference \citep{2mass-psc}.  The refined
astrometric solutions are used to update the world coordinate system (WCS) for each image prior to coaddition. The resulting
fits typically have an internal astrometric residual of 25-35 milliarcseconds (RMS) between the
individual images/exposures and an external astrometric residual of ~250 milliarcseconds with
respect to the 2MASS catalogue.

\subsection{COSMOS Data}
\label{sec:cosmos}

For several simulations and validation tests we make use of a galaxy catalogue from Advanced Camera for Surveys (ACS) imaging of HST's COSMOS field \citep{koekemoer07,scoville07}.
The catalogue of $\sim$73,000 objects has been ``whitened" (correlated noise removal; see \citealt{galsim2015}),
and is a deeper superset of the galaxies used in the GREAT3 challenge\footnote{http://great3.jb.man.ac.uk/}.
It extends significantly beyond the Y1 detection limit of $M_\mathrm{r,lim} =  23.4$ \citep{y1gold}, 
reaching $\sim 25.2$ mag in the HST F814W filter and $\sim27.9$ mag in the DES $r$-band. 

\subsection{Blinding}
\label{sec:blinding}

The DES Y1 shear catalogues were blinded to mitigate experimenter bias, in which analysis methodology may be intentionally or otherwise tuned so that results match expectations.  The 
blinded catalogues have all ellipticities $e$ as defined below in equation (\ref{eq:edef}) transformed via $|\eta| \equiv 2 \arctanh{|e|} \rightarrow f |\eta|$, with a hidden value $0.9<f<1.1.$  This mapping preserves the confinement of the $e$ values to the unit disc while rescaling all inferred shears.
DES cosmological analyses making use of these catalogues finalized their analysis methodology before being supplied with the unblinded catalogues.  Cosmological parameter estimation for these projects incorporate further secondary blinding strategies as described in their respective papers.

\changedtext{In the interests of full discolsure we must report that two independent but equivalent errors in the two shape pipelines meant that the multiplicative calibration process was incorrectly applied after the blinding process instead of before, partially undoing its effects.  The transformation described above is not a purely multiplicative one, since it acts on $\arctanh{|e|}$ instead of $e$, but for small ellipticities it is nearly so.  Since the calibration removed a multiplicative bias, This meant that most of the effect of blinding was undone by the calibration process.  Since the mistakes were equivalent the two catalogs remained consistent after blinding, and no errors were caused in any tests or comparisons.  

This fact was discovered during the cosmological analysis, but after the catalogues had been frozen and the tests presented in this paper finalized.  The error was not disclosed to the full analysis team, so most members remained effectively blinded.  Additionally, the individual cosmology analyses in \citet{shearcorr} and \citet{keypaper} included another layer of blinding: all cosmology constraint plots included shifts in the positions of the DES results, so that the absolute position could not be compared to existing results or preconceived expectations.  While these errors could not therefore have resulted in any experimenter bias being possible, they will be corrected in the next DES analysis.
}
\section{PSF Estimation}
\label{sec:psf}

One of the most important aspects of image characterization is accurately estimating the point-spread
function.  The PSF describes how a point source of light in the sky is mapped into a two-dimensional
profile on the image.  The images of galaxies are the convolution of the true surface brightness profile
with the PSF.

Since stars are essentially point sources, observations of stars give us a direct (albeit
noisy) estimate of the PSF at the locations of the stars.  However, the PSF is not constant across the
field of view, so the PSF must be interpolated from
the locations of stars, where it is observed, to the locations of galaxies, where it is needed.

The process for PSF estimation in Y1 is largely unchanged from the procedure used in \citetalias{jarvis16}.
We briefly recap the procedure described therein, emphasizing the changes we have made since SV.

\begin{figure}
\includegraphics[width=\columnwidth]{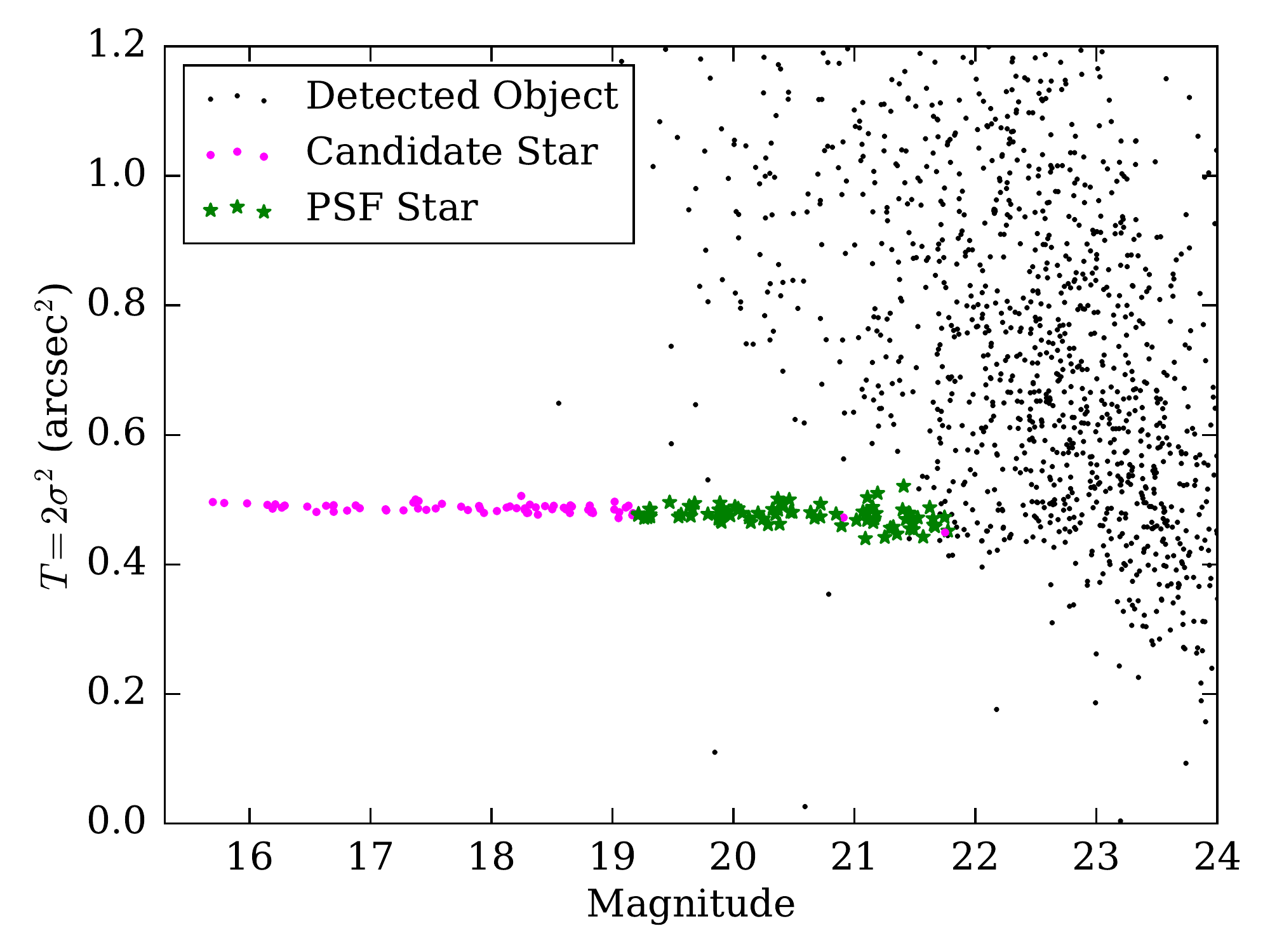}
\caption{An example size-magnitude diagram for a single CCD image,
used to identify stars.
\label{fig:psf:findstars}
}
\end{figure}

\subsection{Selection of PSF Stars}
\label{sec:psf:cuts}

We use the same method for identifying and selecting PSF stars as \citetalias{jarvis16}.  The initial
identification of candidate PSF stars involved
using a size--magnitude diagram of all the objects detected on the image. For the magnitude, we used
the \sex\ measurement \magauto. For the size, we use the scale size, $\sigma$,
of the best-fitting elliptical Gaussian profile using the adaptive moments algorithm HSM \citep{hsm}.

The stars are easily identified at bright magnitudes as a locus of points with size
nearly independent of magnitude. The galaxies have a range of sizes, all larger than the PSF size.
The candidate PSF stars are taken to be this locus of objects from about $m \approx\ 15,$ where the objects
begin to saturate, down to $m \approx\ 22,$ where the stellar locus merges with the locus of faint, small
galaxies.

From this list of candidate stars, we remove objects that are not suitable to use as models of the PSF.
Most importantly, we remove all objects within 3 magnitudes of the faintest saturated star in the same CCD exposure
in order to avoid
stars whose profiles are affected by the so-called ``brighter-fatter effect'' \citep{Antilogus14, Guyonnet15,
Gruen15} - see \S \ref{sec:psf:measurement}.  This magnitude cutoff varies between $18$ and $19.5$.

In addition, we remove stars that overlap the ``tape bumps''.  The CCDs on DECam each have
six spots where 2 mm $\times$ 2 mm $\times 100\,\mu$m-thick spacers were placed behind the CCDs when 
they were glued to their
carriers \citep[see][]{Flaugher15}.  This alters the electric field and hence the PSF is distorted near each  spacer.
Figure~\ref{fig:psf:findstars} shows such a size-magnitude diagram for a representative CCD image. The stellar locus is easily
identified by eye, and the stellar sample identified by our algorithm is marked in pink and green. The pink
points are stars that are removed by our various selection cuts, while the green points
are the stars that survive these cuts. 

\begin{figure}
\center\includegraphics[width=\columnwidth]{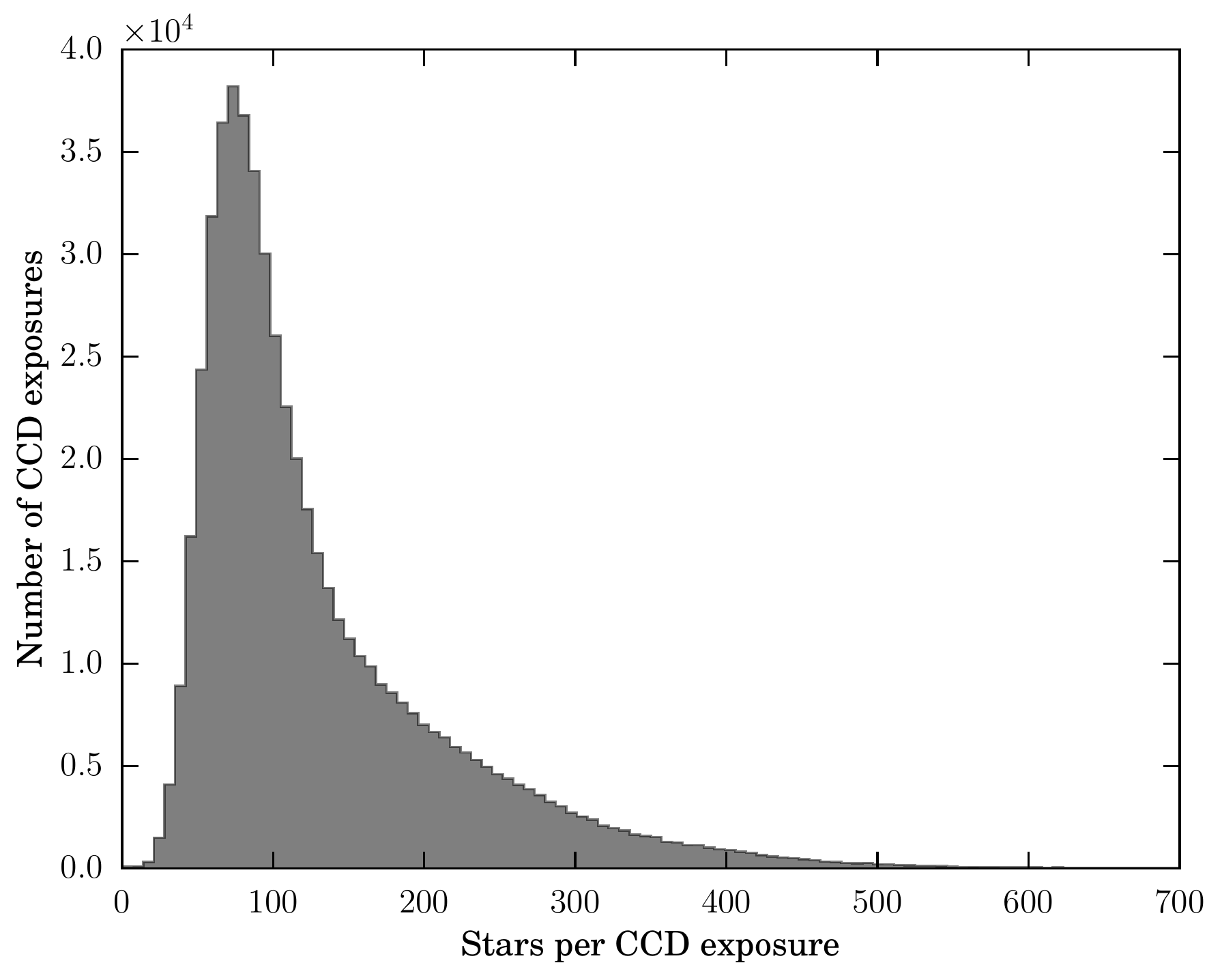}
\caption{The distribution of the numbers of stars used to constrain the PSF model per CCD image.
\label{fig:psf:nstars}
}
\end{figure}

\begin{figure}
\center\includegraphics[width=\columnwidth]{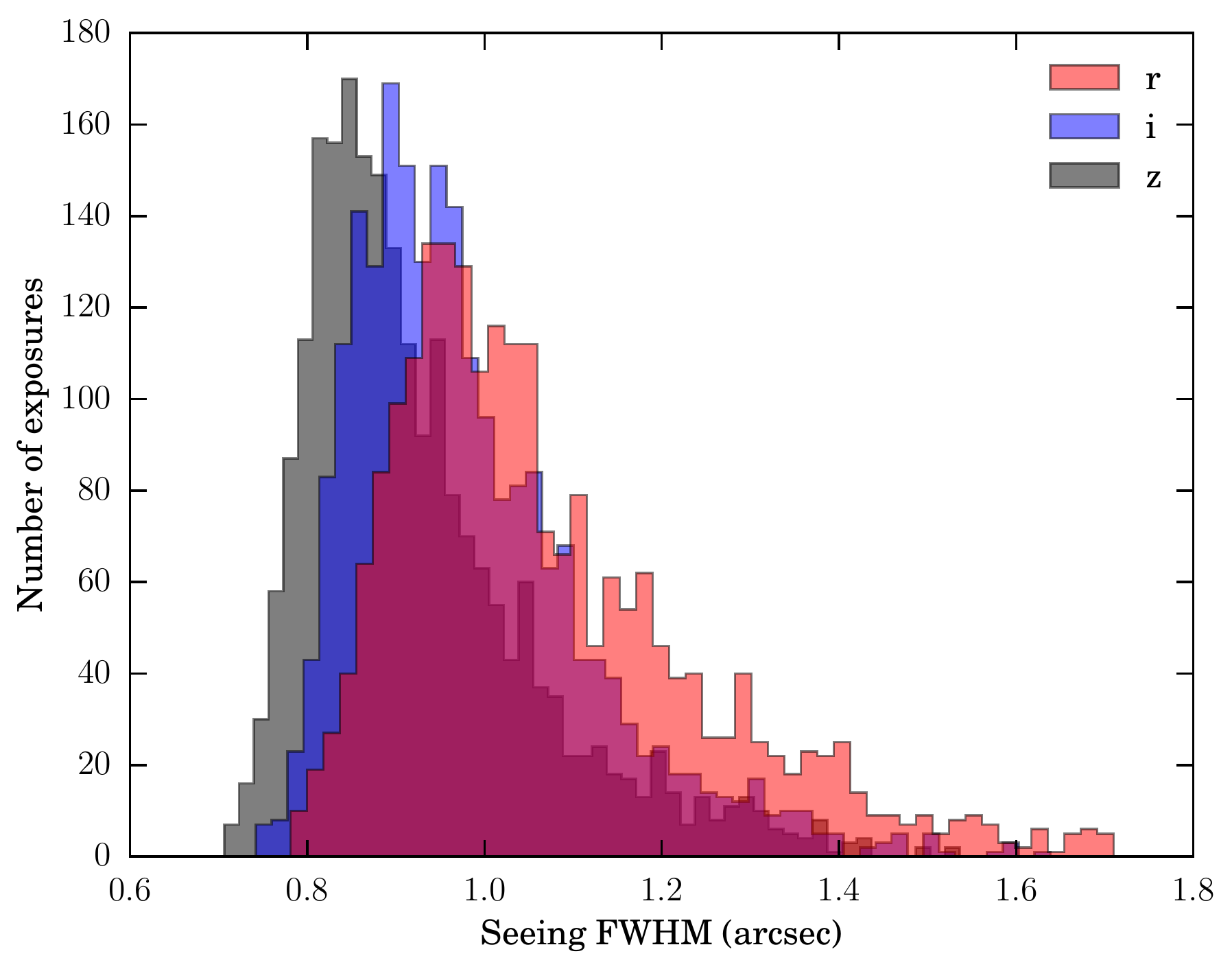}
\caption{The distribution of the median seeing FWHM of the stars used to model the PSF in the $riz$-bands. The median seeing of these distributions is 1\farcs03 in the $r$-band, 0\farcs95 in the $i$-band, 0\farcs89 in the $z$-band, and 0\farcs96 in the three bands overall.
\label{fig:psf:seeing}
}
\end{figure}

We find a median of 115 useful stars per CCD image, which we use to
constrain the PSF model. The distribution of PSF stars per CCD exposure is shown in Figure~\ref{fig:psf:nstars}.
In Figure~\ref{fig:psf:seeing}, we show the distribution of the median measured full-width half-maximum (FWHM) for the PSF stars used in our
study, restricted to the exposures used for shear measurements. The overall median seeing is 0\farcs96, which
is significantly better than we obtained in the SV observations (1\farcs08), but still somewhat larger than the
original target of 0\farcs90.

Occasionally, this process for selecting stars fails, in which case we add the CCD's image to
a ``blacklist'' of those not used for shear estimation.  For instance, if fewer than 20 stars 
are identified as PSF stars (e.g. because there is a very bright star or galaxy dominating a large fraction of 
the CCD area), then we blacklist the CCD image.  Sometimes the star-finding algorithm finds the wrong
stellar locus and ends up with far too many ``stars'' or finds a very large FWHM ($> 3\farcs6$).  These
CCDs are similarly excluded from consideration.
These PSF blacklist entries are added to the {\sc Gold}-catalogue blacklist, which includes 
CCDs with large ghosts, scattered light, satellite trails, or other apparent defects \citep{y1gold}.

\subsection{PSF Measurement and Interpolation}
\label{sec:psf:measurement}

We used the software package \psfex\ \citep{BertinPSFEx2011} to measure the surface brightness profile $I(x,y)$
of the PSF stars selected above as well as to interpolate between the locations of the stars.  We used
the following configuration parameters for \psfex:
\begin{verbatim}
BASIS_TYPE      PIXEL
PSF_SAMPLING    0.5
PSF_SIZE        101,101
PSFVAR_KEYS     XWIN_IMAGE,YWIN_IMAGE
PSFVAR_GROUPS   1,1
PSFVAR_DEGREES  2
\end{verbatim}

The one change from the procedure described in \citetalias{jarvis16} is to switch the
\code{BASIS\_TYPE} from \code{PIXEL\_AUTO} to \code{PIXEL}.
With \code{PIXEL\_AUTO}, there was an overall mean residual in the
size of the PSF models compared to the measured sizes of the stars.  Switching to \code{PIXEL}
yields near-zero size residual for faint stars (i.e. those unaffected by the brighter-fatter effect).

\begin{figure}
\includegraphics[width=\columnwidth]{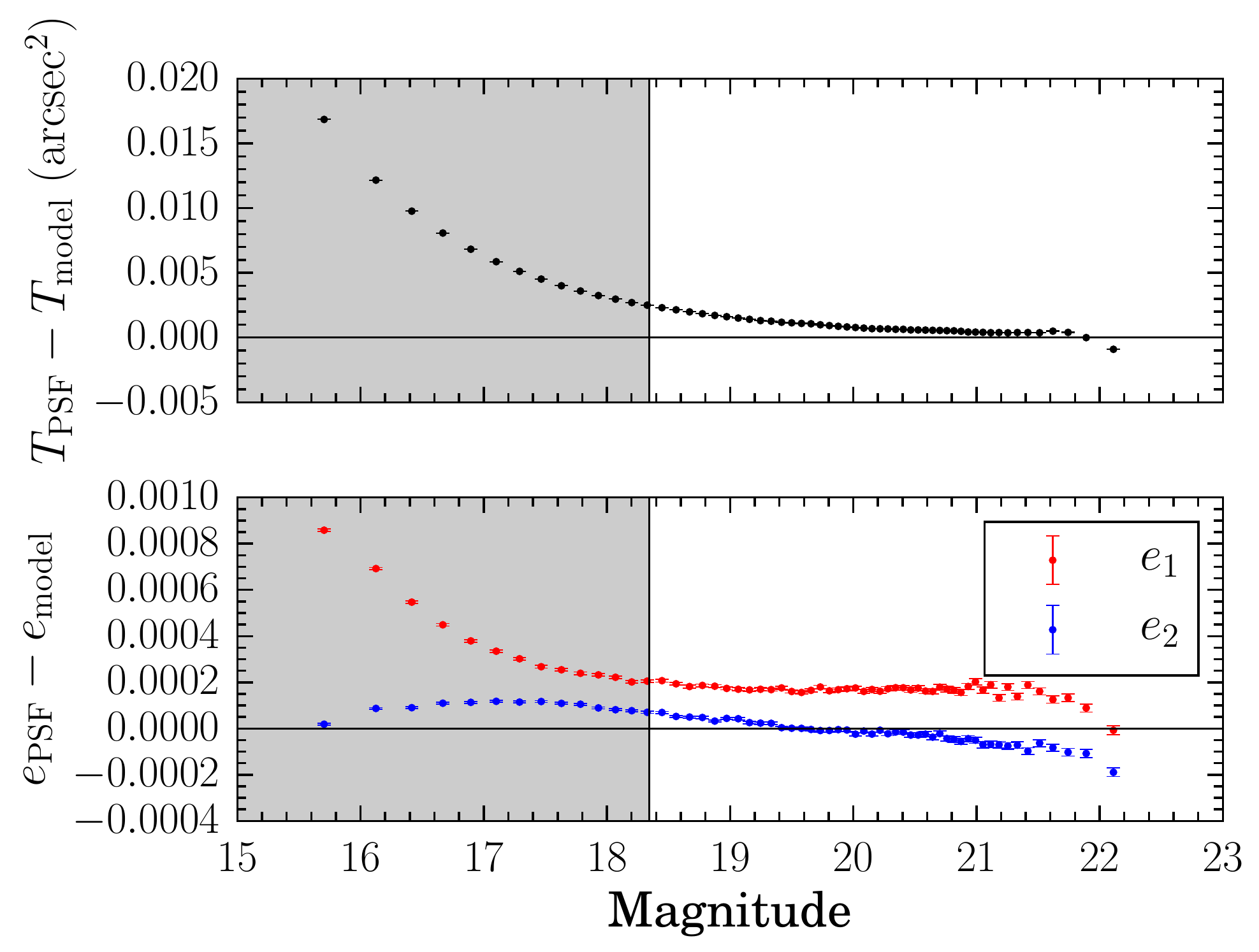}
\caption{The relative model size (top) and shape (bottom) residual of stars. 
To reduce the impact of the brighter-fatter effect bright stars are excluded from our PSF models;
the cut-off varies between CCD exposures but the shaded region shows a typical example.
\label{fig:psf:bfe}
}
\end{figure}

In Figure~\ref{fig:psf:bfe} we show size and shape residuals of all identified stars, relative to our standard \psfex\ model, which uses only the faint ones. The sizes and shapes are
defined in terms of the second moments of the surface brightness profile \citep{Seitz97}:
\begin{align}
T &= I_{xx} + I_{yy} \label{eq:Tdef} \\
e &= e_1 + i e_2 = \frac{I_{xx} - I_{yy} + 2i I_{xy}}{I_{xx} + I_{yy} + 2\sqrt{I_{xx}I_{yy} - I_{xy}^2}} \label{eq:edef}
\end{align}
where the moments are defined as
\be
I_{\mu\nu} = \frac{\int \mathrm{d}x \mathrm{d}y I(x,y) (\mu-\bar\mu)(\nu-\bar\nu)}{\int \mathrm{d}x\mathrm{d}y I(x,y)}.\ee
The moments are measured using HSM \citep{hsm}.  The quantity $T$ is one measurement of the square of the object radius.

The brighter-fatter effect is seen at bright magnitudes to lead to biases in both the size and shape 
(especially $e_1$).  This motivates the cut described above in \S \ref{sec:psf:cuts} and shown in the shaded region.
There is a small residual error in $e_1$ even at the faintest magnitudes, and we are unable to find settings to \psfex\ that eliminate this bias.  However, the size
residual is now seen to be consistent with zero at faint magnitudes, which was not the case for the
SV analysis.

\begin{figure*}
\includegraphics[width=\textwidth]{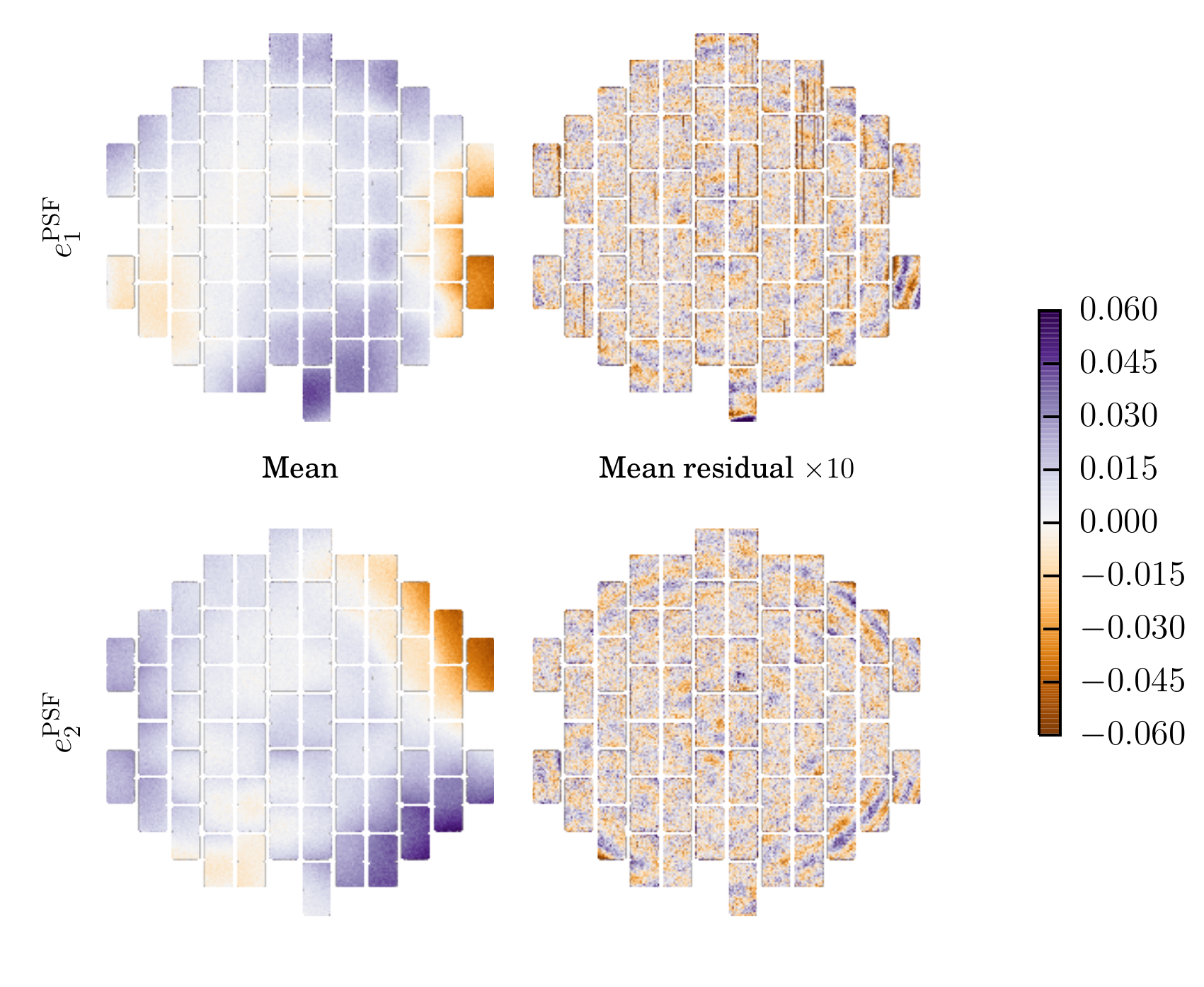}
\caption{The mean PSF ellipticity (left) and mean residual after subtracting the \psfex~ model ellipticity (right), binned by position in the focal plane.
The residual is multiplied by a factor of 10 to be visible in the same colour scale.
\label{fig:psf:fov}
}
\end{figure*}

In Figure~\ref{fig:psf:fov}, we show both the raw PSF shape and the residuals as a function of position on the
focal plane. The residuals are small, but not quite zero, and there is an evident rippling pattern.
The impact of these spatially correlated residuals is investigated below in \S\ref{sec:psf:rho}.

\subsection{PSF Model Diagnostics}
\label{sec:psf:rho}

\begin{figure*}
\includegraphics[width=0.48\textwidth]{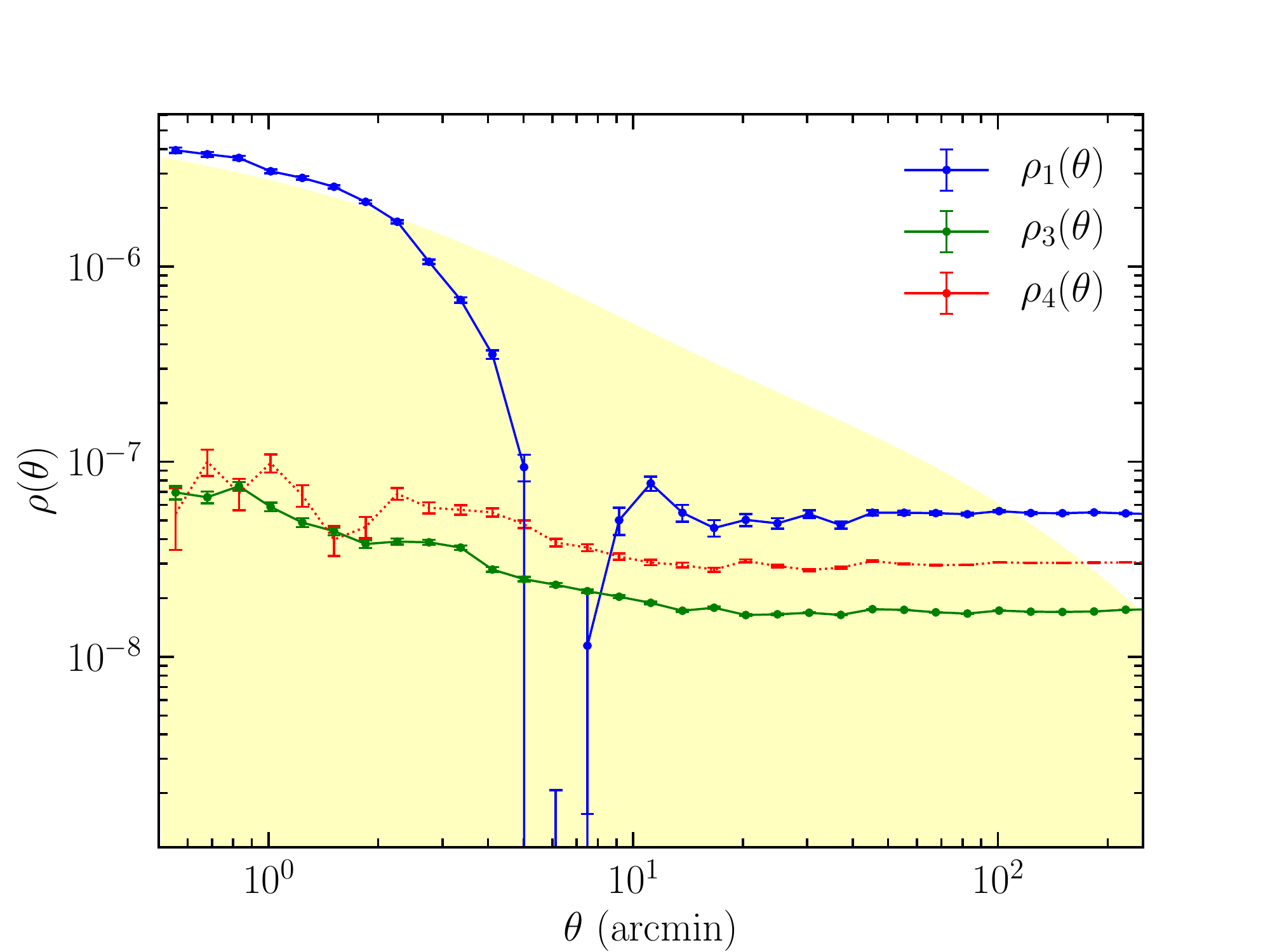}
\includegraphics[width=0.48\textwidth]{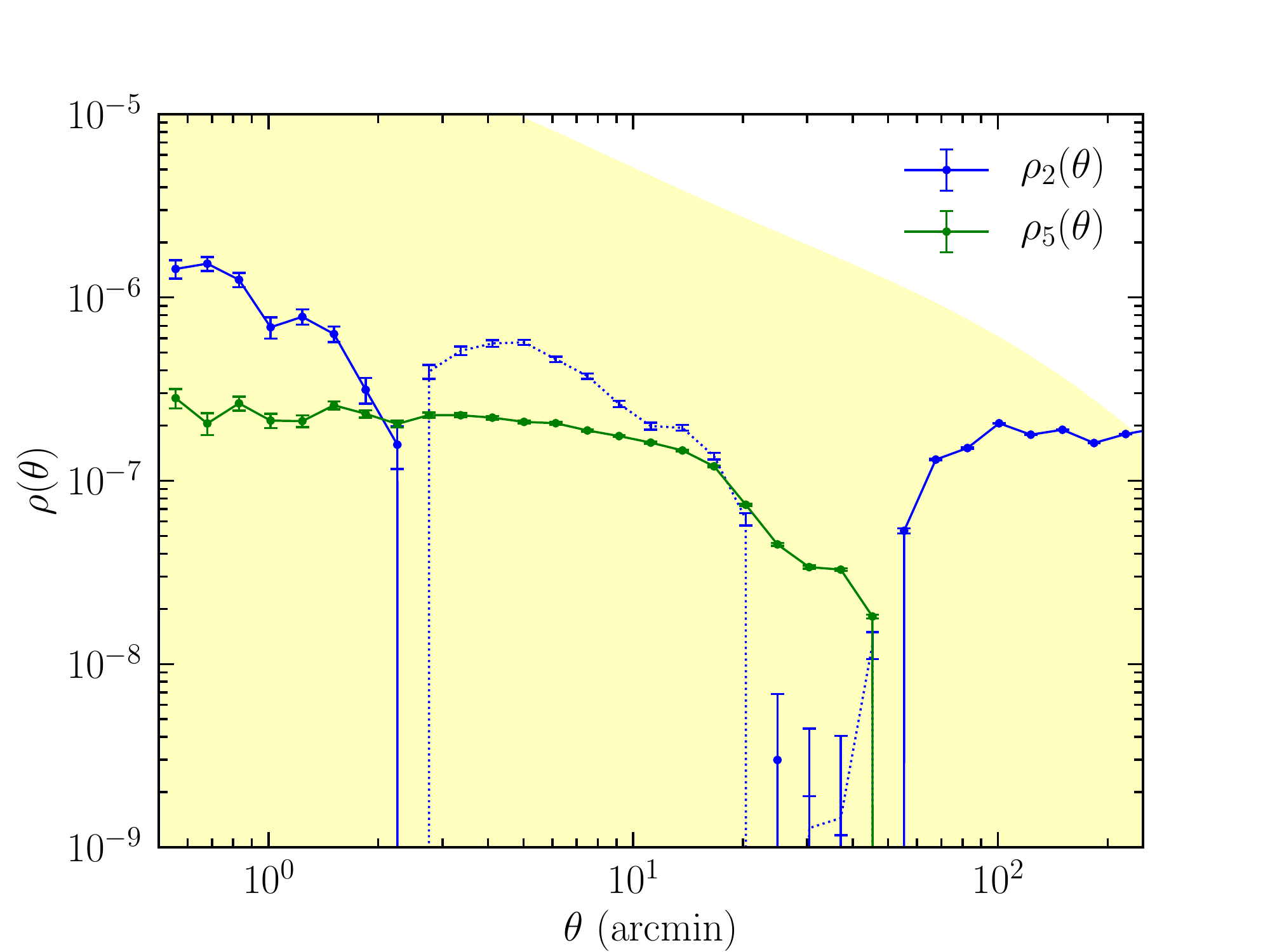}
\caption{The $\rho$ statistics for the PSF shape residuals.
Negative values are shown in absolute value as dotted lines.
Requirements on the $\rho$ statistics are specific to individual science cases; the yellow fill is a general guide, rather than a requirement, and is ten percent of the value of the weakest cosmic shear $\xi_+$ signal, which is from the lowest redshift tomographic bin (for this bin only scales above $\theta \approx 7$ arcmin were used in the analyses in \citealt{shearcorr} and \citealt{keypaper}).  It pessimistically assumes $\alpha=0.1$ and $\Tpsf/\Tgal=1$.  Contributions to the signal from the flat regimes at large scales will be absorbed by the marginalization over the mean shear discussed in \S\ref{sec:tests:meanshear}.
\label{fig:psf:rho}
}
\end{figure*}

The errors in the PSF model seen in Figure~\ref{fig:psf:fov} will propagate
into the galaxy shapes and inferred lensing shear.  
To estimate the impact of PSF modelling errors on the shear two-point correlation function,
$\xi_+$ we turn to the $\rho$ statistics \citep{Rowe10,jarvis16}:
\begin{align}
\label{eq:req:rho1def}
\rho_1(\theta) &\equiv \left \langle \delta \epsf^*(\bfx) \delta \epsf(\bfxpt) \right \rangle \\
\label{eq:req:rho2def}
\rho_2(\theta) &\equiv \left \langle \epsf^*(\bfx) \delta \epsf(\bfxpt) \right \rangle \\
\label{eq:req:rho3def}
\rho_3(\theta) &\equiv \left \langle \left(\epsf^* \frac{\delta\Tpsf}{\Tpsf}\right)\!(\bfx)
   \left(\epsf \frac{\delta\Tpsf}{\Tpsf}\right)\!(\bfxpt) \right \rangle \\
\label{eq:req:rho4def}
\rho_4(\theta) &\equiv \left \langle \delta\epsf^*(\bfx)
   \left(\epsf \frac{\delta\Tpsf}{\Tpsf}\right)\!(\bfxpt) \right \rangle \\
\label{eq:req:rho5def}
\rho_5(\theta) &\equiv \left \langle \epsf^*(\bfx)
   \left(\epsf \frac{\delta\Tpsf}{\Tpsf}\right)\!(\bfxpt) \right \rangle,
\end{align}
where $\epsf$ and $\delta\epsf\equiv \epsf - e_\mathrm{model}$ are the measured ellipticity of the
PSF model at the locations of stars and its measured residual, respectively;
$\Tpsf$ and $\delta\Tpsf \equiv \Tpsf - T_\mathrm{model}$ are the size of the model and its residual; 
the asterisk denotes complex conjugation; 
and the averages are taken over pairs of stars separated by angle ${\boldsymbol \theta}$.  These statistics
neglect anisotropy in PSF errors, but will indicate the first order effects on the correlation functions. There
is no equivalent effect on $\xi_-$, where such additive effects are negligible.

The values $\delta \Tpsf / \Tpsf$ as measured from sizes of reserved stars are typically positive, meaning stars are slightly larger than smooth polynomial PSF models predict.  In DES data we find mean size errors from this effect $\langle \delta \Tpsf / \Tpsf \rangle \sim 8.3 \times 10^{-4}$
and a much larger size variance:
$\langle (\delta \Tpsf / \Tpsf)^2 \rangle^{1/2} \sim 3 \times 10^{-2}$.

For these tests, we constructed PSF models using only 80\% of the PSF stars that were selected
as described in \S\ref{sec:psf:cuts}.  The PSF model was then interpolated to the locations
of the other 20\% of the stars that had been reserved from the modelling step.  This is an
improvement over the procedure used by \citetalias{jarvis16} where the same stars that were used
for making the PSF model were used in the $\rho$ statistics.
The statistics are shown, binned by $|{\boldsymbol \theta}|$, in Figure~\ref{fig:psf:rho} averaged over single-epoch
stellar observations in \emph{r, i} and \emph{z} bands.  The averages thus include pairs
of observations from different exposures as well as those from the
same exposure, thus corresponding to the way these residuals impact the two-point shear
correlation of the shear estimates of the galaxies.

The $\rho$ statistics for individual exposures in Y1 are similar to
those obtained for SV in \citetalias{jarvis16}.
The SV data, however, have a mean of 19.7 usable
exposures per galaxy in the \emph{riz} bands, while the Y1 data presented here have a mean of
only 8.4 exposures. 
When the modelling errors are uncorrelated between exposures for a
given star or galaxy target, the survey-averaged statistics scale as 
$\rho \propto 1/N_{\rm exposures}$.
As such, the amplitude of the $\rho_1$ statistic is significantly larger
for Y1 than reported for SV in \citetalias{jarvis16}.  In addition, the fact that we are using reserved stars
this time also increased the measured correlations somewhat compared to SV, especially at large scales. In the
SV statistics the mean residual was close to zero by construction, so the statistics were probably spuriously low.
The mean residual of the reserved stars is expected to be a better estimate of the actual error in the fitted PSF models.

The PSF modelling residuals constitute the largest known additive systematic error on the estimated shear values.
For two-point shear statistics such as $\xi_+$, we expect the additive error due to these statistics
to be \citepalias[eqn. 3.17]{jarvis16}:
\begin{align}
\delta \xi_+(\theta) &= 
 \left\langle\frac{\Tpsf}{\Tgal} \right\rangle^2 \left(\rho_1(\theta) + \rho_3(\theta) + \rho_4(\theta) \right)
    \nonumber\\& \quad
- \alpha \left\langle\frac{\Tpsf}{\Tgal} \right\rangle \left( \rho_2(\theta) + \rho_5(\theta) \right)
\label{eq:req:psfmodel}
\end{align}
where $\alpha$ is the amount of ``leakage'' of the PSF shape into the galaxy shape (see \S\ref{sec:tests:alpha}). We discuss the impact of this contribution to $\xi_+$ further in \citet{shearcorr}. 
For other analyses of these data that are sensitive to additive errors, we also recommend explicitly accounting for the potential impact of additive systematics due to the PSF model residual.

\section{The \metacal\ Catalogue}
\label{sec:metacal}
\subsection{\metacal\ Overview}
\label{sec:metacal:overview}

Our primary catalogue uses \emph{metacalibration}, a new method for shear measurement that derives shear calibrations
directly from the available imaging data.  Metacalibration is decribed in detail in \citet{HuffMandelbaum2017} and \citet{SheldonHuff2017}, hereafter \citetalias{SheldonHuff2017}. 

The principle behind \metacal\ is to measure the response of a shear estimator
\vest\ to shear.  Unlike in most methods this response is not estimated from a suite of 
simulated images, but rather calculated directly for each observed image, using the scheme described below. 

 Any estimator that has sensitivity to
shear can be used with metacalibration, and here we use measurements of galaxy ellipticity.  For small shears, ellipticity estimators
can be written as a Taylor expansion:
\begin{align} \label{eq:Eexpand}
    \vest &= \vest|_{\gamma=0} + \frac{ \partial \vest }{ \partial \vecg}\bigg|_{\gamma=0} \vecg  + ... \nonumber \\
          &\equiv \vest|_{\gamma=0} + \mbox{\mcalRg}\vecg  + ...
\end{align}
where we have defined the shear response matrix \mcalRg.  The shear response is
calculated by artificially shearing the images and re-measuring the
ellipticity.  \changedtext{We do this by directly deconvolving the PSF (by dividing the Discrete Fourier Transform (DFT) of the image by the DFT of the PSF image), applying a shear, and then
reconvolving by a symmetrized version of the PSF (the latter steps using the \galsim\ package, \citealp{galsim2015})}.  We then form a numerical derivative: for a given
element of the response matrix, we calculate
\begin{equation} \label{eq:Rnum}
    \mbox{\mcalRg}_{i,j} = \frac{\est_i^+ - \est_i^-}{\Delta \gamma_j},
\end{equation}
where $\est^+_i$ is the measurement of component $i$ made on an image sheared by $+\gamma_j$,
$\est^-_i$ is the measurement made on an image sheared by $-\gamma_j$, and $\Delta
\gamma_j = 2\gamma_j$.  We used an applied shear $\gamma_j=0.01$.

When measuring a shear statistic, such as mean shear or a shear two-point
function, these responses can be averaged appropriately to produce a calibrated
result.   For the example of mean shear, we can take the expectation value
of equation (\ref{eq:Eexpand}). Keeping terms to first order
in the shear, and assuming the mean ellipticity is zero in
the absence of shear, we find
\begin{align}
    \langle \vest \rangle &= \langle \vest \rangle |_{\gamma=0} + \langle \mbox{\mcalRg} \vecg \rangle
                          \approx \langle \mbox{\mcalRg} \vecg \rangle,
\end{align}
With estimates of \mcalRg\ for each galaxy, we can form a weighted average:
\begin{align} \label{eq:rcorr}
    \langle \vecg \rangle_w
    & \approx \langle \mbox{\mcalRg} \rangle^{-1} \langle \mcalRg \vecg \rangle
    \approx \langle \mbox{\mcalRg} \rangle^{-1}  \langle \vest \rangle 
    ,
\end{align}
where the subscript $w$ implies this is a weighted average over the true
shears.
The generic correction for two-point functions was also derived in
\citetalias{SheldonHuff2017} \changedtext{as
\begin{equation}
  \xi = (\langle R^{\alpha}\rangle \langle R^{\beta} \rangle)^{-1} \langle \mathbf{e}^{\alpha} \mathbf{e}^{\beta} \rangle
\end{equation}
for two samples of objects (e.g. tomographic bins) $\alpha$ and $\beta$ where $\langle \mathbf{e}^{\alpha} \mathbf{e}^{\beta} \rangle$ is a standard two point correlation function estimate.}
The application of this method to other specific statistics
should be worked out carefully, as the details of the
averaging are important.

We can also correct for selection effects, for example shear biases that
may occur when placing a cut on signal-to-noise ratio \snr.
This is accomplished by measuring the mean response of the estimator to the selection,
repeating the selections on quantities measured on
sheared images.  Again taking the example of mean shear, a given element
of the mean selection response matrix \mcalRSmean\ is
\begin{align} \label{eq:RSmean}
    \mbox{\mcalRSmean}_{i,j} &\approx
    \frac{\langle \est_i \rangle^{S+} - \langle \est_i \rangle^{S-}}{\Delta \gamma_j},
\end{align}
where $\langle \vest \rangle^{S+}$ and $\langle \vest \rangle^{S-}$ represent the means of 
ellipticities measured on images without artificial shearing, but with selection based on parameters
from positively and negatively sheared images respectively.
The full response for the mean shear is then given by the sum
of the shear response and selection response
\begin{align} \label{eq:Rmean}
    \mbox{\mcalRmean} &= \langle \mbox{\mcalRg} \rangle + \langle \mbox{\mcalRS} \rangle.
\end{align}

\changedtext{For the ellipticity estimators used here we have found that the response
matrix $R$ is on average diagonal, and that $R_{11} \approx R_{22}$, so that a single
scalar value characterizes the response.}

\metacal\ was tested using an extensive set of simulations, and proved to be
unbiased for galaxy images with realistic properties matching the deep COSMOS
data, and noise and PSFs similar to DES data \citepalias{SheldonHuff2017}.
Furthermore, \metacal\ was shown to be robust to the presence of stars in the
sample if the PSF is well determined.  There are additional challenges for real
data, which we will discuss below.

\subsection{\metacal\ in DES Y1} 
\label{sec:metacalcat}

For DES we ran \metacal\ in a mode similar to that used in
\citetalias{SheldonHuff2017}, using the \metacal\ implementation available in the
\ngmix\ software package\footnote{The ngmix package is freely available at
\url{https://github.com/esheldon/ngmix}} \citep{ngmix}.  We used an estimate of
each object's ellipticity as the basis for shear estimation, to be calibrated
using \metacal.  The total time for a run of \metacal\ on DES Y1 data was about 
150,000 CPU hours for the full set of 139M detected objects. The 
calculations were performed using computational resources
at the SLAC National Accelerator Laboratory. 

To determine the ellipticity, we fit the images associated with each object to
a simple parametric model using the \ngmix\ code. For efficiency reasons, we
chose a single Gaussian to model the object, convolved by a model of the PSF.
As described above, we model the PSF in each image using the \psfex\ code. We
then reconstructed an image of the PSF at the location of each object from \psfex\
output using a separate package\footnote{\url{https://github.com/esheldon/psfex}}.

This full PSF image was used for the devonvolution step in the metacalibration
process.  For the shape fitting stage itself we represented the PSF as a single
Gaussian, for efficiency reasons.   A Gaussian is not a good description of the
DES PSF, but \citetalias{SheldonHuff2017} find that this does not limit our
ability to calibrate the shear estimate, because the response accounts
for any mismatch between the actual PSF and the model used for fitting.

The full model for each galaxy image was the analytic convolution of the object
Gaussian with the Gaussian representation of the PSF.  We then found the
parameters that maximized the likelihood, as calculated across all available
imaging epochs and bands $r,i,z$.

We simultaneously fit images from all available observing epochs, and all
available bandpasses, with a free flux in each band.  
In general the r-band is the most powerful band and gives the most shape signal, though the others do add significant information.  The total Gaussian model
has five structural parameters, shared between all bandpasses, plus a free flux
for each bandpass. The structural parameters are two for the center in sky
coordinates, two for ellipticity components, and one for the size $T$ 
as in equation (\ref{eq:Tdef}) (equal to the trace of the Gaussian's covariance matrix).
These quantities were measured for the unsheared
as well as artificially sheared images discussed in \S \ref{sec:metacal}.

We applied priors on all model parameters.   These priors were uninformative,
except for the prior on ellipticity, which we found necessary to provide a
stable fit for faint objects: we used the isotropic unlensed distribution presented in
\cite[][equation 24]{bernsteinarmstrong}, with $\sigma=0.3$. The details of this prior are not important,
because \metacal\  can accurately calibrate
this shear estimator as long as the fitting is stable \citepalias{SheldonHuff2017}.


Real data present significant challenges that were not tested in the
simulations of \citetalias{SheldonHuff2017}.  For this work we tested the
following additional issues:
\begin{enumerate}
    \item \label{mitem:me} Shear estimation using multi-epoch data.
    \item \label{mitem:deblending} Effects of neighbouring objects.
    \item \label{mitem:psf} PSF modelling and interpolation errors.
\end{enumerate}
We show tests using simulations for \ref{mitem:me} in \S \ref{sec:mcal:me}, for
\ref{mitem:deblending} in \S \ref{sec:mcal:deblending}, and for \ref{mitem:psf} in \S
\ref{sec:mcal:wrongpsf}.  We also show tests of \ref{mitem:deblending} using real
data in \S \ref{sec:budget:metacal}.  The behaviours of the response functions $R\gamma$ 
and $R_s$ are described in \S \ref{sec:mcal:response-example}.

\subsection{\metacal\ and Multi Epoch Data} \label{sec:mcal:me}

\metacal\ was tested in \citetalias{SheldonHuff2017} in the simple case where each
object is observed once.  However, in DES we simultaneously fit to 
images from multiple observing epochs in each of multiple bands.
Each object was thus imaged in
different seeing and noise conditions, and was observed at different locations
within the focal plane.  Furthermore, galaxies do not in general have
the same morphology in every band.

We generated simulations to mimic this scenario using the \galsim\ simulation
package \citep{galsim2015}.  For each object we generated a set of 10 images
with different PSF, noise, and position offset within the image.  The PSF size
was drawn from a distribution similar to DES data, with mean FWHM$=0\farcs9$
(see Figure~\ref{fig:psf:seeing}).  The PSF ellipticity was drawn from a truncated
Gaussian with center $(0.0, 0.01)$ and width $(0.01,0.01$).  The noise was
varied by 10\%, and the object was offset randomly within a pixel.

The galaxy morphology was chosen to be the model used in
\citetalias{SheldonHuff2017}: a combination of exponential disk and \devauc\ bulge
profiles, with additional simulated knots of star formation.   Size and flux
distributions were matched to the 25.2 mag limited sample from COSMOS.  The
fraction of light in the disk was chosen uniformly from 0 to 1, and the
fraction of disk light in knots was also chosen uniformly from 0 to 1.  To
simulate morphological differences between bands we varied the flux and size of
the object in each image by 10\%, and assigned a random ellipticity.   The noise
was chosen such that the minimum \snr\ for the sample was approximately 5 in
the combined set of 10 images.  The \snr\ definition was that used in \galsim,
which is the same definition used in the GREAT3 simulations \citep[][equation
16]{great3}.  We applied a constant shear to each image of $(0.02, 0.00)$.

We fit the images using the same code used for fitting DES data.  We applied a
selection similar to that in our data: $\snr > 10$ and size $T/\Tpsf > 0.5$,
where $\Tpsf$ is the size of the PSF determined from a Gaussian fit. We found
no multiplicative bias $m$ or additive bias $c$, with limits $m < 1 \times 10^{-3}$ (95\%)
and $c < 2 \times 10^{-5}$ (95\%).  We also applied various threshold
and range cuts in \snr, flux, and size and again found no bias.

\subsection{\metacal\ and Blending} \label{sec:mcal:deblending}

In this section we explore the effect of blending on \metacal\ shear
estimation.

In principle, \metacal\ will correctly infer the shear applied even to complex
blends of multiple objects.  However, even if the calculated shear
response is accurate, there will be uncertainty in the {\em interpretation} of
the inferred shear for overlapping objects.  Consider first a scenario where
two overlapping objects are at the same redshift.  The same lensing shear has
been applied to both,  as well as to the apparent separation of the pair, so
the entire scene has been transformed with a single shear and the \metacal\
shear inference should be an unbiased estimate of this shear as long as the
model fitting process is stable.

Now consider the case that the two objects are at very different redshifts.  In
this case, the two sources have been sheared by different amounts, and the
separation between the pair depends on the different deflections applied to the
light from each object.  To facilitate a correct interpretation, a detection
algorithm must determine that there are two objects present, and a deblending
algorithm must accurately assign a fraction of the light in each pixel to each
of the blended objects.  Then the response must be measured separately for each
object based on this flux assignment. 

This flux assignment can in principle be done given some a priori assumptions
on the morphology allowed for each galaxy, but can be exceedingly difficult
when galaxies are highly irregular or heavily overlapped.  In addition,
the response of the deblending algorithm to shear can cause selection effects
that may be significant.

In DES we have utilized two different approaches to dealing with blends.  The
first, and simplest, is the \uberseg\ algorithm \citepalias{jarvis16}, which
masks pixels close to neighbours rather than assigning a fraction of the light
in each pixel to them.  

We have also developed a second algorithm, called multi-object fitting
\citep[MOF,][]{y1gold}, which does attempt to assign a fraction of the light
in each pixel to sets of blended objects.  In brief, MOF finds groups of
overlapping galaxies using a friends-of-friends algorithm.  Within a group, MOF
sequentially applies forward-modelling of simple bulge/disk models to each
source, subtracting the models for all other sources in the group. MOF fits use
all available imaging data in the $g,r,i,z$ bands.  Once convergence is
achieved, a fraction of the flux in each pixel can be assigned appropriately to
each object. The light of neighbours using the MOF models were subtracted off
when running \metacal, in addition to using the \uberseg\ masking. Note the
\metacal\ fitting only uses the MOF models to subtract neighbours.

We suspect that the \uberseg+MOF method may be more accurate than
\uberseg-only, based on the performance in simulations (see \S
\ref{sec:mcal:deblending:sims}). However, we use the \uberseg-only measurements
in our fiducial catalogue, because \photoz\ measurements based on flux
measurements from the artificially sheared images used in \metacal\ were not
available for \uberseg+MOF at the time of writing.  These are required in order
to correct for selection effects associated with placing galaxies into redshift
bins, for example when studying shear correlations in tomographic bins.

We tested these two methods for flux assignment using both simulations and real
data, as described in detail in the following sections. In brief, we find
indications of relative shear biases at the 1-2 per cent level between our
fiducial \uberseg\ and the MOF method of flux assignment, consistent between
data and simulations.

\subsubsection{Deblending Tests in DES Data} \label{sec:mcal:deblending:data}

We compare tangential shear profiles around foreground galaxies, measured with
our two different treatments of near neighbours described in \S
\ref{sec:mcal:deblending}.  We use a simple ratio of the measurements as our
test.  As foreground galaxies we use the redMaGiC high density sample
\citep{rozo16} cut to the redshift range of $z=0.2\ldots0.4$.

To reduce shape noise in the shear ratio, we perform the measurements with an
identical source sample from the two catalogues.  Note that with \metacal\ we can
correct for selections as long as those selections are repeatable on quantities
measured in artificially sheared images.  We therefore apply the recommended
cuts on flags, signal-to-noise ratio and pre-seeing size described in \S \ref{sec:cats:flags}, 
demanding that they be met in both catalogues and correcting for
the selection bias induced by the joint selection criterion.  One selection in
which we do not strictly adhere to this policy is source photo-$z$ selection.
As our source sample, we use galaxies with a BPZ photometric redshift estimate
\citep{photoz} based on their MOF photometry of $\langle z\rangle$ of
0.4 and above.  While this could induce a small selection bias in our
measurement, we do not expect it to significantly differ between the two
catalogues, and we have confirmed that \emph{not} applying a photo-$z$ selection
makes the resulting estimate of bias more noisy but consistent with our
estimate. 

We measure the tangential shear profiles from both catalogues in a set of
jackknife resamplings of the full, matched source catalogue.  For each
jackknife resampling, we determine the best-fit re-scaling $r$ required to
bring the fiducial, \uberseg\ masked catalogue to the same amplitude as the MOF
subtracted catalogue by minimizing 
\begin{equation}
\chi^2=\sum_i \frac{(g_{\mathrm{t},i}^{\rm MOF-subtracted}-r\times
g_{\mathrm{t},i}^{\rm fiducial})^2}{\sigma^2_i} \; ,
\label{eqn:budget:mcalneighbours} 
\end{equation} 
where $g_{\mathrm{t},i}$ is the mean tangential shear measured in a set of
angular bins $i$, logarithmically spaced between 0.2 and 30~arcmin, from which
the tangential shear around redMaGiC random points was subtracted to remove
additive systematics, and $\sigma_i$ is the jackknife uncertainty of
$g_{\mathrm{t},i}^{\rm MOF-subtracted}$. While the two estimates of tangential
shear are highly correlated, this is meant to give an appropriate
\emph{relative} weighting to each radial bin. We test that swapping
$g_{\mathrm{t},i}^{\rm MOF-subtracted}$ and $g_{\mathrm{t},i}^{\rm
fiducial}$ in Equation \ref{eqn:budget:mcalneighbours} or variation of the binning
scheme does not change our constraints on $r$ significantly, except for the
inclusion of very small scales, on which fiducial shears fall off while MOF
shears continue on a power-law like profile, suggesting that MOF may be
correcting for effects of blending with the lens galaxies.

\begin{figure}
\includegraphics[width=\columnwidth]{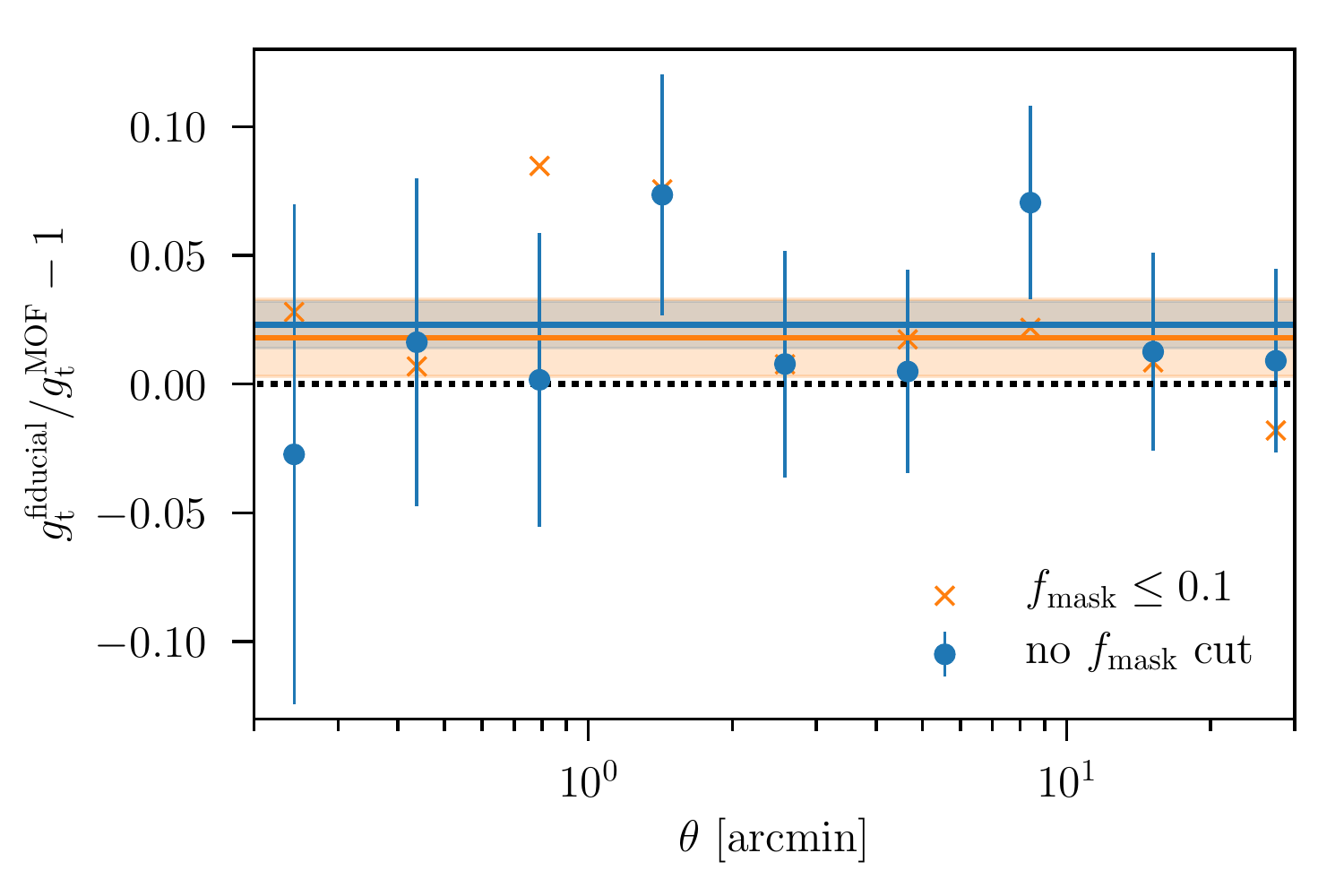}

\caption{Ratio of shears measured with our fiducial \metacal\ pipeline to
    shears measured after subtracting neighbouring galaxies with model from the
    multi-object fit, a proxy for the multiplicative shear bias in either method due to the
    presence of neighbours. Blue data points show measurements for the full
    sample, including jackknife error bars of the fiducial run. Orange data
    points show measurements that only use objects with less than 10\% of area
    in their postage stamp masked for missing data or \uberseg\ neighbours. Blue
    and orange horizontal lines and shaded regions are the best-fit $m$ and
    statistical $1\sigma$ uncertainties for both cases. 
    Marginalizing over an unknown multiplicative bias starts to dominate the 
    posterior uncertainty in our main science use cases at about 2\% uncertainty.
    }
    \label{fig:budget:metacalmof}

\end{figure}

These measurements are shown in \S \ref{fig:budget:metacalmof}. For a
measurement involving all objects (blue data points), we find a relative
multiplicative bias of the \uberseg\ shear catalogue of $m=1-r=0.023\pm0.009$
(blue shaded region).

If all of this difference is due to neighbour bias, we would expect this $m$ to
approach 0 as we limit the measurement to objects without many pixels
contaminated by neighbours.  We find that the ratio drops to $m=0.018\pm0.015$
among galaxies with little contamination, selected as those with less than 10\%
of area in their postage stamp masked for missing data or \uberseg\ neighbours
(orange data points and shaded region).  This ratio is no longer detected with high significance, but is consistent with both zero and the 0.023 ratio measured for the
full sample.

We will test the effects of neighbours further using
simulations in \S \ref{sec:mcal:deblending:sims}.
Note that there will also be an effect from objects that are below the detection
threshold, and thus not included in the catalogue.  Studies using image
simulations have shown that, when using simulations to calibrate the signal,
neighbouring galaxies can indeed have a significant effect on multiplicative
bias \citep{hoekstra15,hoekstra16,des_sim_2017}.  We test this
effect as well in \S\ref{sec:mcal:deblending:sims}.

\subsubsection{Deblending Tests in Simulations} \label{sec:mcal:deblending:sims}

We used simulations to test the effect of blended objects, as well as objects
fainter than the detection threshold. The motivation is to examine
the relative performance of \uberseg\ and \uberseg+MOF, not to
determine the numerical value of bias or to calibrate the shear
measurements.

The simulations are similar to those used in \S \ref{sec:mcal:me} in the type
of galaxy used, but include additional complexities. We used the same galaxy
models described in \S \ref{sec:mcal:me}, with a maximum magnitude for the
COSMOS catalogue of approximately 25.2.  In addition, we also added a lower
flux population by simply scaling the flux such that the faintest magnitude was
about 27.5, and scaled the sizes by a factor of 0.5.  We also added a small
number of big and high flux objects by scaling the sizes by a factor of 5 and
flux by a factor of 50.  Approximately 85\% of the objects were in the low flux
category, and 0.1\% were in the high flux category.  Half the objects were
sheared by 0.01 and half were sheared by 0.02, such that the mean shear was
close to 0.015.

All images were convolved by a PSF modelled as a Moffat profile
\citep{Moffat1969} with FWHM = 0.9 arcsec, and ellipticity 0.025 in the reduced
shear convention. We added noise appropriate for 5-year DES depths, such that
the 5-sigma detection limit was about magnitude 24.  Note that most of the
objects in these images are much fainter than the detection limit.  Also note
that these images are ``deeper'' than Y1 data, with higher galaxy density.
Thus, if the images matched real data exactly, we would expect the effects of
neighbours to be larger than in Y1 data.

We generated images similar to DES coadds.  We verified that the number density
of objects in the resulting \sex\ catalogue matched that expected for DES data.
We placed objects in the images randomly, with no spatial correlation.  We
found that the number of blends of detected objects is similar to that found in
typical DES coadds field, but is not representative of fields with relatively
low redshift galaxy clusters.     Thus these images are appropriate for testing
cosmic shear measurements, but not necessarily for testing shear
cross-correlations such as cluster lensing studies.  Note that most of the
galaxy images were well below the detection threshold, so there are a large
number of undetected blends.

We then ran \sex\ on the images with settings similar to those used in DES,
created \meds\ files, and spatially matched the \sex\ catalogue to the input
simulation catalogue, in order to associate a ``true'' shear with each detection.

We ran the \metacal\ shear code on the \meds\ files with identical settings
used for the real data, including cuts on \sex\ flags $\leq 3$, which only
removes objects near edges or with incomplete apertures.  Note that we cannot
correct for this flag selection within the \metacal\ formalism.  We then
calculated the mean shear, which we compared to the mean ``true'' shear from
the matched catalogue.  We ran in two different modes: one using the \uberseg\
algorithm only, and one subtracting the light of neighbours as measured using
the MOF algorithm in addition to \uberseg\ masking.

The results are shown in Table~\ref{tab:mcal:deblending} for a few different
\snr\ thresholds, in addition to our fiducial size cut $T/T_\mathrm{PSF} > 0.5$ and
\sex\ flags $\leq 3$.  When using \uberseg\ only we detect a $\sim 2$\% bias in
all cases. We detect no bias when subtracting the light from neighbours.

\begin{table}
    \centering
    \begin{tabular}{ l c c}
        \hline
        Method         & \snr\ Cut & m            \\
                       &           & $[10^{-2}]$  \\
        \hline
        \hline

        \uberseg       & \snr$ > 10$ & 2.18 $\pm$ 0.16  \\
        \uberseg       & \snr$ > 15$ & 1.73 $\pm$ 0.17  \\
        \uberseg       & \snr$ > 20$ & 1.74 $\pm$ 0.18  \\

        \hline

        \uberseg+MOF   & \snr$ > 10$ & 0.10 $\pm$ 0.20  \\
        \uberseg+MOF   & \snr$ > 15$ & -0.24 $\pm$ 0.21 \\
        \uberseg+MOF   & \snr$ > 20$ & -0.04 $\pm$ 0.22 \\

        \hline
    \end{tabular}

    \caption{Shear calibration bias $m$ for \metacal\ in the deblending
    simulations.  Shown are results when using the \uberseg\ algorithm to mask
    neighbours, as well as results when subtracting the light of neighbours using
    MOF models.  Results are listed for three different threshold cuts in \snr,
    in addition to cuts at $T/T_\mathrm{PSF} > 0.5$ and \sex\ flags $\leq 3$.
    \label{tab:mcal:deblending}}

\end{table}

For
our default cuts ($\snr\ > 10$, $T/T_\mathrm{PSF} > 0.5$ and \sex\ flags $\leq 3$) the
ratio of biases for the two methods is approximately
\begin{align}
    \frac{1+m_{\mathrm{useg}}}{1+m_{\mathrm{MOF-sub}}} \approx 1.02.
    \label{eq:mcal:mofvsuseg}
\end{align}
This is consistent with the empirical bias ratio found in
\S \ref{sec:mcal:deblending:data}, but as noted above this simulation
has a higher density of galaxies than Y1 data, so this may be
an upper limit.

Because these simulations do not match the real data perfectly, we do not use
these results to predict the systematic bias that may exist in either method
for dealing with neighbours.  Rather we treat these results as independent
confirmation of the presence of a bias between the two methods.  It is
suggestive that the bias may be higher in the fiducial
\uberseg-only measurements.

\subsubsection{Multiplicative bias prior due to blending}

From the tests done on DES data and on simplified image simulations in the
previous sections, we conclude the following
\begin{itemize}

    \item The shear calibrations $m$ of the MOF-subtracted and
        non-MOF-subtracted runs of \metacal\ differ at a level
        of $\Delta m\approx0.023$, measured by the ratio of shear around lens
        galaxies with a matched version of both catalogs. 

    \item A $\Delta m$ of the same sign but somewhat smaller amplitude remains
        when limiting the comparison to galaxies with little masking.  The
        measured ratio $0.018\pm0.015$ is consistent with both zero and the
        0.023 detected for all galaxies, and is thus inconclusive.

    \item  Although the image simulations do not match the data perfectly, and
        thus should not be used to estimate a numerical value for the bias, the
        results do give a bias difference of the same sign and comparable
        amplitude to the $1.02$ value seen in equation (\ref{eq:mcal:mofvsuseg}).

\end{itemize}

The relative difference in multiplicative bias of the two runs of 0.023 can be
interpreted in multiple ways.  It could be that one of the methods is unbiased
and the other is biased by 0.023.  It is also plausible that neither method is
unbiased, in which case the 0.023 is divided between the two methods.  In order
to encompass both scenarios, we adopt a Gaussian prior on the multiplicative
bias of $0.012 \pm 0.012$, which is consistent with a bias of both zero and
0.023 at the $1\sigma$ level. We continue to use the \uberseg-only shear
because photoz\ information is not available for the MOF-subtracted catalog.

We note these numbers were derived directly from the data. For this reason we
do not artifically increase the width of the prior as we did for numbers based
on simulations.

\subsection{\metacal\ and PSF Modelling Bias} \label{sec:mcal:wrongpsf}

The Y1 PSF modelling and interpolation exhibit small biases
both in the size and shape (see \S\ref{sec:psf}), which result in additive and
multiplicative errors.

The additive errors come about due to PSF mis-estimation, which results in
inaccurate deconvolution during \metacal\ process, resulting in some remnant of
PSF ellipticity remaining in deconvolved images of circular sources.  We have
calculated empirical estimates for the additive bias in the shear
two-point correlation function as a function of angular separation (see \citealp{shearcorr}), 
and shown that they are negligible once the mean shear is corrected.

We discuss multiplicative bias from PSF modelling errors in
detail in \S\ref{sec:mcal:psfmult}. Additional biases
due to stellar contamination in the source sample are discussed
in \S\ref{sec:mcal:starbias}.

\subsubsection{\metacal\ Shear Bias from PSF Modelling Errors} \label{sec:mcal:psfmult}

As discussed in \S\ref{sec:psf:measurement}, our PSF model does not perfectly model the
true PSF.  Most aspects of the PSF modelling errors manifest as additive shear errors
(see \S\ref{sec:tests:psf});
however, the mean error in the size estimate of the PSF, described in \S\ref{sec:psf:rho}, manifests as a multiplicative error.
To measure the shear biases caused by these effects, we created further
bulge+disk+knots simulations with a range of \snr, similar to those presented
in \S\ref{sec:mcal:me} but using only a single simulated image per object.

We chose the minimum galaxy $\snr$ used for these simulations to be
\snr $\approx 50$ in order to measure the shear bias at high precision. We saw
no evidence that the magnitude of the effect depends on the galaxy $\snr$, so
we expect these results to hold for our full galaxy sample including
lower \snr\ objects.

We kept the nominal PSF model identical
in all cases, using a Moffat profile with $\beta=2.5$,
a size corresponding to $0\farcs9$ FWHM, and $(e_1,e_2) = (0,0.03)$.
We then created several versions of the simulation with different true PSFs:
(1) the same as the nominal PSF, 
(2) a constant PSF that was larger by $\Delta T/T = 8.3 \times 10^{-4}$,
(3) a variable PSF with the correct mean size but varying in a normal distribution
with $\sigma(T)/T = 3 \times 10^{-2}$,
and (4) a variable PSF with both the larger mean and this variance.

The shears were measured with the same code used to process the DES data. We
used the nominal PSF, not the true PSF, when performing the \metacal\ image
manipulations.  We applied various size cuts, and applied the appropriate shear
and selection responses.  For case 1 we found the bias to be less than 0.001 at
95\% confidence. This was expected because these simulations are similar to
that presented in \citetalias{SheldonHuff2017}, for which no bias was found.
For case 2 with an overall PSF size error but no variance, we found a
multiplicative shear bias of $m\approx0.001,$ independent of the size cut.
With the additional size variance (cases 3 and 4) the mean bias increased,
reaching nearly $m=0.002$ for case 4. For case 4 the bias was independent of
galaxy size for $T/\Tpsf > 0.3$, but we saw some variation for smaller sizes.
This partly motivated the choice to cut at $T/\Tpsf > 0.5$ for our final shear
catalogues.

We do not expect this simulation to produce an exact measure of the shear bias
present in the real data, because it undoubtedly depends on the details of the morphology
distribution of the galaxies and the precise distribution of the PSF errors around the mean value.
We therefore attempt no correction for this effect.  Rather, we
take the results of this simulation as an estimate of a systematic
uncertainty $\sigma_m$ in the multiplicative error from this effect.  We conservatively take the
value of $\sigma_m = 0.003$.  See \S\ref{sec:budget:metacal} for a summary of
all systematic uncertainties.

\subsubsection{\metacal\ Shear Bias From Stellar Contamination} \label{sec:mcal:starbias}

\begin{figure}
\center\includegraphics[width=\columnwidth]{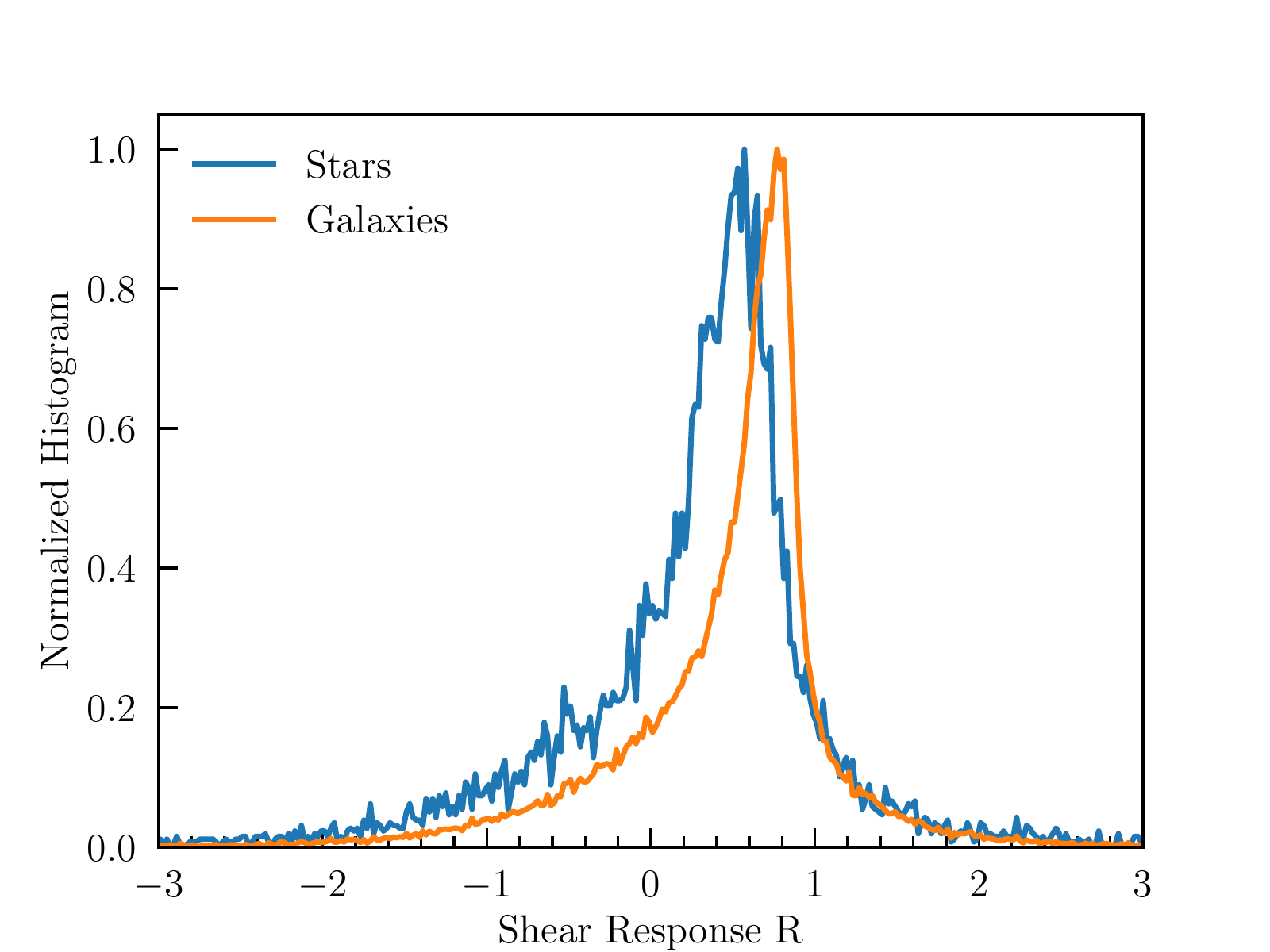}
\caption{The distribution of responses measured in the COSMOS
    field for objects flagged as galaxies or stars.  The non-zero response of the stars is due to 
    noise in the measurement and interpolation of the PSF, and is accounted for in our error budget.
  \label{fig:mcal:cosmosresponse}
}
\end{figure}

Stars do not bias \metacal\ shear recovery when the PSF is accurately known,
since they should yield $\langle e \rangle = \langle \mcalR \rangle=0$
\citepalias{SheldonHuff2017}.  If, however, the PSF model is biased, stars will
not have zero mean shear response \mcalRgmean, which can potentially result in
a shear bias.  In figure~\ref{fig:mcal:cosmosresponse} we show the decidely
non-zero response for known stars and galaxies in the COSMOS field.  Such
a distribution of responses for point sources will occur when the PSF estimate
is biased, but the exact distribution depends intimately on the details of
the PSF errors.

We can calculate the expected bias from these stars given our fiducial cuts
\snr$ > 10$ and $T/\Tpsf > 0.5$.  Based on COSMOS we found this cut results in
0.4\% stellar contamination.  Averaging over the stellar response distribution,
and taking into account a re-scaling from the COSMOS stellar density to the mean density inside the DES Y1 footprint, 
results in a shear bias of $|m|<0.001.$  Because the COSMOS field
may not be representative of the wider DES survey data, we take this
measurement as indicative of the uncertainty in this systematic effect.  We
conservatively take $\sigma_m = 0.002.$


\section{The \imshape~ Catalogue}
\label{sec:im3shape}
\subsection{Overview}
\label{sec:im3shape:overview}

Our second DES Y1 catalogue was generated with the maximum likelihood code \imshape~\citep{zuntz13}, which uses \levmar~ minimization to find the maximum-likelihood (ML) fit of two S\'ersic models, with power-law indices $n=1$ and $n=4$, to all the exposures of each galaxy simultaneously. Each galaxy is then identified as a bulge or a disc, depending on which model returned the superior likelihood. 

The \imshape~ code\footnote{https://bitbucket.org/joezuntz/im3shape-git}  is largely unchanged from the version used for SV, though the simulations used to calibrate it have been upgraded significantly. We refer the reader to \citetalias{jarvis16} and the original code release paper \citet{zuntz13} for code details.

The code fits six parameters:
two ellipticity components ($e_1$, $e_2$), a half light radius $r$, a centroid offset $(x_0, y_0)$ and an amplitude $A$.
For each fit we also compute a signal-to-noise ratio (\snr) using the convention of \citet{great3} and \citetalias{jarvis16}.
As we point out in \citetalias{jarvis16}, this signal-to-noise measure is analogous to a matched filter,
favouring maximal agreement between the model fit and the image pixel fluxes. \imshape\ also defines a size metric $R_{gp}/R_p$, the ratio of the convolved galaxy FWHM to the PSF FWHM, where the former is measured from a circularized version of the best-fit galaxy profile.

A small number of changes have been made to the code to improve internal organization, human readability, and tools for running it on high-performance computing systems. After these, the mean time taken to analyze a galaxy was $1.6$ seconds per exposure, which was dominated by a small number of difficult objects.  The total time was approximately $200,000$ CPU hours for a single full analysis (not including the time taken for calibration simulations).

As noted in the introduction, maximum-likelihood methods such as
\imshape\ accrue
\emph{noise bias} when the pixel values are a non-linear function of the model parameters, as is true in galaxy model fits. 
Along with all other sources of systematic measurement bias in our shear estimates, this effect must be calibrated. The most common approach, which we also adopt for Y1 of DES, is to do this using a suite of image simulations. We describe these simulations and the calibration process used to generate the corrections needed for the data in \S \ref{sec:sims}.

All \imshape~ measurements presented here were carried out at the National Energy and Scientific Research Computing Center\footnote{http://www.nersc.gov/} (NERSC) and the GridPP grid computing system\footnote{https://www.gridpp.ac.uk/} \citep{gridpp}.
The calibration simulations were generated entirely using the NERSC facility.

\subsection{The \hoopoe~ Image Simulations}
\label{sec:sims}

We use a suite of simulations, called \hoopoe, to calibrate biases in the \imshape\ shape measurements. They account for noise bias, model bias, PSF leakage, mask effects, and selection biases. These simulations were used to model the $m_i$ and $c_i$ terms in equation (\ref{eq:mc_bias}).
Previous studies have found no evidence of off-diagonal multiplicative bias when fitting \sersic\ models,
and we see no evidence of a systematic difference between $m_1$ and $m_2$ in any region of parameter space. 
Our calibration therefore uses the simple arithmetic mean $m=(m_1+m_2)/2$.

Our Y1 simulations differ significantly from those used in SV.  The
latter started with postage-stamp images of isolated galaxies, to
which we applied simulations of observational features. For Y1 we
start with reduced image data from the survey, and create an object-for-object simulacrum, preserving as much of the original detail as possible. The differences are listed in detail in Table~\ref{tab:sims:sv_y1_comparison}.

Variants of the \hoopoe\ simulations specifically to explore biases from blending are discussed further in \citet{des_sim_2017}.


\begin{table*}
\begin{center}
  \begin{tabular}{c|cc}
  \hline
   & \blockfont{GREAT-DES} (DES-SV) & \hoopoe~ (DES Y1) \\
   \hhline{=|==}
   Multiple Exposures & Coadd Only & Multi-Exposure \\
   Point Spread Function & Discrete, Kolmogorov &  PSFEx, Image Plane Variations from Data \\
   Pixel Noise & Gaussian Random per Pixel, Fixed $\sigma_n$ & Gaussian Random per Pixel, $\sigma_n$ from Weight Maps \\
   Blending/Neighbours & None (postage stamp simulations) & Simulate Full Image Plane \\
   Galaxies Below Detection Limit & None & Random Positions, Drawn from Faint COSMOS Cache \\
   Source Detection & None  & Rerun \sex~on Simulated Images \\
   Masking & None  & Spatial Masks and PSF Blacklists from {\sc Gold} Catalogue\\
   Input Galaxy Selection & COSMOS ($ < 23.5$ mag) & COSMOS Deep ($ < 25.2$ mag)  \\
   Magnitudes/photometry & HST F814W Magnitudes & SDSS (DES) $r$-band Magnitudes \\
   Stars & None & Point Sources \\
   Simulated Galaxies & 48M & 68M \\
   Input Shears & $|g|=0.05$, 8 Discrete Rotations & Continuous Uniform Random $-0.08<g_{1,2}<0.08$ \\
   \hline
   Galaxy Morphology & No Morphology Split & Bulges/Discs Calibrated Separately \\
   Interpolation/Fit  & Polynomial Fit$^{*}$ &  Grid Nodes$^{*}$ \\
   & Radial Basis Functions & Radial Basis Functions \\
   & & Polynomial Fit \\
   PSF Measurements & \imshape~ Weighted Moments & HSM Adaptive Moments \\   
  \hline
  \end{tabular}
  \caption{Comparison of \imshape\ shear calibration schemes used in DES-SV and DES Y1.
  The upper portion of the table itemizes differences in the calibration simulations \hoopoe~ and \blockfont{GREAT-DES}.
  Entries below the dividing line pertain to methodological choices rather than systematic differences between simulations.
  In the case of the interpolation scheme we compare three methods in this study,
  as described in the text below.
  The asterisks highlight the two interpolation methods used as the fiducial schemes in SV and Y1.
  }
  \label{tab:sims:sv_y1_comparison}
\end{center}
\end{table*}

\subsubsection{Simulating DES Y1: The Image Pipeline}
\label{sec:sims:y1}

The simulation pipeline for the \hoopoe~image simulations is shown in the blue (left-hand) part of Figure~\ref{fig:flowchart}. 
The analysis of the resulting simulations was closely matched to the equivalent process used on real data,
although we do not repeat the single-epoch data processing or PSF estimation stages.  The position, noise levels, and PSFs
of each simulated galaxy are taken from the real observations. The mask is made by combining the bad-pixel
map, which is imported directly from real data, with the object segmentation map, which is remade on the simulations using \sex.

The \hoopoe~image simulator 
begins by choosing one of the  $0.73 \times 0.73$ degree coadd tiles output
by the DESDM pipeline, each of which is generated by coaddition of around 70 partially overlapping exposures. For each tile we require 
(a) a source catalogue generated by \sex~or similar object detection algorithm,
(b) a WCS specifying the image bounds and the transformation between pixel and world coordinates per exposure,
(c) a model describing the PSF variation across the image plane,
(d) a noise variance weight map per exposure.

With these basic inputs the simulation then proceeds as follows for each sky region:
\begin{enumerate}
    \item{Generate a set of noise images from the \sex\ weight maps, matched to the bounds of each data image. A simulated coadd-image is also generated in the same way.}
    \item{Import the true detection catalogue for the region, and add to it a population of fainter undetected galaxies (see \S\ref{subsubsection:faint_galaxies}).}
    \item{Iterate through positions, selecting a random COSMOS profile (from a rolling cache designed to make the expected number of unique profiles per coadd tile 2000) and simulating it with additional shears and rotation angle drawn from $g_{1,2} \in [-0.08,0.08]$,  and $\theta \in [-\pi, \pi]$.}
    \item{Convolve the profile with the PSF at the position in each image and draw it into each image (including the coadd).}
    \item{If a faint galaxy is associated with this position (see \S\ref{subsubsection:faint_galaxies}) then draw one from a secondary cache of faint profiles. It is placed in some point in the region formed by the overlap of all the exposures that contain the current galaxy, so that it will be in approximately the same
    geographic region as the primary galaxy but is not guaranteed to overlap it. It is sheared and convolved as in the previous steps.}
    \item{Once the full image is simulated, run \sex\ on the simulated image, generating a new detection catalogue.}
    \item{Iterate through the detection positions a second time, building the \sex\ mask for each and extracting a postage stamp cutout. In the version of the simulations presented here the stamp size was not recomputed for each object, but came from the size of the original object in the real images. Later code versions corrected this, but re-running the full simulations was deemed too expensive. The impact of this error is discussed in \S\ref{sec:imbudget}}.
    \item{Store and stack the cutouts in the MEDS format \citepalias{jarvis16}.}
    \item{Run \imshape\ on the \hoopoe\ MEDS files, blinding using the prescription described in \S \ref{sec:blinding}
    with the same factor $f$ as applied to the data.}
\end{enumerate}

\subsubsection{Galaxy Sample}

To capture the range of morphologies found in a photometric survey like DES the Y1 \hoopoe~simulations use real galaxy profiles rather than analytic constructions. 
In order to obtain an accurate calibration the profiles used as input should extend to at least the same depth as the data and have sufficiently low levels of noise and seeing to allow them to be degraded to match DES precisely, which limits the available data.  We make use of the COSMOS sample described in \S \ref{sec:cosmos}, which meets these requirements.

We do not use the 
quality flags supplied with the COSMOS sample, which were not available at the time the code was run. Instead we 
visually inspected the sample, as described in Appendix~\ref{app:eyeball}.
The publicly available HST data are limited to wide band photometry in the optical F814W filter. In order to obtain the desired magnitudes in 
the DES bands, we match the sky position of each of these galaxies to the COSMOS mock catalogue of \citet{jouvel09}, which includes photometry specific to the transmission curves of the DES filters.

The input sample for a tile is then generated by splitting the COSMOS catalogue about $M_\mathrm{r,lim}$ 
and discarding objects too faint for detection.
Each of these galaxies is simulated at its original COSMOS magnitude,
rescaled to the zero-point of the DES images.

\subsubsection{Simulated Stars}
\label{subsubsection:sims:stars}

The mock images also contain stars, simulated at the positions of objects classified as stars in the real data.
Stars are rendered as point sources and account for around $10\%$ of simulated objects.  
This should capture any effect they may have as
a source of neighbour bias, including changes they induce in the galaxy selection. 
We do not re-run star/galaxy separation in the simulations, so do not account for any mis-classification 
bias. The cuts to the \imshape\ catalogue in size and \snr, however, will remove the majority of the ambiguous 
objects, so we expect the impact of this decision to be small. For reference, the residual number of objects not 
removed by internal \imshape\ flags, but flagged as stars with the {\sc Gold} star-galaxy classifier in the data, 
is only about 1.5\% of objects.

\subsubsection{Galaxies Below the Detection Limit}
\label{subsubsection:faint_galaxies}

In addition to simulating objects detected in real data we wish to simulate a population of fainter (undetected) galaxies. We choose a number of faint galaxies for each tile $N_\mathrm{faint}$ by integrating the full distribution of COSMOS magnitudes,
\begin{equation}\label{eq:number_faint}
N_\mathrm{faint}=\frac{ f_\mathrm{faint} }{ (1 - f_\mathrm{faint})} \times N_\mathrm{det},
\end{equation} 
where $f_\mathrm{faint} \equiv \int ^{\infty} _{M_{r, \mathrm{lim}}} p(M_\mathrm{r}) \mathrm{d} M_\mathrm{r}$ is the fraction 
of the weight of the normalized magnitude distribution $p(M_\mathrm{r})$ above the nominal DES detection limit, 
and $M_\mathrm{r}$ is the aperture magnitude.  In reality faint galaxies undetected by DES will include objects 
brighter than the nominal DES limiting magnitude, since the survey is really surface-brightness limited; 
the simple model here does not include this population, but should account for the leading order effect 
of faint neighbours.

Each of these extra objects is randomly assigned a companion
from the detections within its coadd tile.
This faint object is randomly placed into the same exposures as its detected companion,
but does not replace it, nor are their properties linked in any other way. 

In the real data the flux from these galaxies enters the images prior to reduction,
and would affect the background subtraction.
We choose not to simulate thermal sky emission and rerun the background subtraction. 
To gauge the impact of the extra background flux, we reran a small subset of the simulations with the same
random seed settings, but without faint galaxies.
The background estimation algorithm was then applied to the two sets of images, which were identical
apart from the omission of the faint objects.
To first order we find the sub-detection galaxies produce a uniform shift in the mean of the estimated sky background. 
To correct for this effect, we subtract the average per-pixel flux of faint objects drawn into our simulated images.  

The tests described here have neglected clustering between faint and bright galaxies, and between the faint 
galaxies themselves.  Clustering would enhance the amount of blending, and would also make the sky 
subtraction effect more heterogeneous.  Based on the variation of $m$ with the density of faint objects we 
expect both of these effects to be smaller than the basic faint object effect described here, but not 
generally negligible.  Future data sets will require simulations that include careful galaxy correlation 
modelling.

The impact of sub-detection galaxies on shear measurement
is explored in more detail in Section 5 of \citet{des_sim_2017}.
In the tests presented therein we find a net contribution to the multiplicative bias
we correct for of $m\sim-0.01$ due to these sub-detection galaxies.

\subsubsection{Comparing Simulations \& Data}
\label{sec:sims:comparisons}

Given the sensitivities of measurement bias to the observable parameters of an image,
most notably signal-to-noise, size, and ellipticity, it is important that the simulations
should cover the same parameter space as the data.  We explicitly calibrate over \snr~and the ratio \rgp~of
the galaxy image FWHM to PSF FWHM,
so exactly matching simulations and data in these parameters is of secondary importance;
matching the distributions of 
ellipticity, PSF size and shape, and other properties is more important for an accurate calibration.

\begin{figure*}
\begin{center}
\includegraphics[width=0.69\columnwidth]{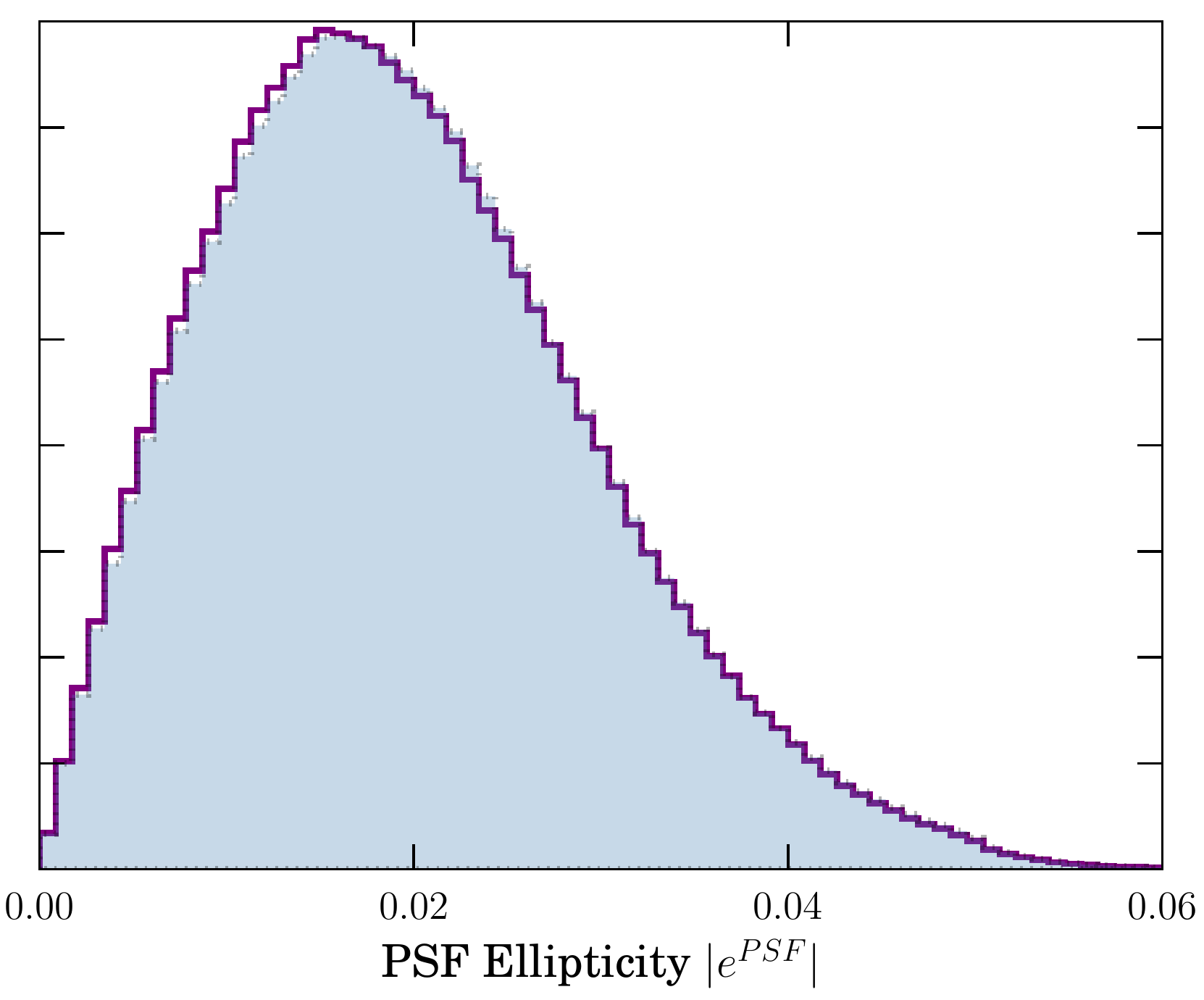}
\includegraphics[width=0.69\columnwidth]{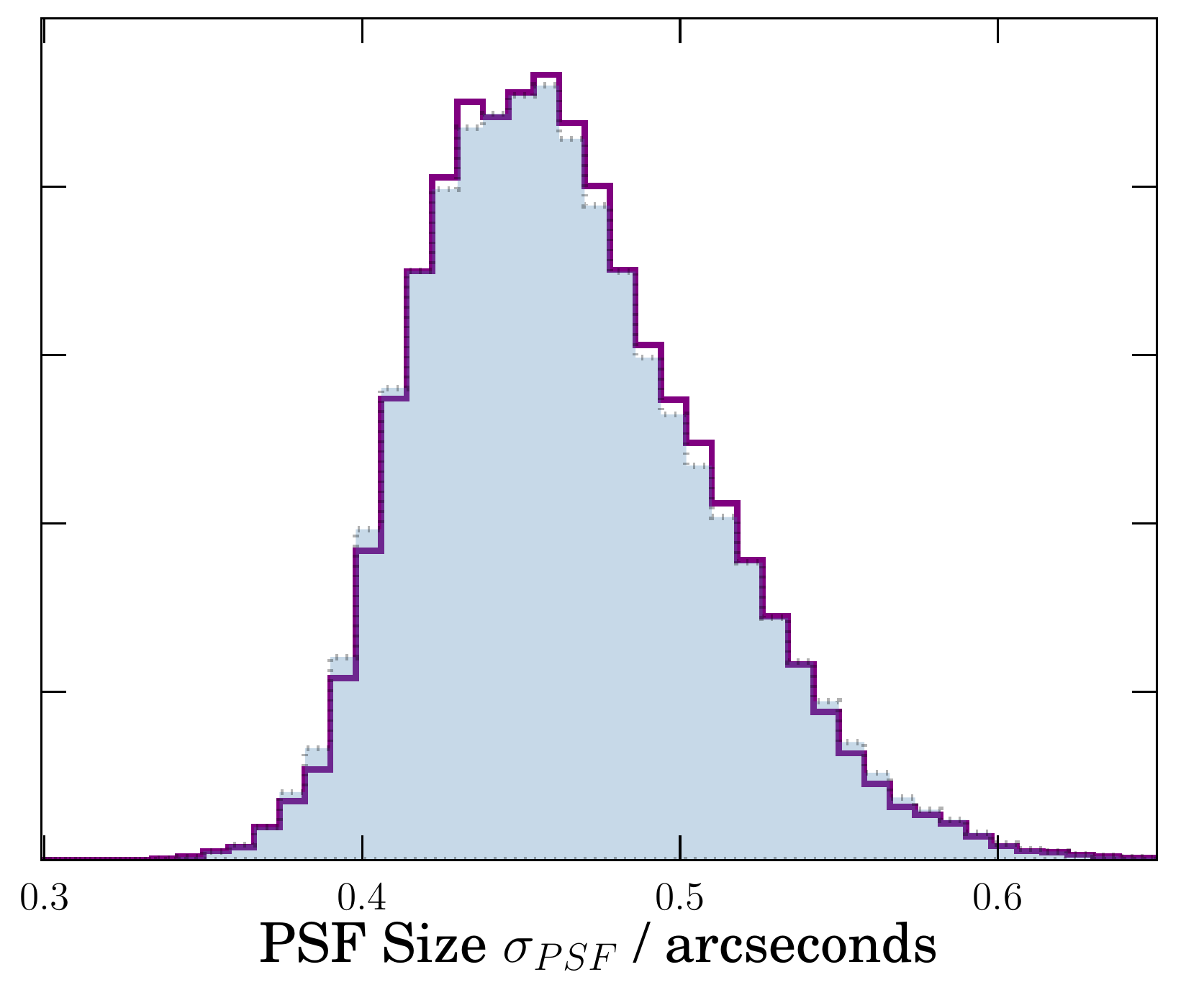}
\includegraphics[width=0.69\columnwidth]{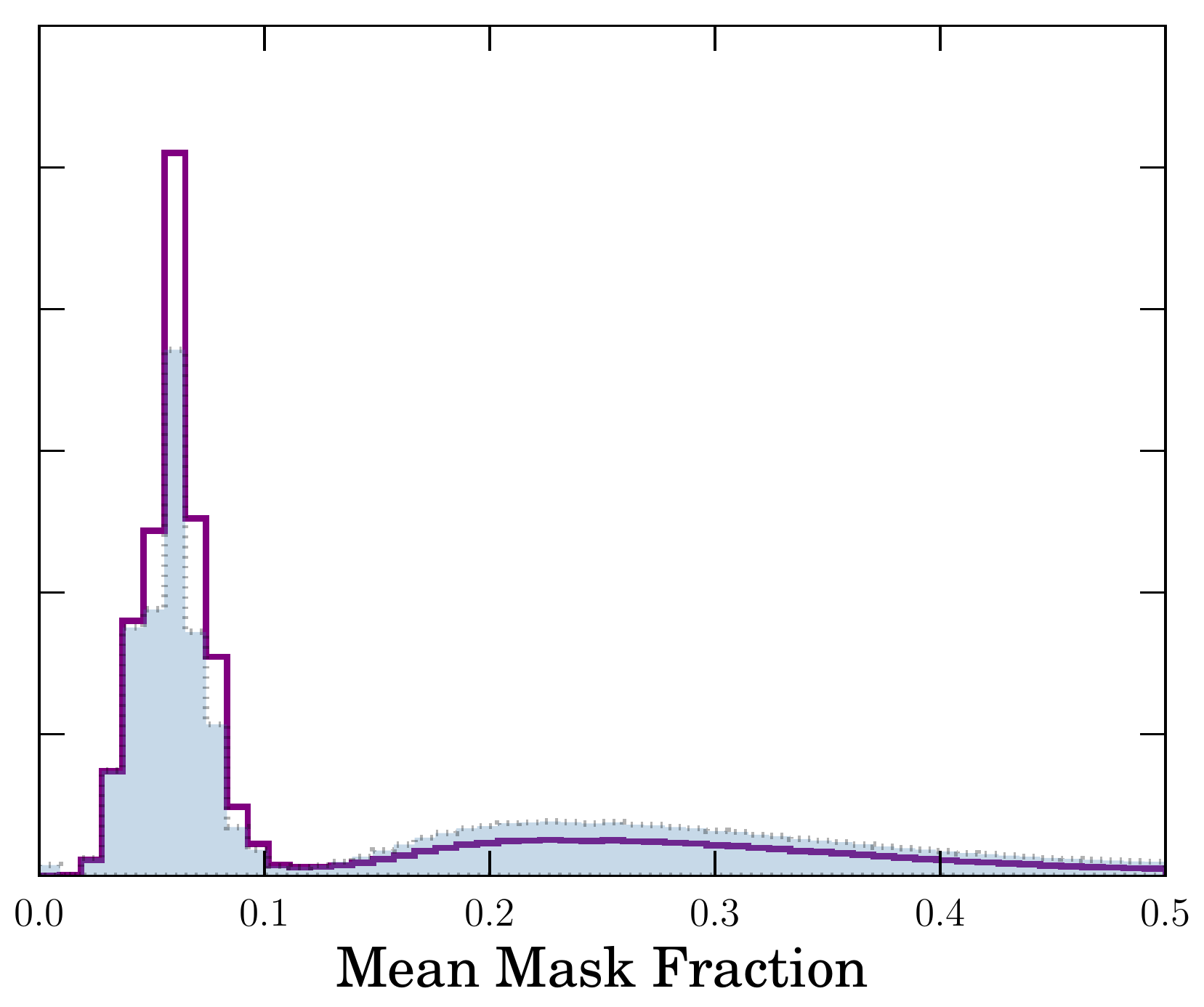}

\includegraphics[width=0.69\columnwidth]{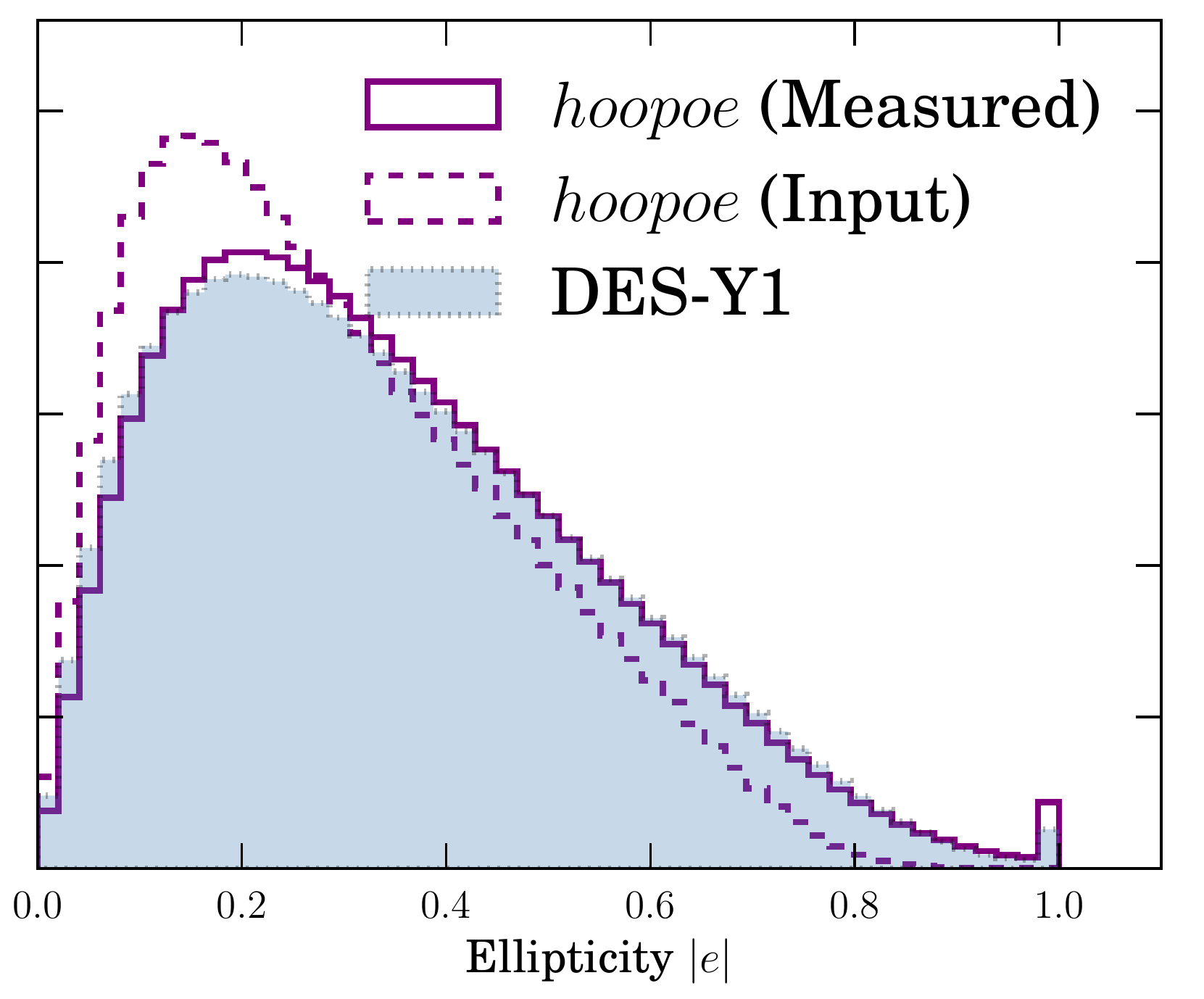}
\includegraphics[width=0.69\columnwidth]{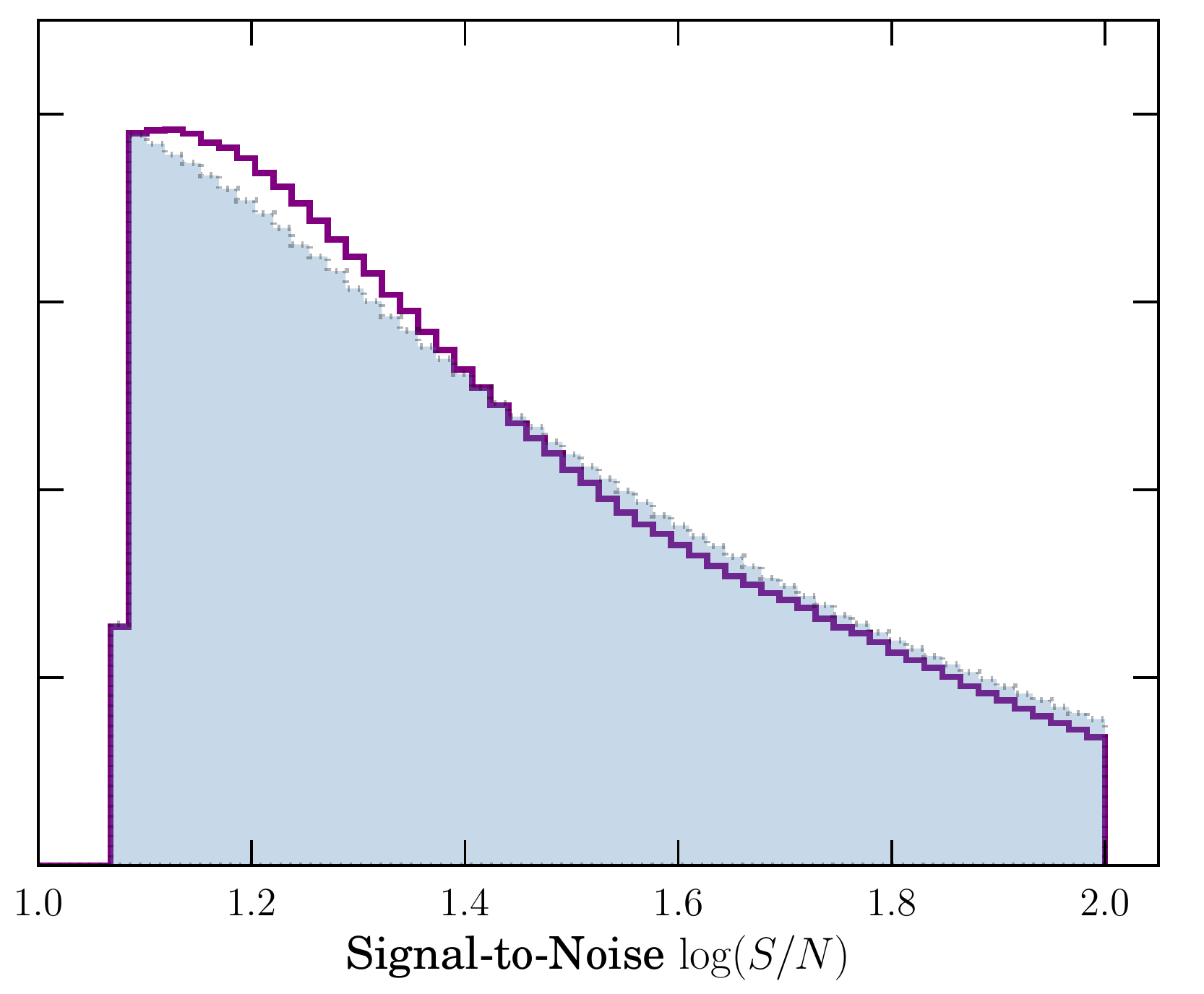}
\includegraphics[width=0.69\columnwidth]{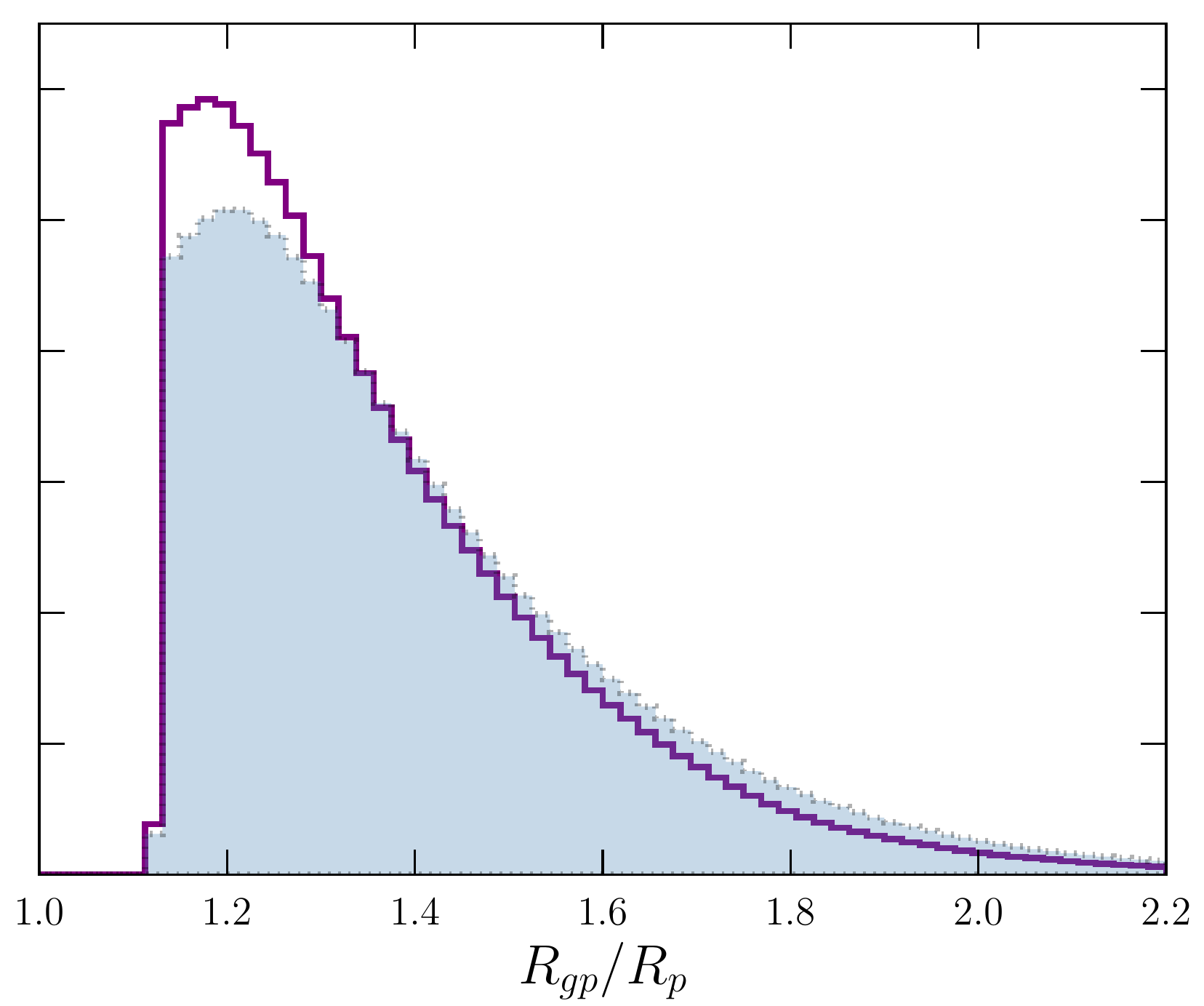}

\includegraphics[width=\columnwidth]{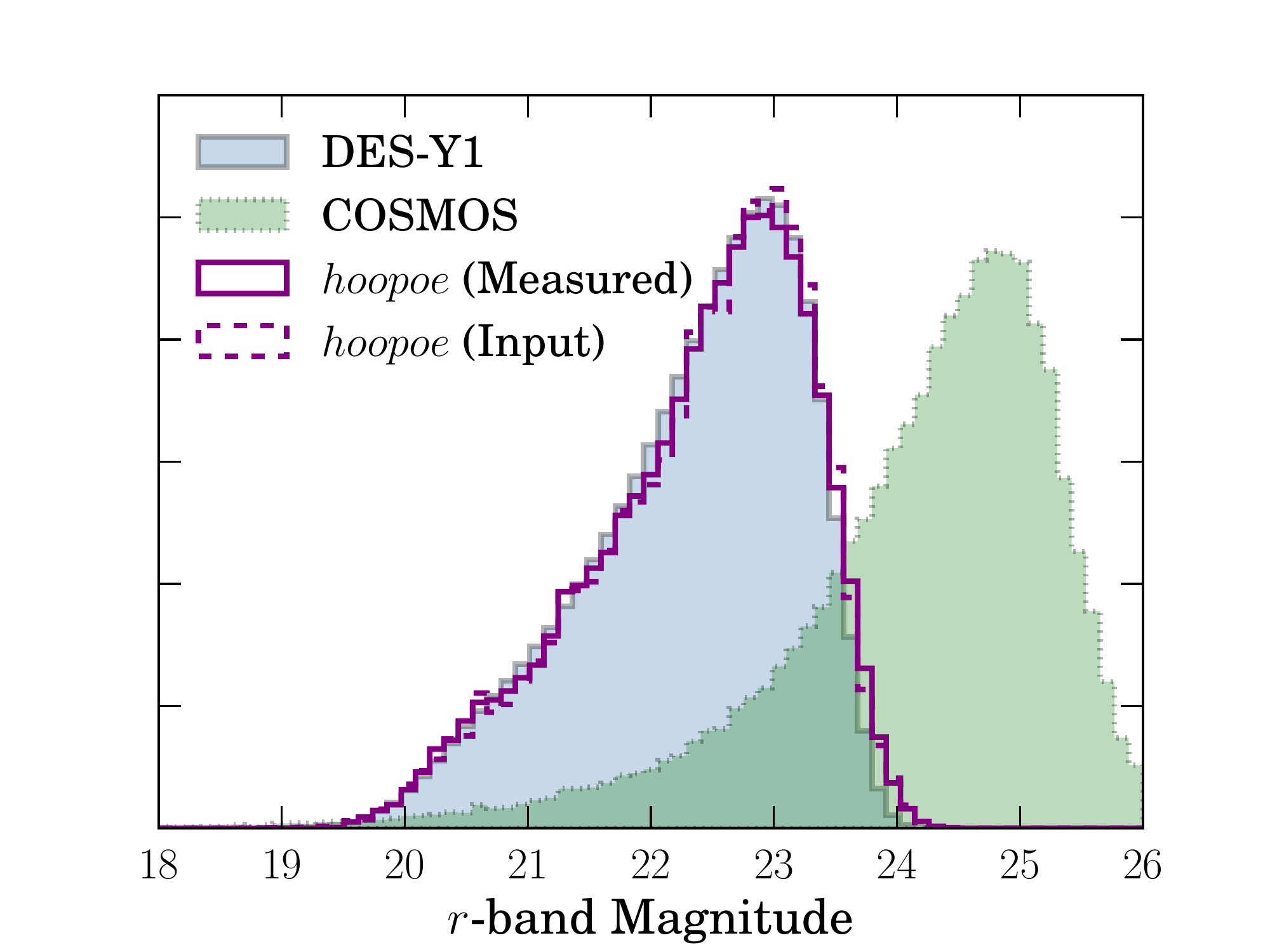}
\caption{A comparison of normalized distributions in the Y1 simulations used for \imshape~calibration (purple) and data (blue). The upper panels show (clockwise from upper left) PSF ellipticity; PSF size, as measured using HSM; the fraction of pixels masked out, averaged across each object's exposures; \imshape's measure of galaxy size relative to the PSF, \rgp; flux signal-to-noise; and total galaxy ellipticity. In the latter we show both the input and remeasured distributions to the simulations as dashed and dot-dash lines respectively.
The lower-most panel shows the distribution of measured and input magnitudes from the simulation,
in addition to the data.
The shaded green (dotted) curve shows the equivalent $r$-band magnitudes for the full COSMOS catalogue
from which we draw our input sample.}
\label{fig:sims:histograms}
\end{center}
\end{figure*}

The distributions of a selection of salient properties are shown in Figure~\ref{fig:sims:histograms}.
Unlike in previous studies, we are convolving simulated galaxies with the measured PSF at each position on the sky, thus more
directly matching the PSF variation across the sky compared to real data.
Since we apply shape quality cuts it is not automatic that the PSF properties still match well to the data 
after those cuts - any significant difference in PSF properties \emph{after} these cuts would imply a 
different selection behaviour with respect to PSF in simulations compared to the data. Unlike in real
data the simulations do not include PSF errors, so we will be susceptible to the kinds of biases 
described in \S \ref{sec:mcal:psfmult}, which is also accounted for in our error budget (see \S \ref{sec:imbudget}).

The distribution of \emph{input} simulated ellipticities in Figure~\ref{fig:sims:histograms}
is notably narrower than the \emph{measured} distributions in both simulations and data.
As well as the expected effect of noise, this arises due to blending.
An interesting comparison can be made with a new set of simulations,
identical to \hoopoe, but with neighbour
light removed (described in \citealt{des_sim_2017}). In those simulations
we find that the measured ellipticity histogram
is much closer to the input distribution.

As in \citetalias{jarvis16}, the difference in \rgp~ increases at small sizes. This may be due to the COSMOS sample used, or the similar PSF estimation methodology.

Finally in the lower panel we compare the input and output magnitudes from the simulations. 
We do not find a significant bias in the remeasured magnitudes, nor serious disagreement with the data.

Though most properties match well, there are obvious inaccuracies in the simulated \rgp~and flux distributions.
In Appendix \ref{sec:sims:sensitivity} we test their impact
by reweighting the simulation to match the data, and find no significant change in the final calibration.  In Appendix \ref{sec:sims:validation} we describe tests of the impact on the calibration of other features in the simulations that differ from the data.

\subsection{Bias Calibration \& Diagnostics}
\label{sec:cal}

\subsubsection{Multiplicative Bias Scheme}

We now define a scheme to correct for the multiplicative bias measured in the simulations,
which must interpolate among the very noisy individual measurements. Both on theoretical grounds
for noise bias \citep{refregier12} and in practice for general biases ({\citetalias{jarvis16}, \citealt{fc16}),
the galaxy size and $\snr$ parameters are the dominant factor in determining bias. We therefore
build a calibration model in terms of those parameters.

The first step in this process is to decide on the most relevant parameters 
upon which the measurement biases depend, and calculate 
$m$, $c_i$ as a function of those parameters.
To this end, we sort the simulated \hoopoe~ data into a $16\times16$ grid 
according to the measured \snrw~ and \rgp, 
allowing the bin width to vary such that each grid cell contains
roughly the same number of
galaxies. 
A multiplicative bias is derived within
each cell by subdividing the galaxies into bins of 
$g^\mathrm{tr}$ and fitting a linear function to 
the bin-averaged shear response 
$\left < e_i \right > - \left < g_i^\mathrm{tr} \right >$ (see equation \ref{eq:mc_bias}).
The resulting bias surface $m^{ij}$ is shown in Figure~\ref{fig:sims:m_interpolation_surface}.

It is important here to define a well motivated gridding scheme in terms of bin numbers along each axis; too coarse a grid will result in real structure in this parameter space being washed out, while an overly fine sampling will inflate the statistical variance on our grid nodes. We have verified that varying our fiducial $16 \times 16$ grid between 
$6\times6$ and $20\times20$ does not lead to a significant change in the results.

We compare three methods for interpolating between grid nodes.
In the first scheme, we follow \citet{fc16}, and compute a fine grid in $m$.
If a galaxy falls within cell $ij$, we simply take the mean $m$ in that cell as our bias estimate.
The accuracy of such an approach will depend on the resolution of the grid. 

In the second scheme we interpolate with radial basis functions. 
The bias at a point is a linear combination of radial basis functions,
each centred on one of the grid nodes:
\begin{equation}
m(x,y) = \frac{\sum_i m_i  f((x-x_i)^2 + (y-y_i)^2)}{\sum_i f((x-x_i)^2 + (y-y_i)^2)}
\end{equation}
where
\begin{equation}
f(r^2) = \left(r^2/\epsilon^2 + 1\right)^{-\frac{1}{2}}
\end{equation}
and the $(x,y)$ coordinates are \snrw~and \rgp~suitably weighted to give the two dimensions parity,
$\epsilon$ is a fixed smoothing parameter, and the sums are over the grid nodes.

Finally, we fit the polynomial basis used in \citetalias{jarvis16}. 
We will not write out the entire functional form here, but note that it consists of a
linear combination of 18 terms of the form 
$(\snrw)^{- \alpha} (\rgp)^{- \beta}$, 
where the indices 
$\alpha, \beta \in (1.25, 1.5, 1.75, 2, 2.5, 3, 4)$.
We will refer to these three methods respectively as 
\emph{grid}, \emph{RBF} and \emph{polynomial} calibration schemes.
Owing to slightly better performance in diagnostic tests
the grid scheme is our fiducial choice. 

The relative performance of the three schemes is shown in 
Figure~\ref{fig:sims:m_calibration_schemes},
where we show the residual bias after calibration as a function of signal-to-noise and galaxy size. 
The grid model is constructed using two sets of equal-number cells, defined for bulge and
disc galaxies independently.
The bin edges are used to evaluate it are defined by the full catalogue, 
and so are not identical.

\begin{figure*}
\includegraphics[width=\columnwidth]{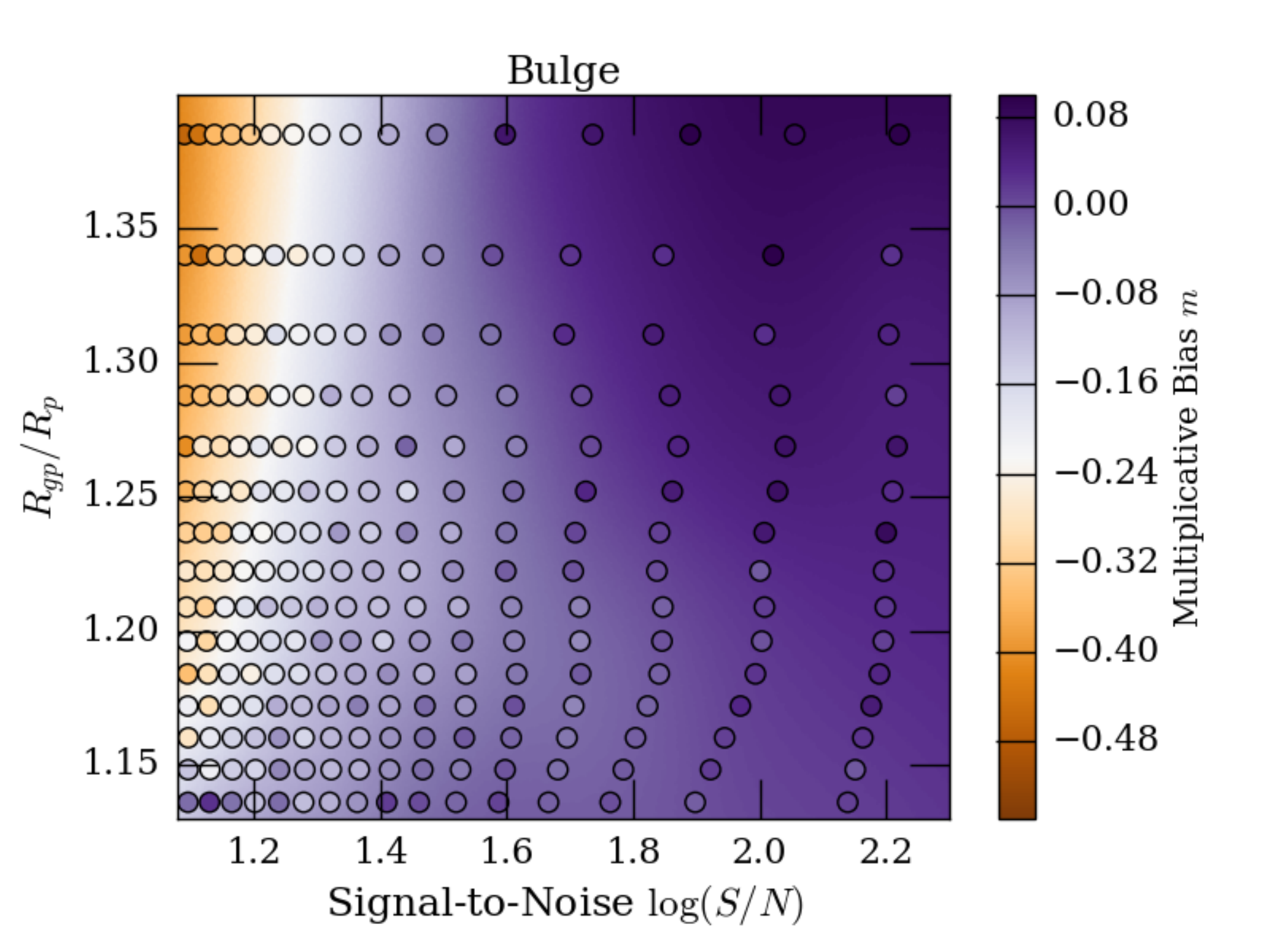}
\includegraphics[width=\columnwidth]{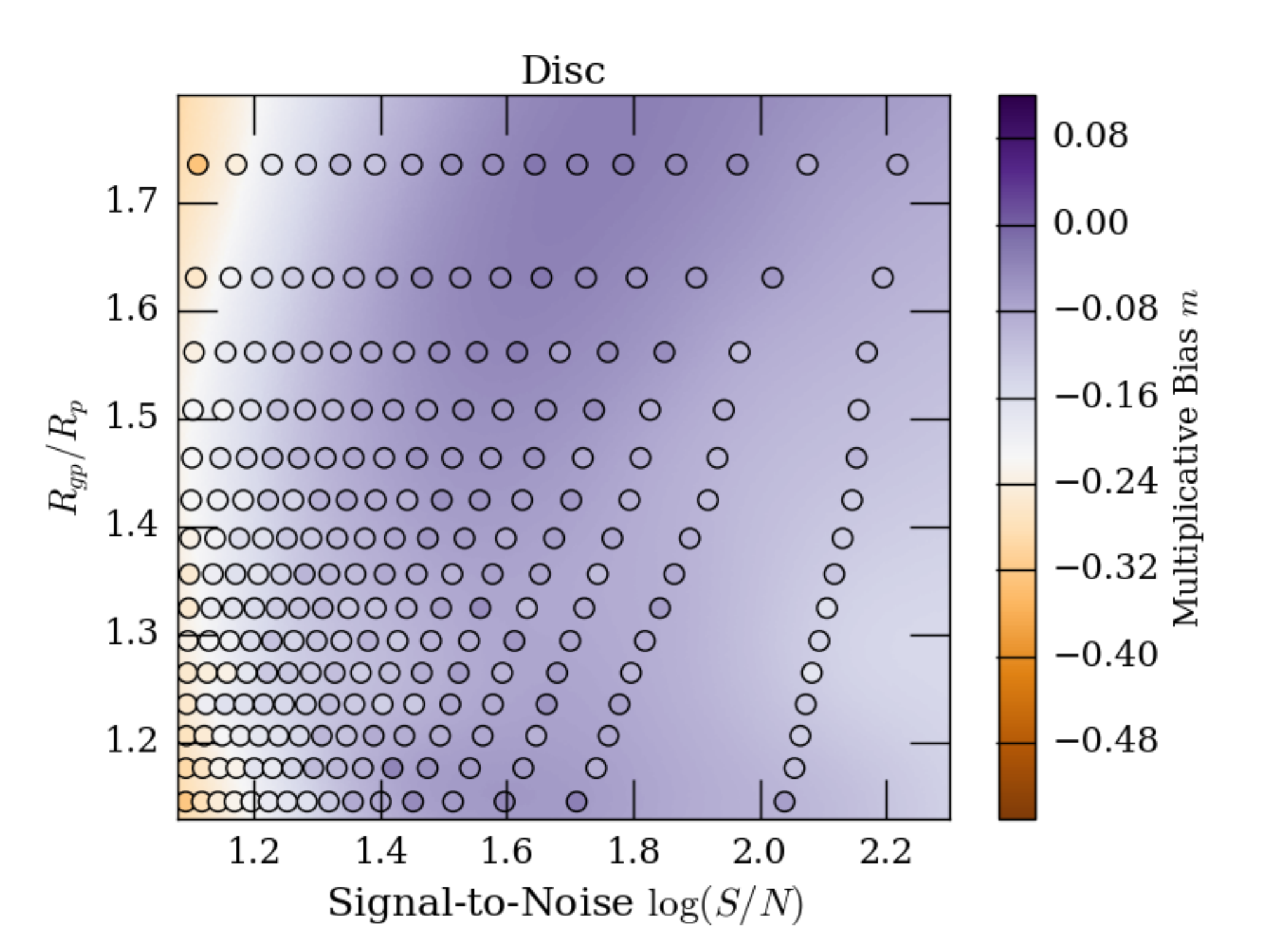}
\includegraphics[width=\columnwidth]{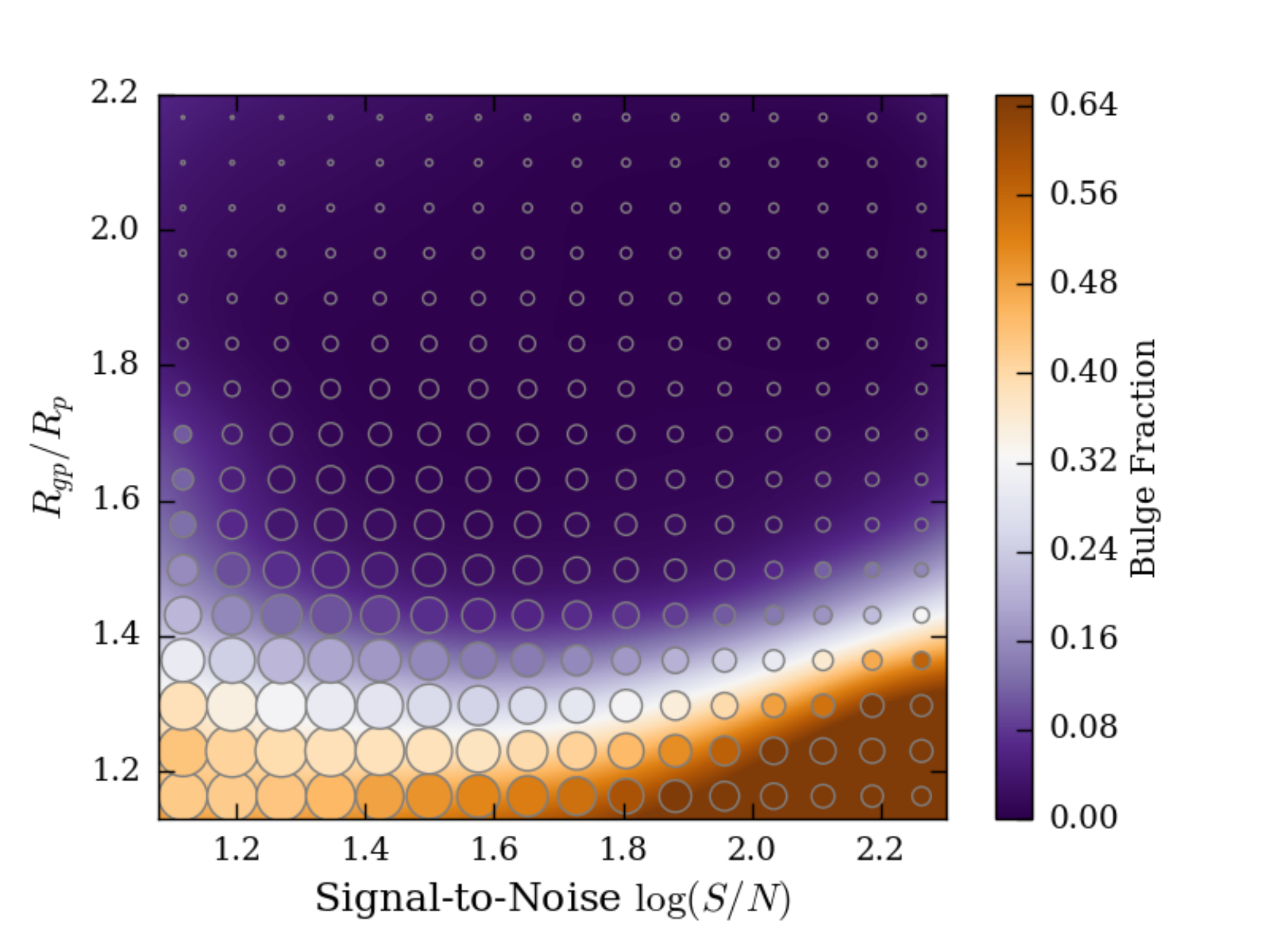}
\caption{\textbf{Top:} Multiplicative bias estimates for Y1 \imshape, using the \hoopoe~image simulations
for objects fitted using bulge profiles (right) and disc profiles (left) . 
The coloured circles represent the grid of directly evaluated $m$ described in the text. 
The underlying colour map is generated using radial basis functions to interpolate between nodes,
and is for illustrative purposes only. 
\textbf{Bottom:} Bulge fraction as a function of galaxy signal-to-noise and size.
The bulge fraction is calculated on a $16\times16$ grid and interpolated to generate the smooth map shown.
The circles represent the grid cell positions, and are drawn at a size proportional to the
total \imshape~lensing weight of galaxies contained.
}\label{fig:sims:m_interpolation_surface}
\end{figure*}

\begin{figure*}
\begin{center}
\includegraphics[width=\columnwidth]{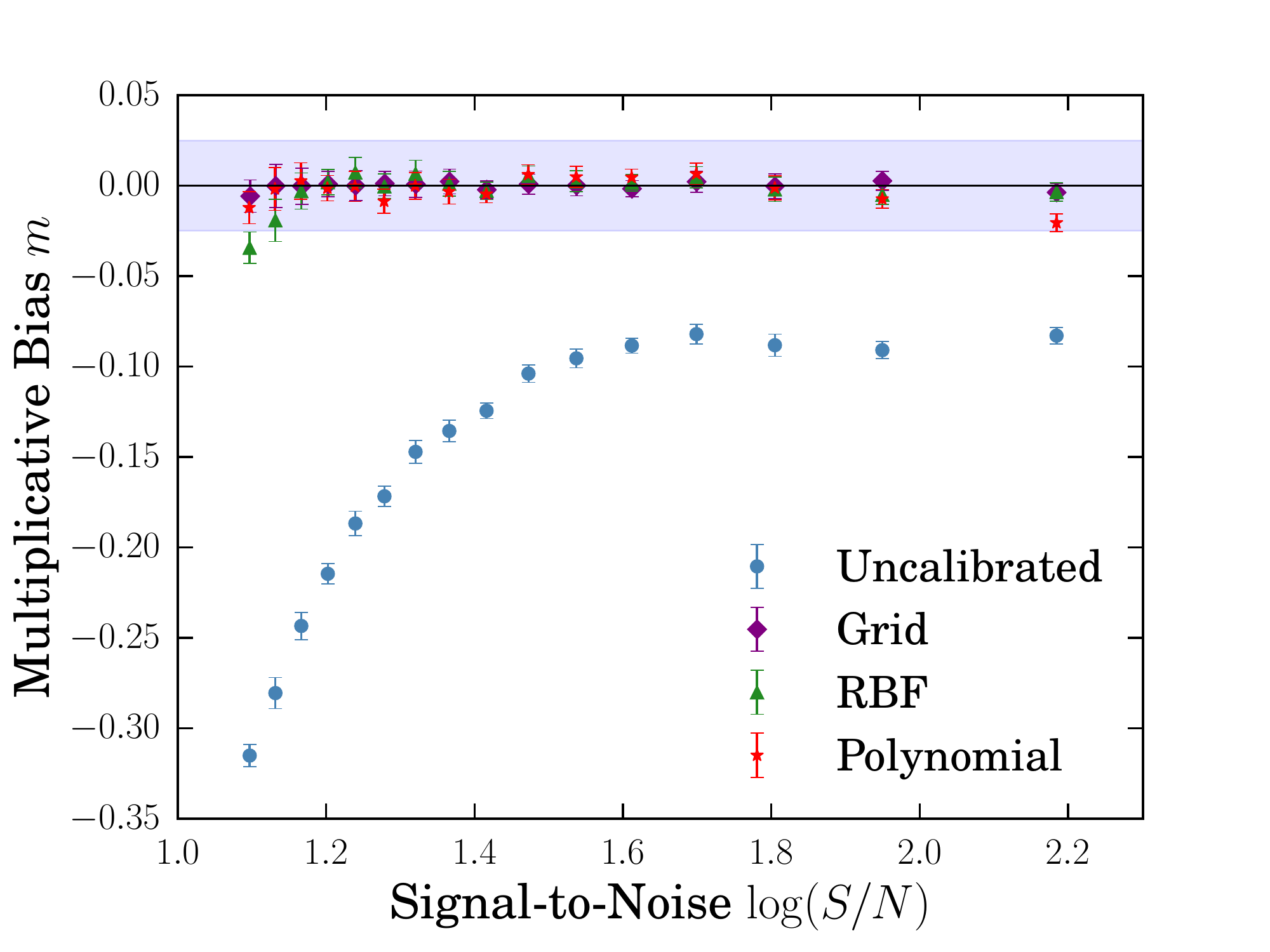}
\includegraphics[width=\columnwidth]{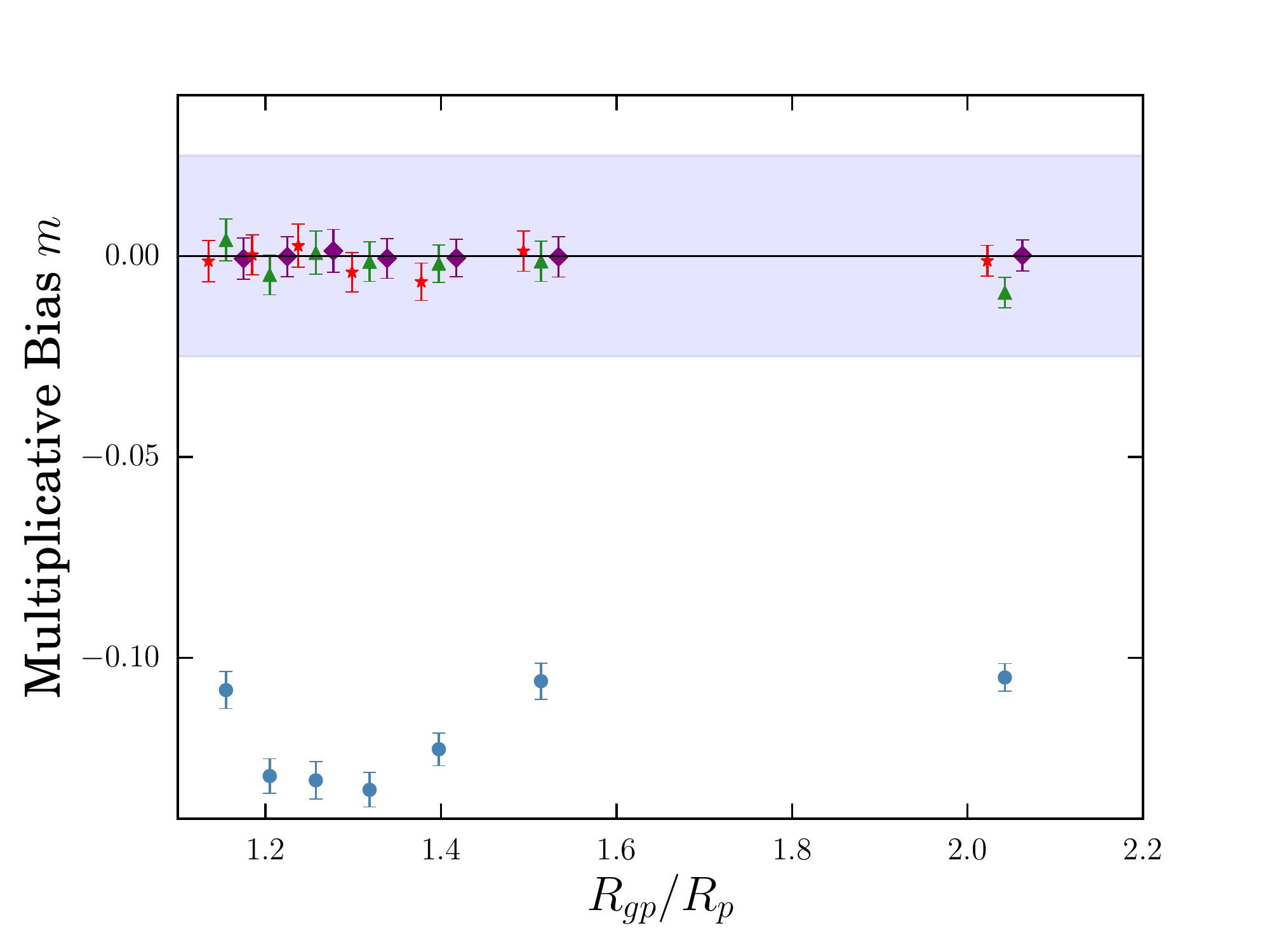}
\caption{Multiplicative bias for \imshape\ measured from the full Y1 simulations, as a function of galaxy signal-to-noise and size. The blue circles in both panels are the measured biases prior to calibration. The other points, labelled grid, RBF and polynomial are the result of correction using the three methods described in the text. The shaded band marks the $\pm 1 \sigma$ Gaussian width of the recommended $m$ prior for the Y1 \imshape\ catalogue.
}
\label{fig:sims:m_calibration_schemes}
\end{center}
\end{figure*}


\subsubsection{Robustness to Tomographic Binning}
\label{sec:im3shape:tomo}
A simulation-based calibration of the sort presented here may be valid for the full dataset,
and yield residual biases within tolerance,
but it does not trivially follow that this is true for all sub-divisions of the data. 
It is perfectly possible that there are competing sources of biases in the catalogue, which by chance cancel to zero. It is also possible to induce biases by introducing extra post-calibration selections based on quantities which correlate with galaxy shape. We will show an explicit example of this in \S\ref{sec:apply:im3shape}. 

Many science applications of the Y1 shape catalogues require a calibration 
that is robust to selection in bins of redshift and angular scale. 
\hoopoe~uses input galaxies with redshifts and generates images in sky coordinates, allowing us to test both of these. 
In this section we focus on the tomographic selection; 
we refer the reader to \citet{des_sim_2017} for discussion of scale-dependent selection effects.

The redshift information we use for each COSMOS galaxy has the form of single point-estimate photo-$z$, 
as estimated using the ACS 30-band photometry.
In the following we assume this measurement is of sufficient quality to allow us to treat it as an input ``true" redshift $z_\mathrm{tr}$.

We build two sets of tomographic bins for the simulated dataset.
For the first set we use the COSMOS measurement
$z_\mathrm{tr}$ for each object; this corresponds to an ideal situation
in which we have no redshift error and sharp-edged (top-hat) redshift bins. 
In the second set we mimic the scatter in photometric redshift that will
inevitably be present in DES.
Each \hoopoe~galaxy is stochastically allocated to one of the four Y1 redshift bins
as follows.
First we construct a realistic set of DES Y1 redshift estimates using the Y1 \imshape~catalogue.
The per-galaxy redshift PDFs obtained from the \blockfont{BPZ} code are stacked in four bins
$z=[(0.2-0.43), (0.43-0.63), (0.63-0.9), (0.9-1.3)]$,
resulting in four normalized distributions $n^{i}(z)$.
We assign each galaxy with true redshift $z_\mathrm{tr}$ to a bin 
$i$, with probability $n^{i}(z_\mathrm{tr})/[\sum \limits_{j=1}^{4} n^{j}(z_\mathrm{tr})]$.
The resulting histograms of $z_\mathrm{tr}$ in each bin cover the full range 
$z \in [0.2-1.3]$, and approximately match the measured $n(z)$ in that bin from the data.
This random assignment
of redshifts is a simplified model;
it does not simulate systematic correlation between photometric redshifts and shear, 
but it does address the smearing out of the estimated redshifts due to noise, 
which we expect to be dominant.

To test the redshift dependence of our calibration, we measure the residual 
bias after splitting into these bins. The results are shown in Figure~\ref{fig:sims:m-vs-z-tomocal} 
and Table~\ref{tab:sims:residual_m} (the latter includes values for the alternative interpolation methods).
The top-hat results show larger RMS scatter than those using the more realistic redshift binning,
which blur out the bias slightly.

The residuals in our top-hat redshift bins demonstrate an important limitation of our current calibration procedure:
namely that galaxy morphology (and thus measurement bias) varies with redshift.
Our calibration assumes that \snr~and size are a sufficient proxy for change,
which will be true only to some level of accuracy. 
The results on DES-like photometric bins suggest this will have less impact on our real data.
We have also neglected noise effects which would induce correlations between both redshift, via fluxes, and ellipticity. 
For higher precision calibrations on future data both of these issues must be addressed.

\begin{table}
\begin{center}
\begin{tabular}{c|cccc}
  \hline
  Method       & $\Delta m^{(1)}$ & $\Delta m^{(2)}$ & $\Delta m^{(3)}$ & $\Delta m^{(4)}$ \\
  \hline 
  Uncalibrated & -0.0886   & -0.0981 & -0.1200 & -0.1547 \\
  Grid         &  0.0069   & -0.0014 & -0.0074 &  0.0013 \\ 
  Radial Basis &  0.0056   & -0.0024 & -0.0082 & -0.0022 \\ 
  Polynomial   &  0.0049   & -0.0028 & -0.0078 & -0.0000 \\ 
  \hline
\end{tabular}
\caption{Residual multiplicative bias in the \imshape~ calibration simulations, after calibration using different methods for 
interpolating $m^{ij}$ nodes onto individual galaxies. The calibration is derived globally, and the residuals are computed for the 
redshift bins used in the cosmic shear analysis in \citet{shearcorr}.
\label{tab:sims:residual_m}
}
\end{center}
\end{table}

As a further test, we split the calibration sample into halves, and then use each half to generate 
a calibration model for the other.  We perform this test twice, once completely at random, such that each part contains 
an equal number of \hoopoe~galaxies, and once by profile, such that each part contains half of the unique COSMOS
profiles used.
Since the biases will depend on both the distribution of galaxy morphologies and the specific observing 
conditions in the calibration sample, both these tests are relevant. The results are shown in Table 
\ref{tab:sims:catalogue_splits}.

\begin{table*}
\begin{center}
\begin{tabular}{c|cccc}
  \hline
  Split Type       & $\Delta m^{(1)}$      & $\Delta m^{(2)}$      & $\Delta m^{(3)}$     & $\Delta m^{(4)}$ \\
  \hline 
  None             & $0.0069 \pm 0.0044$     & $-0.0014 \pm 0.0046$    & $-0.0074 \pm 0.0030$   & $0.0013 \pm 0.0034$ \\
  At random           & $0.0021 \pm 0.0046$     & $-0.0018 \pm 0.0039$    & $-0.0095 \pm 0.0039$   & $-0.0027 \pm 0.0054$ \\
  By COSMOS profile   & $0.0034 \pm 0.0062$     & $-0.0006 \pm 0.0060$    & $-0.0048 \pm 0.0037$   & $0.0073 \pm 0.0039$ \\
  \hline
\end{tabular}
\caption{Residual multiplicative bias in the \hoopoe~simulations under various divisions.
For reference the top line shows the result of applying the fiducial calibration to the whole catalogue,
and is identical to the ``grid" line in Table~\ref{tab:sims:residual_m} and the purple diamonds in Figure~\ref{fig:sims:m_calibration_schemes}.
The other lines show the remeasured biases when using disjoint calibration and validation subsets of the simulation. 
We split first at random, such that there are equal numbers of \hoopoe~galaxies in each subset, and then such that 
there are equal numbers of COSMOS profiles in each.}
\label{tab:sims:catalogue_splits}
\end{center}
\end{table*}

Though subdominant to the other forms of systematic bias discussed in this paper, the residual bias
in the third redshift bin is statistically significant.
Some residual biases might be expected, given that we are using a rigid two parameter grid to
describe complex morphology-dependent biases.
Unfortunately it is not possible to predict the magnitude or sign of these residuals, 
which depend on the details of the COSMOS sample and how they are distributed between redshift bins.
It is thus not guaranteed that the measured residual $m$ in the third redshift bin
implies an equivalent bias in the data.

To account for this uncertainty we widen our prior on $m$ after calibration.
The maximum amplitude of the residual bias in Figure~\ref{fig:sims:m_calibration_schemes} is $|\Delta m^{(i)}| = 0.0074$.
We include this amplitude rounded up to $\sigma_m=0.01$
as a systematic contribution to the prior on residual bias in the \imshape~catalogue
(see \S\ref{sec:budget}).
To be conservative, we also widen the $m$ prior to account for the fact that these residual
biases will be correlated between redshift bins (Appendix \ref{app:m:tomography}).

\begin{figure}
\includegraphics[width=\columnwidth]{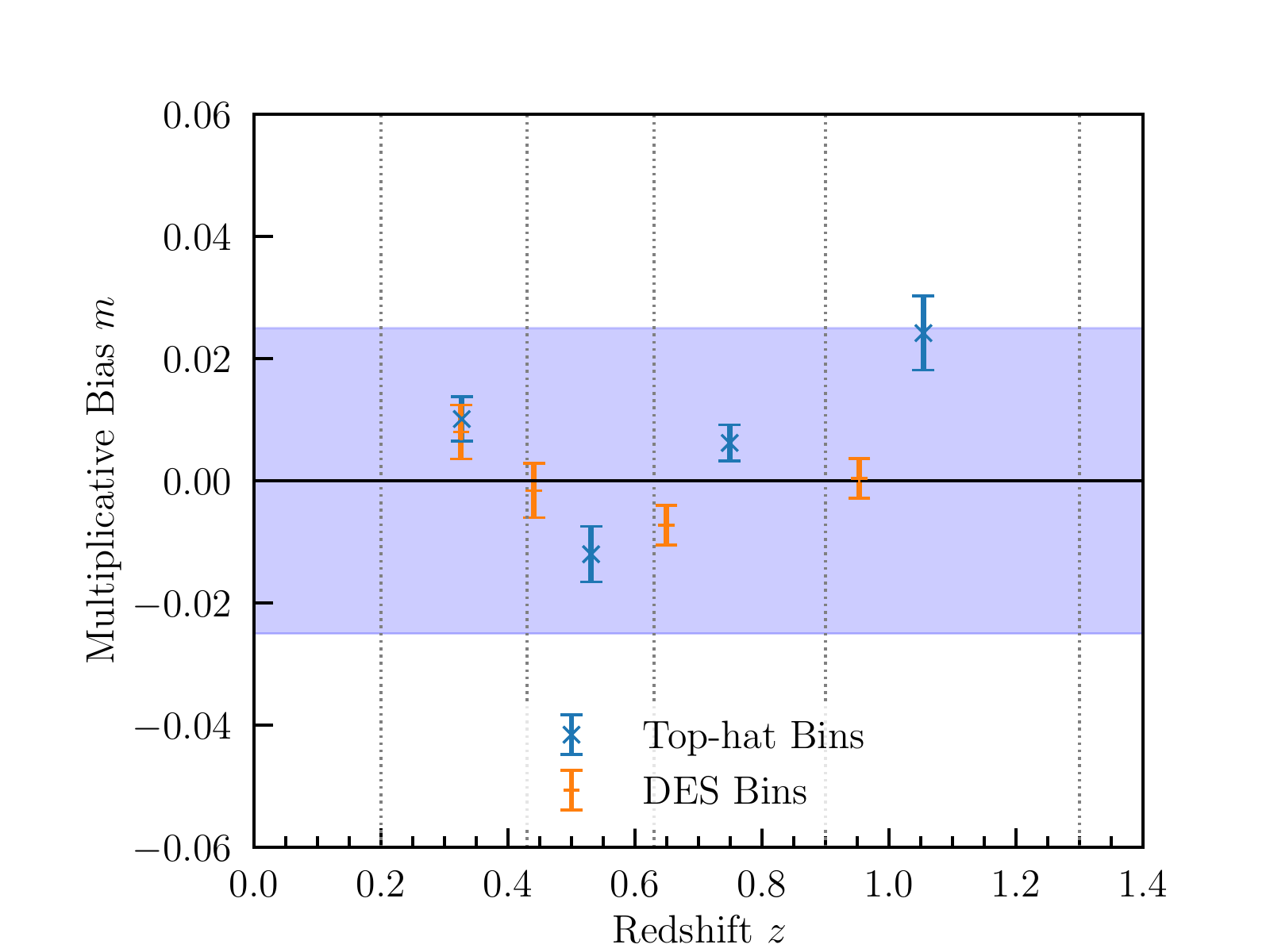}
\caption{Residual multiplicative bias for \imshape\ measured from the full \hoopoe~ catalogue in four tomographic bins after fiducial calibration. For the ``Top-hat'' points objects are binned by their COSMOS redshifts, and for the ``DES'' bins they are assigned to match DES Y1 redshift distributions, partially simulated photometric redshift errors.
As above the shaded band shows the $1\sigma$ width of our Gaussian prior on $m$ in the Y1 \imshape\ catalogue, and the vertical dotted lines show the redshift bin boundaries.
}
\label{fig:sims:m-vs-z-tomocal}
\end{figure}

\subsection{Galaxy Weights for \imshape}
\label{sec:im3shape:weights}
We compute an \imshape\ measurement weight for each galaxy using a very similar calculation to that used in \citetalias{jarvis16}. In summary, we first define a 2D grid of signal-to-noise and size, with each cell containing roughly the same number of galaxies. 
In each cell a zero-centred Gaussian is first fitted to the histogram of the $e_1$ component ellipticity, and the standard deviation is also calculated directly. This yields two similar but non-identical variance estimates, of which we adopt the maximum. The resulting grid is then interpolated using 2D radial basis functions, and the weight allocated to a given galaxy is simply the inverse of the interpolated variance at that position.
This process is designed to estimate the total uncertainty of an ellipticity measurement, including both shape noise and measurement uncertainty, or $(\sigma^2_e + \sigma^2_{SN})$ in the syntax of \citetalias{jarvis16} \S 7.3.
Simulated galaxies were assigned weights by the same process, which were used in constructing the calibration.

\section{Tests of the Shear Measurements}
\label{sec:tests}
Lensing null tests can be difficult to construct, because of 
strong correlations (both inherent and noise-induced) between measured shear
and other measurable observables. 

Nonetheless they remain a powerful tool when correctly understood. These null tests can be broken up into several broad categories:

\begin{description}
\item [{\bf Spatial tests}] check for systematic errors that are connected to the physical
    structure of the camera.  Examples of these are errors in the WCS
    correction, including effects like edge distortions or tree rings
    \citep{plazas14}, and errors related to features on the CCDs such as the
    tape bumps.
    (\S\ref{sec:tests:spatial})

\item [{\bf PSF tests}] check for systematic errors that are connected to the PSF correction.
    This includes errors due to inaccurate PSF modelling as well as
    leakage of the PSF shapes into the galaxy shape estimates.
    (\S\ref{sec:tests:psf})

\item [{\bf Galaxy property tests}] check for
    errors in the shear measurement algorithm related to properties of the
    galaxy or its image.  This can include effects of masking as well, which
    involve the other objects near the galaxy being measured.
    (\S\ref{sec:tests:gal})

\item [{\bf B-mode statistics}] check for systematic errors that show up as a B-mode signal in the
    shear pattern.  The gravitational lensing signal is expected to be
    essentially pure E-mode.
    Most systematic errors, in contrast, affect the E- and B-mode
    approximately equally,
    so the B-mode is a direct test of systematic
    errors.
    (\S\ref{sec:tests:bmode})

\item [{\bf Cross-catalogue comparisons}] check that the two shear catalogues are consistent
    with each other.  It has previously proven extremely challenging 
    to test the agreement 
    between two catalogues directly, because of the calibration corrections that are required 
    when selecting any given subset of a catalogue.  In particular, the
    \metacal\ selection bias correction in \eqn{eq:RSmean} requires
    executing any selection cuts on the sheared renditions of the
    galaxy images.  It is impractical to run \imshape\ in this
    context, and even less so to incorporate \metacal\ cuts into the
    \imshape\ bias correction simulations.
    This makes it 
    impractical to use direct shear comparisons, either object-by-object or on populations, 
    to compare the catalogues.  

The best cross-catalogue comparison we
    can make is therefore to compare the results they yield
    at the ``science'' level, such as cosmological parameter constraints from shear-shear or galaxy-galaxy lensing.
    These tests are described in accompanying papers \citep{keypaper,shearcorr,gglpaper}.
    Considering the large differences between the
    \metacal\ and \imshape\ codes, these are very stringent tests, especially since
    no tuning or modification of any kind was performed to ensure agreement of the results
    from the two codes. Agreement in cosmological parameter constraints demonstrates the agreement 
    of the catalogues for one specific scientific use case, not general agreement in other areas.
\end{description}

\subsection{Spatial Tests}
\label{sec:tests:spatial}

Several sources of error related to the variation in pixel behaviour
and response across the CCDs might, if not properly accounted for,
leave an imprint on the shape catalogue.  These could include silicon ``tree
rings'' \citep{plazas14}, CCD defects and bad columns, and a
``glowing-edge'' effect in which the pixels at the edges of the CCDs
have a different effective size to those in the bulk.  To search for
these effects we can bin the catalogues in pixel and field-of-view
coordinates.  We can also plot mean shears in radial bins around the central points of exposures and CCDs---if all is well these points should have the same signal as randomly chosen points.

Another potential spatial bias comes from the effects of masks in the data, which can have a preferred direction. 
Columns of CCDs, for example, are often masked out together, and diffraction spikes orient with the optics of the telescope. 
The DES focal plane does not rotate, so these effects always correspond to the same orientation in sky coordinates.
This can affect
shape measurement of galaxies near the edge of the mask in two ways---a selection effect on their detection since objects aligned perpendicularly to the mask are more likely to have pixels removed, and on the measurement of their signal-to-noise, for a similar reason.  The latter effect, which is expected to be larger, is included in the \metacal~ response function and the \imshape~ simulations. The exposure dither means that if a galaxy is masked in one exposure it is generally not in others; this reduces the size of the former effect. No detection selection bias is seen in simulations with real masks.

\subsubsection{Position in the Field of View}
\label{sec:tests:fieldofview}

Figure~\ref{fig:tests:fov} shows the mean ellipticity for each pixel in the focal plane, binned across all exposures.  No trends or problematic regions are visible in the plot, which is consistent with noise.

\begin{figure*}
\includegraphics[width=\textwidth]{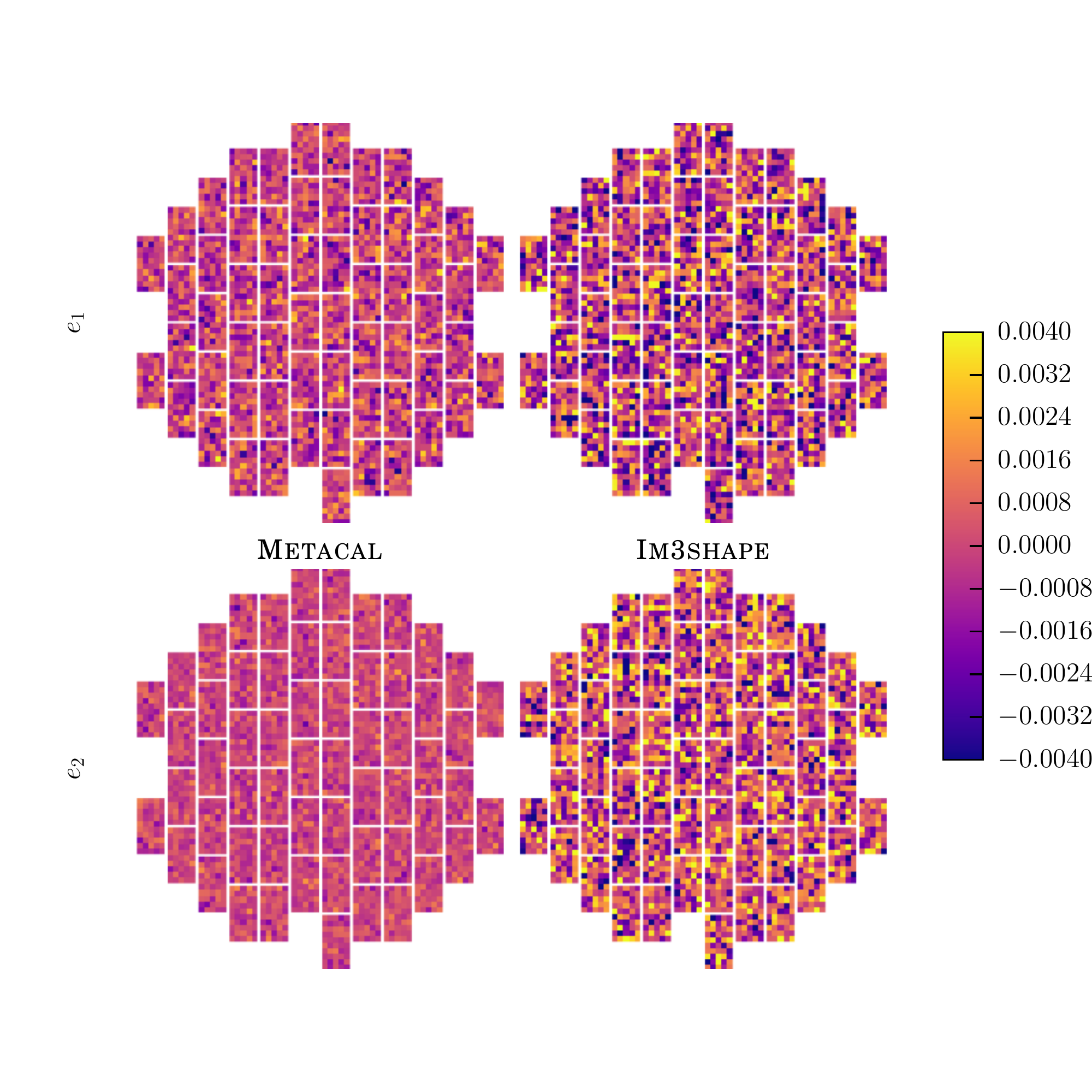}
\caption{The mean ellipticity for \metacal\ (left) and \imshape~(right) binned by position in the focal plane.  Each bin is approximately 400 pixels across. The \imshape~ catalogues use only r-band data and so are noisier.
\label{fig:tests:fov}
}
\end{figure*}

\subsubsection{Tangential Shear around Field Centres}
\label{sec:tests:fieldcenters}

Figure~\ref{fig:tests:fieldcenter} shows the tangential shear binned by radius around field centres (the set of points where the centre of the focal plane is pointing over all exposures) of the Y1 survey.  The mean tangential shear around a comparable number of randomly selected points is subtracted before plotting.
No significant difference is seen at separations $\theta<200$ arcmin, but on larger scales we see a significant deviation of $\gamma_t$ up to $10^{-5}$ around the centres (note that figure shows $\theta \gamma_t$). We verify in \citet{gglpaper} that this contamination is not a significant contaminant to the cosmological $\gamma_t$ signals in our bins; other users of these catalogues should perform similar tests for their science case.

\begin{figure}
\includegraphics[width=\columnwidth]{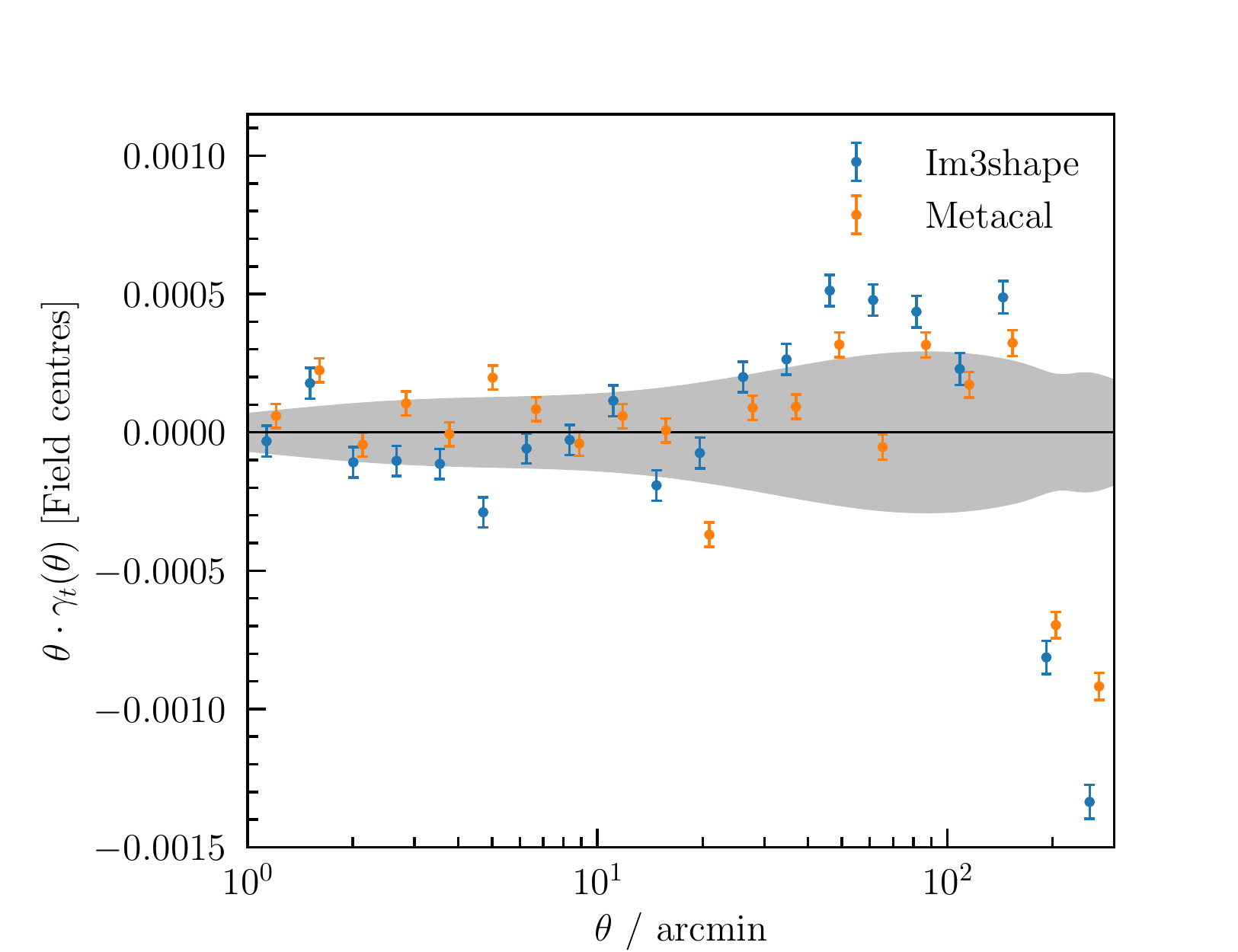}
\caption{The tangential shear of galaxies in the two catalogues around field centres, after subtraction of shear around random points.   The grey band shows ten percent of the weakest tangential shear signal around galaxies in any of the galaxy-galaxy lensing measurements used in \citet{gglpaper}.  This weakest signal combination was from lens bin 3, source bin 4; the data from this combination were used only at $\theta>30$ arcmin.  As noted in the text the deviations from nullity are significant, and a further test was done to ensure that they did not impact our galaxy-galaxy lensing science.
\label{fig:tests:fieldcenter}
}
\end{figure}

\subsection{Tests of the PSF Correction}
\label{sec:tests:psf}
\subsubsection{Shear-PSF Size Correlation}
\label{sec:tests:psfsize}
Figure~\ref{fig:tests:psfsize} shows the mean shear in galaxies in bins of the size of the PSF for the two catalogues, each using its own size metric. In each case a mean shear is visible, which is discussed further in \S \ref{sec:tests:meanshear}. The mild trend in $e_1$ is negligible compared to our measured results.

\begin{figure}
\includegraphics[width=\columnwidth]{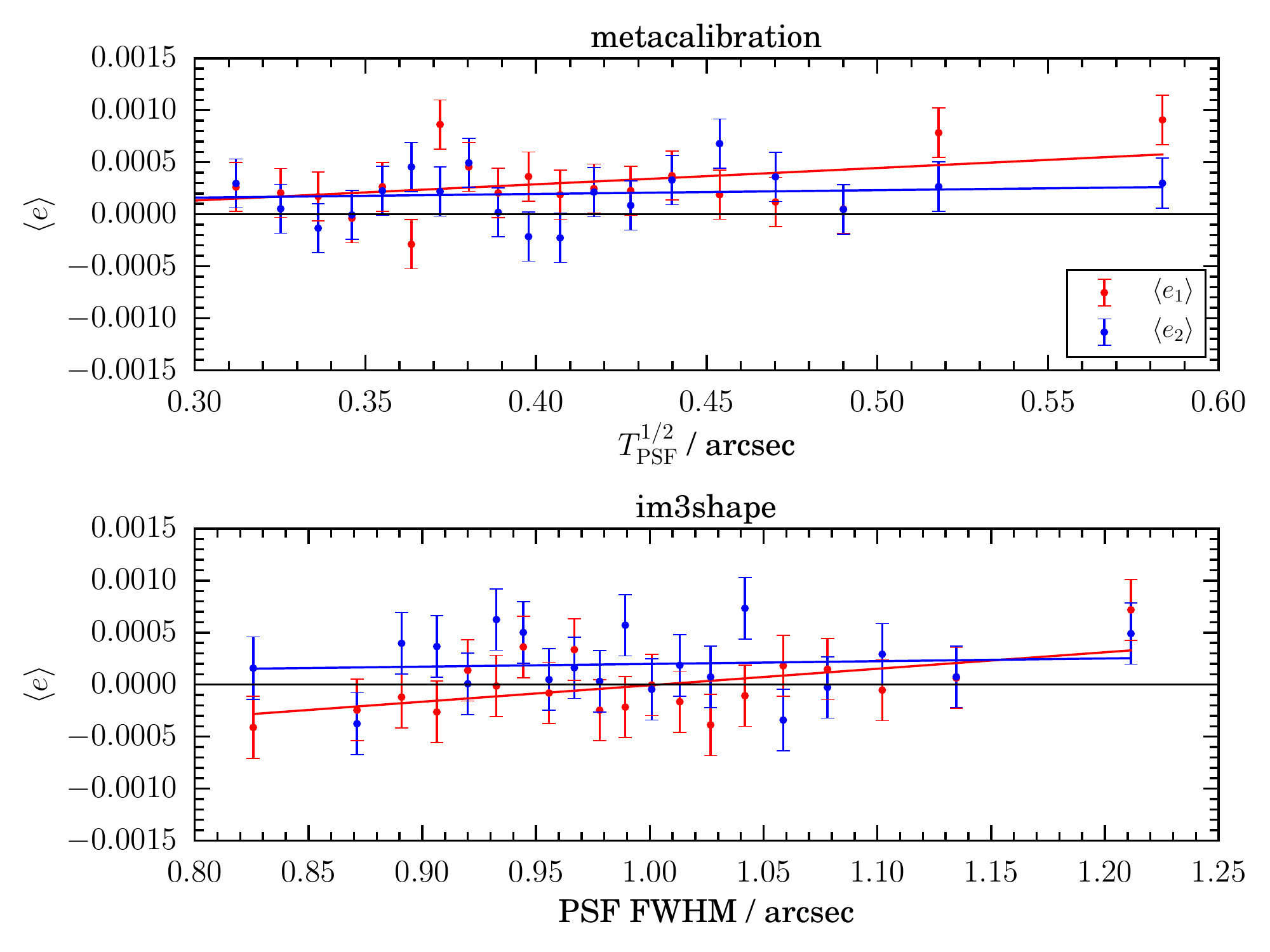}
\caption{The mean galaxy shear as a function of the input PSF size (\metacal\ top, and \imshape\ bottom). The solid lines are linear best fits to the data points.
}
\label{fig:tests:psfsize}
\end{figure}

\subsubsection{Shear-PSF Ellipticity Correlation}
\label{sec:tests:alpha}

Figure~\ref{fig:tests:psfe} shows the mean estimated shear in bins of PSF model ellipticity. The clearly detected correlation between shear and PSF ellipticity can be an indication of imperfect deconvolution of the PSF from the galaxy image, or of simply imperfect modelling of the PSF. \citet{paulinhenriksson08} demonstrate that size errors in the PSF model can potentially produce an additive bias in the shear in the direction of the PSF ellipticity.

\begin{figure*}
\includegraphics[width=\columnwidth]{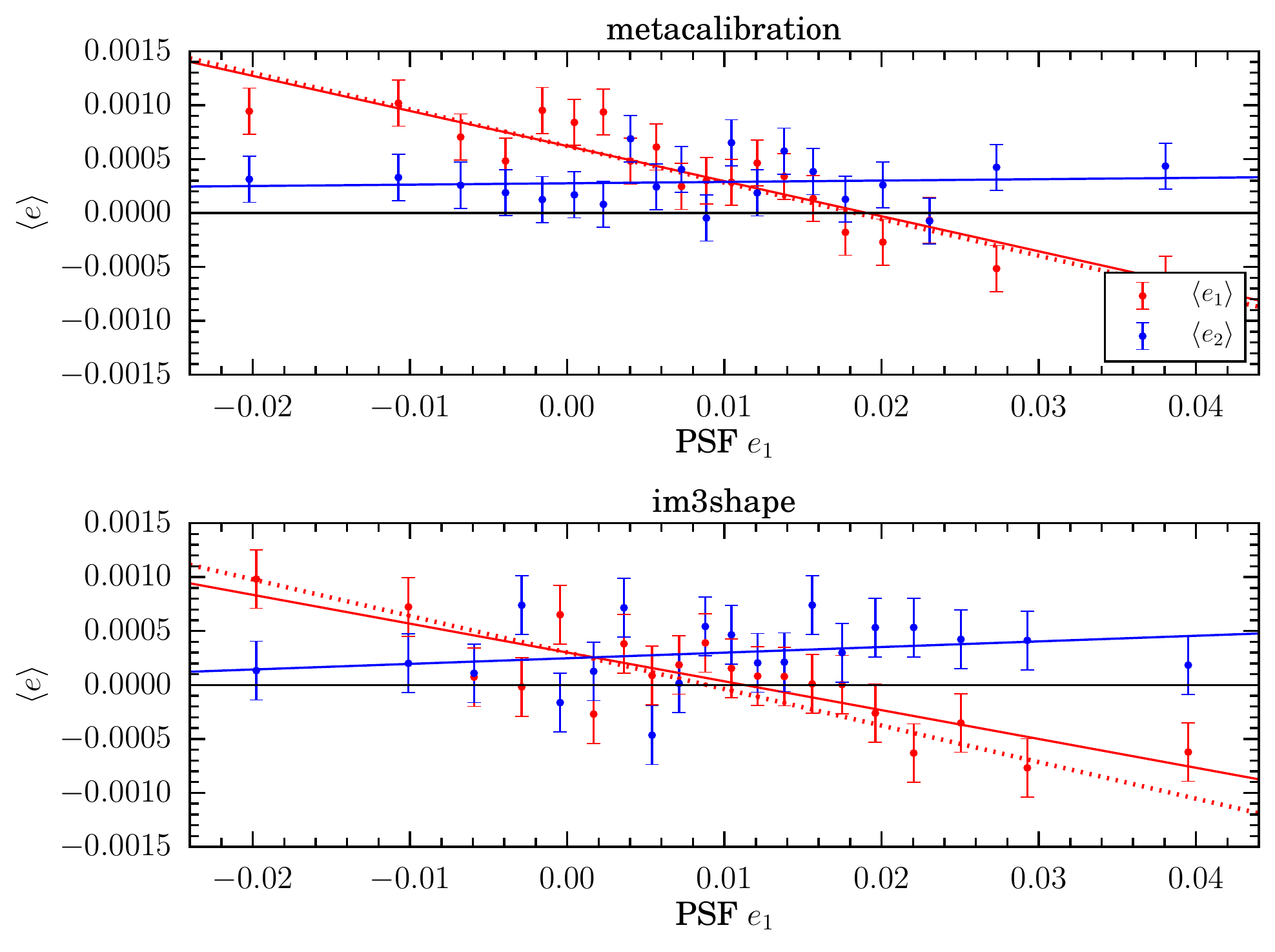}
\includegraphics[width=\columnwidth]{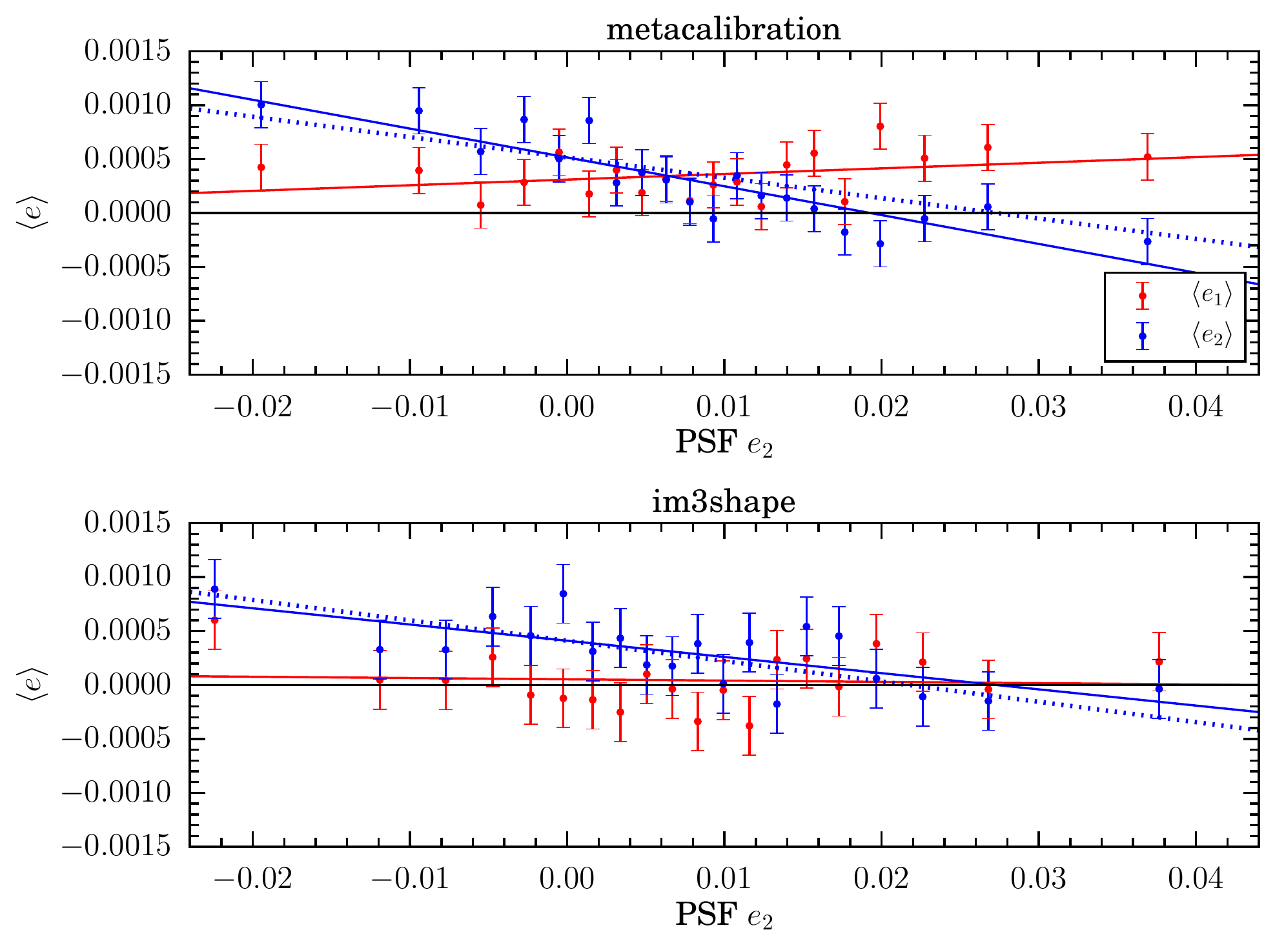}
\caption{The mean galaxy shear as a function of the input PSF ellipticity (PSF $e_1$ left, and PSF $e_2$ right; \metacal\ top, and \imshape\ bottom). The PSF ellipticity means and limits for both catalogues are not identical. This is partly due to \metacal\ using exposures from all three $riz$-bands, while \imshape\ is limited to the $r$-band. Different models are also used to measure the PSF ellipticity in both catalogues. The solid lines are linear best fits to the data points. 
The $e_1(p_1)$ and $e_2(p_2)$ terms have non-zero slopes that are consistent between the two catalogues, and are
consistent with the dotted lines, which come from a model in which all the variation with PSF arises from 
mis-estimation of the PSF, rather than in the shape measurement directly.
The dotted line slopes are $\partial e_1/\partial p_1=-0.030$ and $\partial e_2/\partial p_2=-0.018$.
}
\label{fig:tests:psfe}
\end{figure*}

The trends in Figure~\ref{fig:tests:psfe} can also be produced when there is a correlation between the PSF ellipticity and the PSF model ellipticity errors (i.e. a non-zero $\rho_2$, see \S\ref{sec:psf:rho}). We find that while in this case PSF model size errors do not significantly contribute to Figure~\ref{fig:tests:psfe}, the PSF model ellipticity errors (and their correlation with the PSF model ellipticity) do. We split up the $\alpha$ term in Eq. \ref{eqn:m-c-alpha}, into a ``true'' $\alpha$ from imperfect deconvolution and a term $\beta$ from imperfect measurement:
\begin{equation}
c_i = \alpha_i p_i + \beta_i q_i
\end{equation}
where $p_i$ is component $i$ of the PSF model ellipticity and $q_i$ is component $i$ of the PSF model ellipticity error i.e. $q_i = p_i - p_i^{\mathrm{true}}$.  For perfect deconvolution,  we expect $\alpha=0.$ On the other hand, we expect $\beta$ to be of order $-1$ for any shape measurement algorithm, since an error in the PSF model ellipticity will propagate to an error of the same order of magnitude, but opposite sign, in the inferred shear (see \citealt{paulinhenriksson08} for a theoretical estimate of the linear order effect of PSF size and ellipticity errors).  

While we can estimate the PSF model ellipticity errors $q_i$ at the position of stars, we do not have an estimate at galaxy positions, so we cannot directly estimate the coefficient $\beta$. However, we can use the fact that PSF modelling errors are spatially correlated either in focal plane coordinates (as demonstrated in Figure~\ref{fig:psf:fov}) or in sky coordinates (as demonstrated by the non-zero $\rho_1$ in \S\ref{sec:psf:rho}). We take advantage of the former by computing a PSF ellipticity residual estimate for each galaxy in our sample by interpolating the ellipticity residual maps at Figure~\ref{fig:psf:fov} to the focal-plane positions where the galaxy appears.  We average this quantity over the multiple focal plane positions at which each galaxy was observed; call this $\bar{q_i}$. We can then compute the correlation of this quantity with the inferred shear; this is shown in Figure~\ref{fig:tests:psfeq}. For both components, the slope (which in our model is given by $\beta$) is indeed $O(-1)$ ($\beta_1=-1.08\pm0.08$, $\beta_2=-1.05\pm0.07$).

With this estimate of $\beta$ in hand, we can then estimate the contribution of PSF model ellipticity errors to the correlation between shear and PSF model ellipticity in Figure~\ref{fig:tests:psfe}. Assuming $\alpha=0$ we expect a slope
\begin{equation}
\frac{\partial e_i}{\partial p_i} = \beta_i \frac{\partial q_i}{\partial p_i}
\end{equation}
We estimate the derivative on the right hand side using the ellipticity measurements of the `reserved' stars described in \S\ref{sec:psf:rho}. We find an expected contribution to the slope of $\delta e_i$ vs. $e_{\mathrm{PSF}}$ as shown in Figure~\ref{fig:tests:psfe} of $\frac{\partial e_1}{\partial p_1}=-0.030$ and $\frac{\partial e_2}{\partial p_2}=-0.018$ for two components in \metacal.  
For both the catalogues the overall leakage from PSF to shear is explained well by this term alone.

\begin{figure}
\includegraphics[width=\columnwidth]{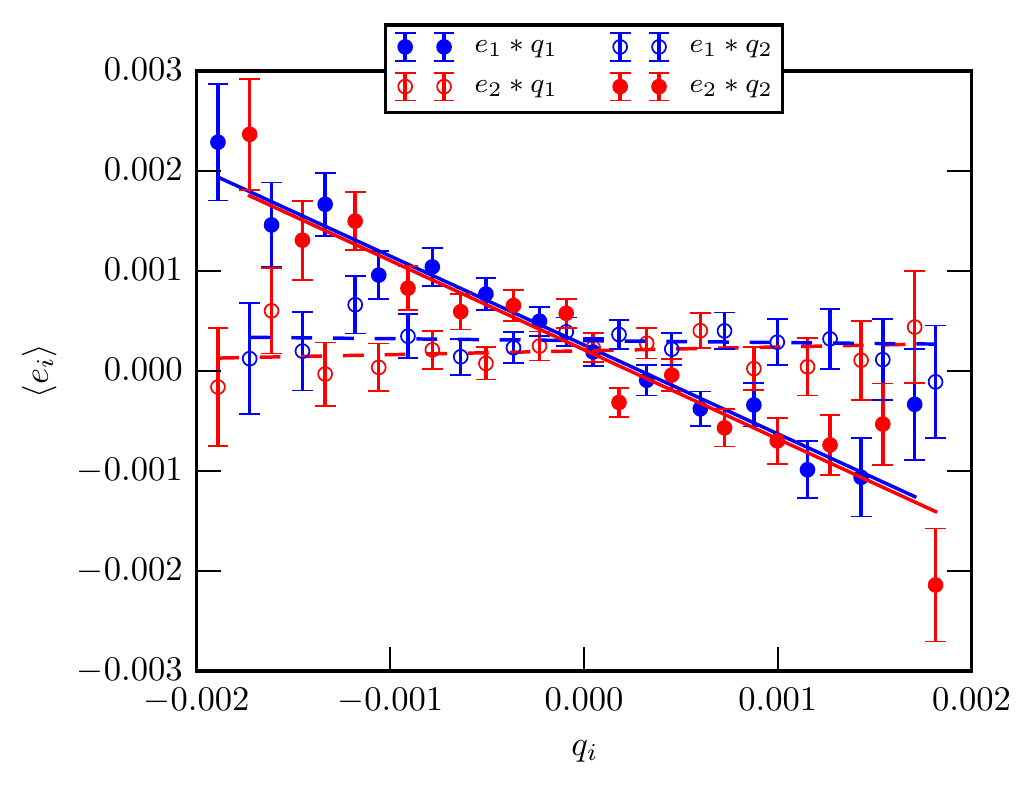}
\caption{The mean shear $\langle e_i \rangle$ as estimated by \metacal\ as a function of focal-plane-position-averaged PSF ellipticity model residual  $q_i$, for each pair $i=(1,2)$ for each quantity. See \S\ref{sec:tests:alpha} for a description of this latter quantity.
}
\label{fig:tests:psfeq}
\end{figure}

\subsubsection{Tangential Shear around Stars}
\label{sec:tests:ggstars}

Since stars will not act as effective gravitational lenses of distant galaxies the measurement of tangential shear around them provides a null test that can reveal problems that could potentially contaminate the galaxy-galaxy lensing signal. In particular, the tangential shear around faint stars, which includes objects used to constrain the PSF modelling, can be used to check issues with PSF modelling and interpolation. On the other hand, bright stars are not used in the PSF modelling but can induce problems around them due to blending and pixel saturation. We define the bright/faint cuts from \citetalias{jarvis16}, with $14<m_i<18.3$ for the bright sample and $18.3<m_i<22$ for the faint one.

The results of these tests are shown in Figure~\ref{fig:tests:tanshearstars}, for both \metacal\ and \imshape. We find the signal to be consistent with zero in all cases, using the covariance from jackknifing the stars.

\begin{figure}
  \includegraphics[width=\columnwidth]{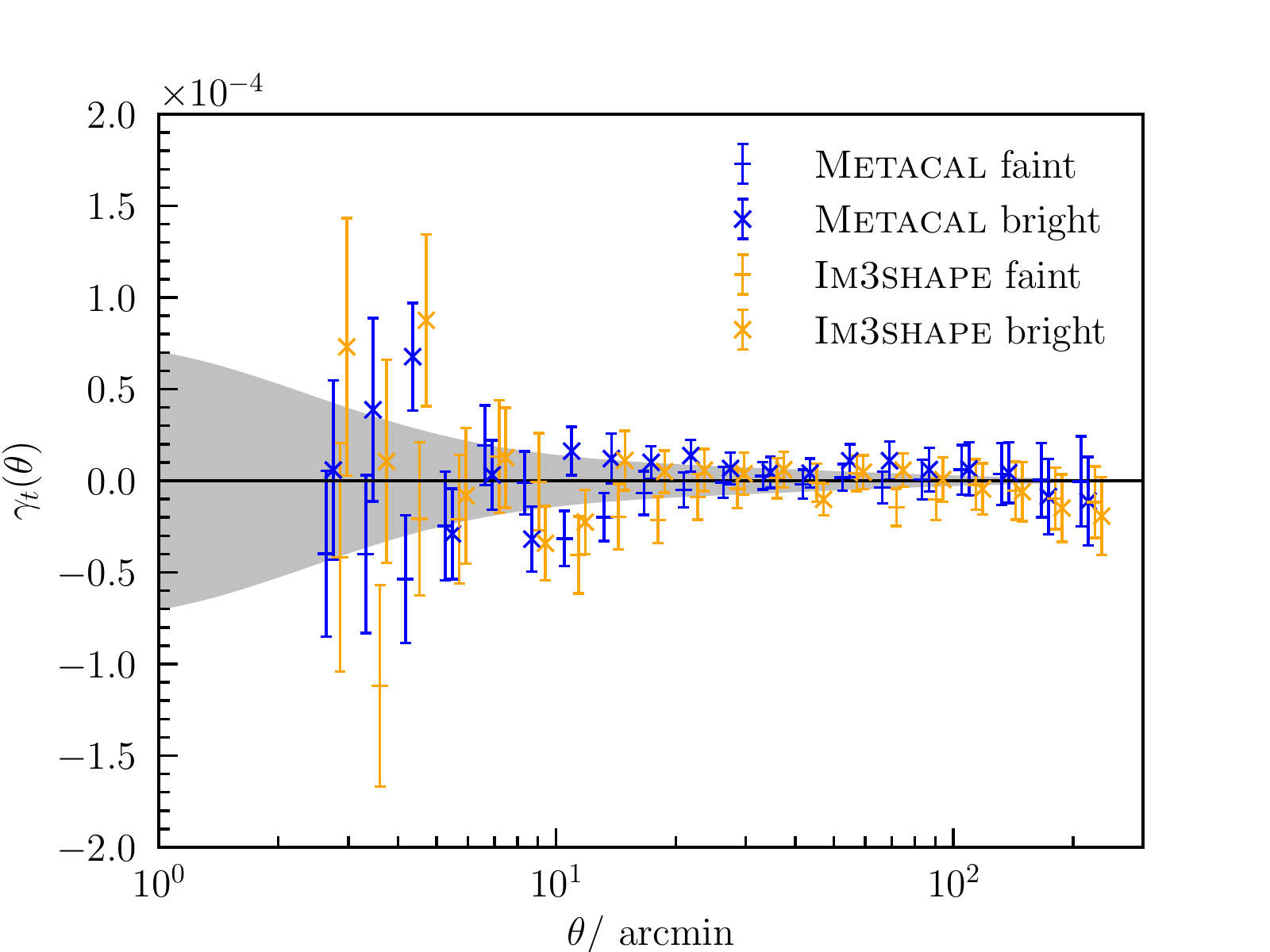}
  \caption{Tangential shear around stars, which have been split into a bright ($14<m_i<18.3$) and a faint sample ($18.3<m_i<22$). The faint sample includes stars used for PSF modelling while bright stars are used to test other effects related to the saturation around them. The error bars come from the jackknife method.  The grey band is 10\% of the weakest expected signal, as in Figure~\ref{fig:tests:fieldcenter}.  The deviations from null in this test at small scales were excluded by the scale cut $\theta>30$ arcmin.
    \label{fig:tests:tanshearstars}
  }
  \end{figure}

\subsection{Galaxy Property Tests}
\label{sec:tests:gal}

\subsubsection{Galaxy Signal-to-Noise}
\label{sec:tests:snr}

Figure~\ref{fig:tests:snr} shows the mean ellipticities $e_1$ and $e_2$ after calibration for the two catalogues in bins of measured signal-to-noise. The $S/N$ value for each catalogue comes from its own measurement process, and different cuts have been applied, so the galaxies in corresponding bins are not identical.

The \imshape\ calibration process uses signal-to-noise as a calibration parameter, so after calibration the mean shape should be uncorrelated with signal-to-noise. The \metacal\ calibration process should also remove any correlation.  Any physical correlations between shape and brightness should have no preferred direction, and therefore should not appear in Figure~\ref{fig:tests:snr}.

\begin{figure}
\includegraphics[width=\columnwidth]{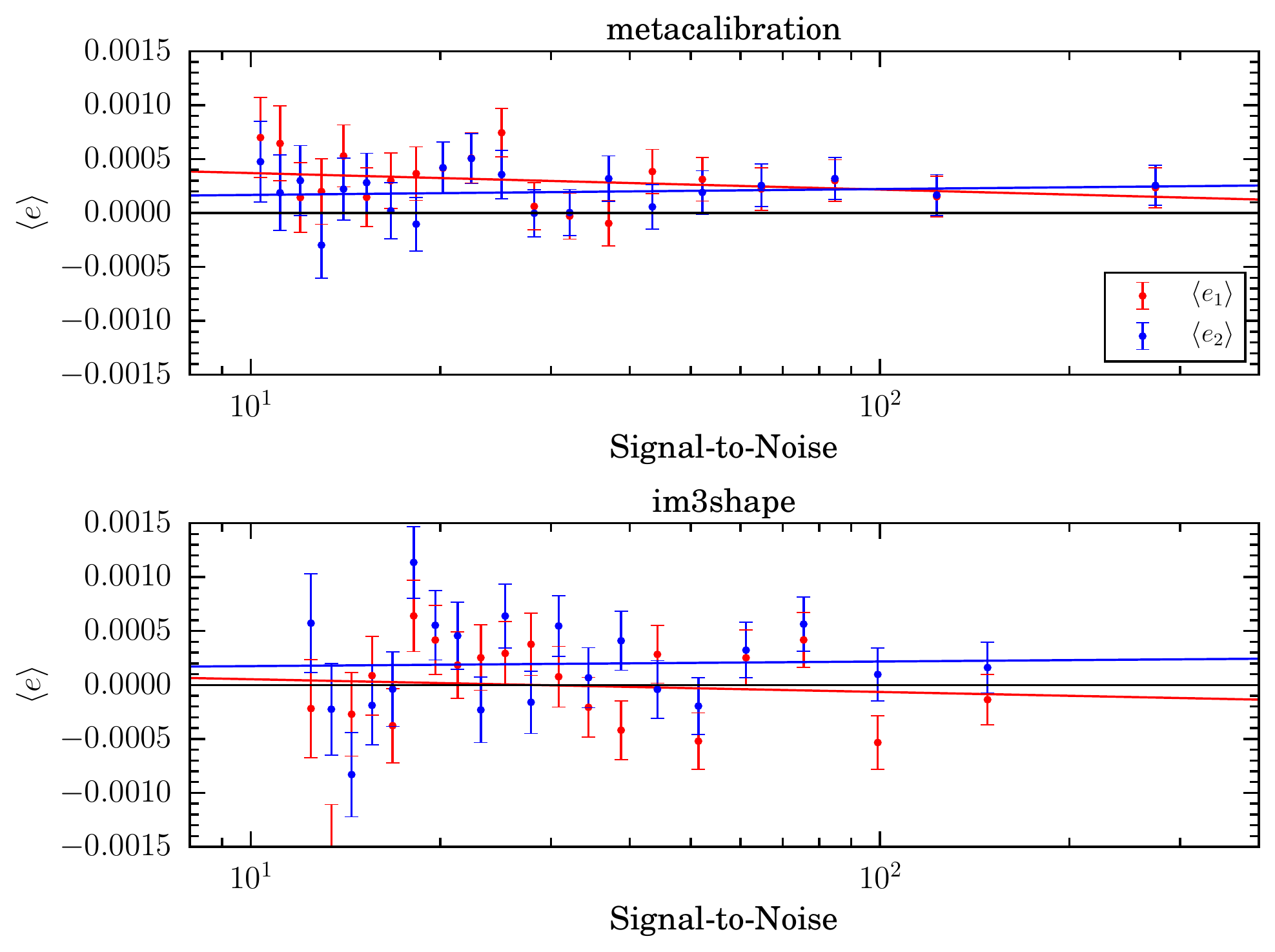}
\caption{The mean galaxy shear as a function of the signal-to-noise (\metacal\ top, and \imshape\ bottom). The solid lines are a linear best-fit to the data points.
}
\label{fig:tests:snr}
\end{figure}

Neither catalogue shows a strong trend in shear with $S/N$.
Both catalogues have a non-zero mean shear which is visible here and discussed in \S \ref{sec:tests:meanshear}.

\subsubsection{Galaxy Size}
\label{sec:tests:size}

In \imshape\ the size of source galaxies is measured by $R_{gp}/R_p$ as described in \S \ref{sec:im3shape:overview}, and in \metacal\ we measure it as $T^{\frac{1}{2}}$, where $T$ is defined in equation (\ref{eq:Tdef}).  In neither case should any correlation between the size and ellipticity be present after applying the calibration process.
Figure~\ref{fig:tests:galsize} shows mean galaxy ellipticity as a function of the size metrics. In neither case do we see any significant trend in ellipticity as a function of galaxy size.

\begin{figure}
\includegraphics[width=\columnwidth]{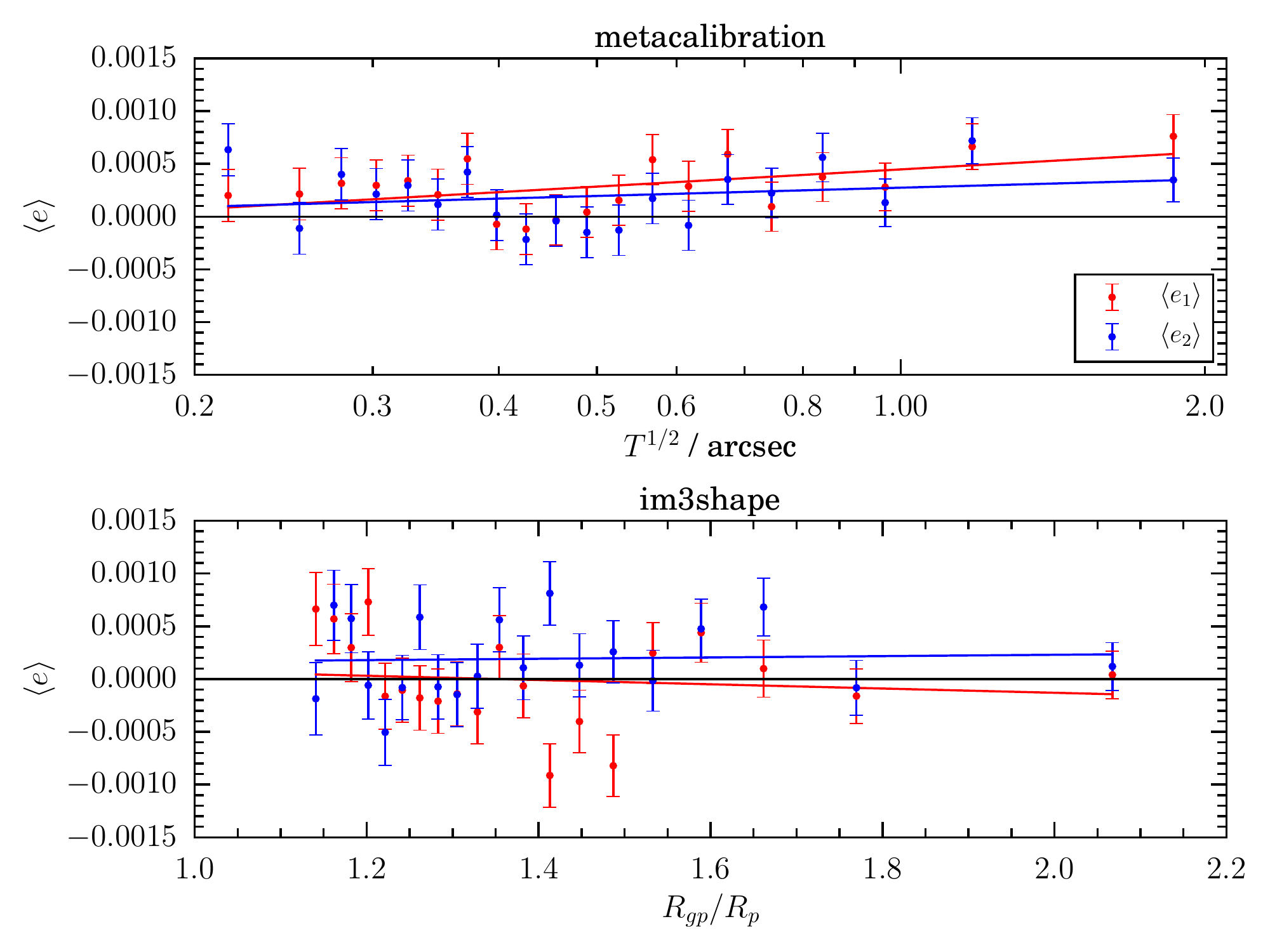}
\caption{The mean galaxy shear as a function of the galaxy size (\metacal\ top, on a logarithmic scale, and \imshape\ bottom). The solid lines are a linear best-fit to the data points.
}
\label{fig:tests:galsize}
\end{figure}

\subsection{B-mode Statistics}
\label{sec:tests:bmode}
In general relativity (GR) lensing produces an E-mode (curl-free) pattern in the shear field, and no detectable B-mode (divergence-free) pattern.  Contaminants to the signal such as PSF or other leakages might produce either mode, so if we assume GR we can use the presence of B-modes as a null test\footnote{Higher order lensing effects and PSF leakage can both generate B-modes, but not at a level detectable here.}.  In Figure~\ref{fig:tests:bmode} we show tomographic B-mode measurements using the redshift bins used for cosmology meaurements in \citet{shearcorr} and \citet{keypaper}. They are computed using a pseudo-$C_\ell$ estimator \citep{hikage2011}. The displayed $\chi^2$ values are for individual bins; the total $\chi^2$ values, which also account for the correlations between bins, are $99.8$ for \metacal\ and $90.8$ for \imshape, which for 90 data points indicates no evidence for B-modes.

\begin{figure*}
\includegraphics[width=\textwidth]{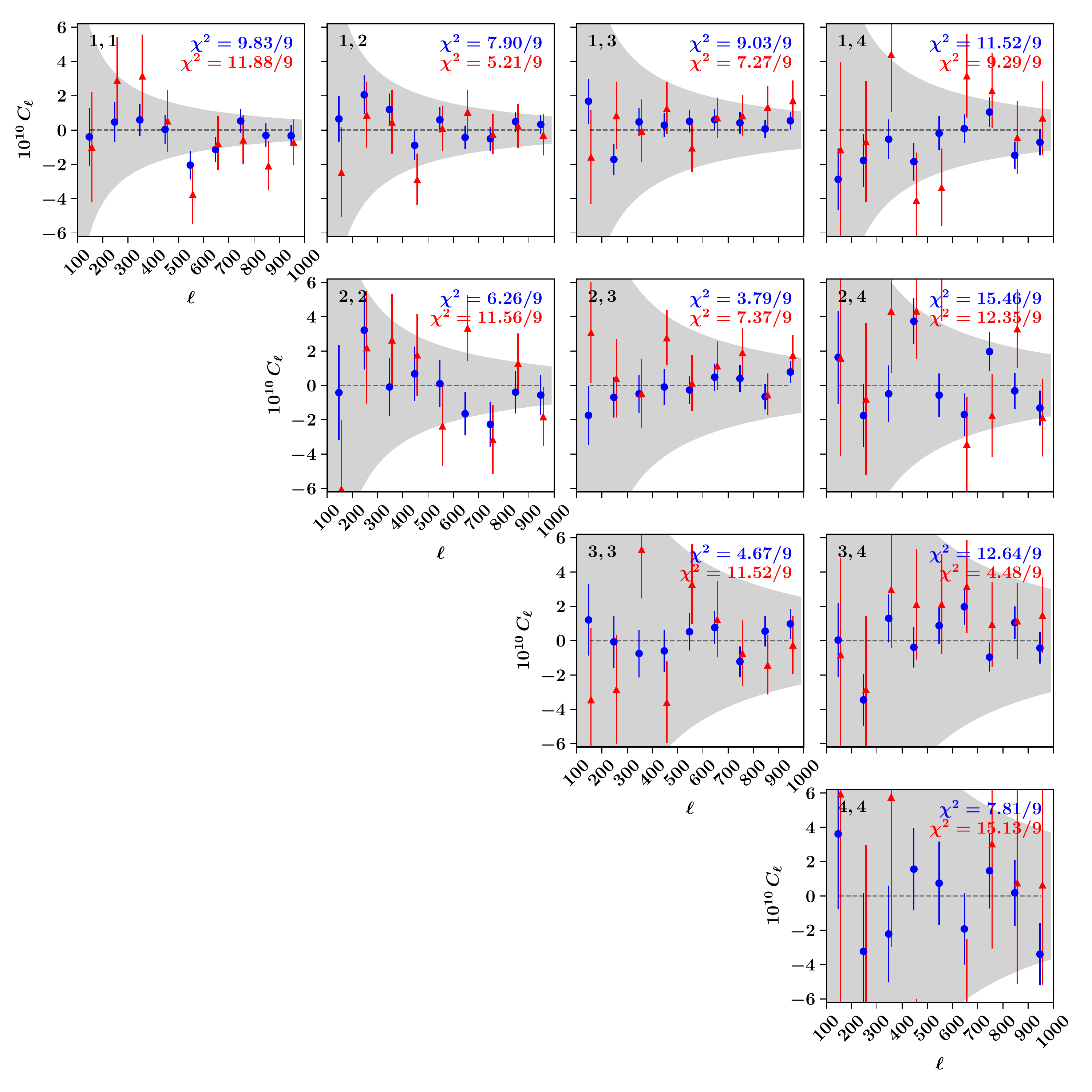}
\caption{The measured B-mode in \metacal~  (blue circles) and \imshape~ (red triangles), and the corresponding detection $\chi^2$ values. The measurements use the tomographic bins 1 to 4 as used in \citet{shearcorr}, and the auto- and cross-correlations between them are shown. The value is expected to be close to zero in the absence of systematics.  Error bars were calculated from a set of log-normal simulations matching the DES-Y1 survey geometry and redshift distributions.  The grey band show $\pm$ the E-mode signal in a fiducial cosmology.
\label{fig:tests:bmode}
}
\end{figure*}

\subsection{Summary of Systematics Tests}
\label{sec:tests:summary}

There are two
additive systematics, a PSF-related term and mean shear, which should each be subtracted, marginalized over,
or demonstrated to be subdominant in precision analyses.  They are described in \S \ref{sec:psf:rho} and \S \ref{sec:tests:meanshear}, respectively.

There is also a residual uncertainty in the overall multiplicative calibration of the two catalogues, which 
should be marginalized over.  This is described in \S \ref{sec:budget}.

We have found no tests that imply any further systematic errors are present at a level significant for 
our cosmological analyses.

\section{Using the Shear Catalogues}
\label{sec:cats}

\subsection{Mean Shear}
\label{sec:tests:meanshear}

Both catalogues show a non-zero mean ellipticity over the entire Y1 survey, with a value $e_{1,2} = (3.5,2.8) \times 10^{-4}$ for \metacal~ and $e_{1,2} = (0.4,2.9)\times 10^{-4}$ for \imshape.  This is marginally too large to be the mean of cosmic shear over the field: in log-normal simulations we find a standard deviation of the mean $e_{1,2}\ \sim 1 \times 10^{-4}$ over our region. An added constant shear will appear as a constant offset in correlation function measurements, so this signal should either be subtracted or marginalized over in cosmological parameter estimation.

The origin of this mean shear is not known definitively, and may be the combination of several effects. Charge self-interaction effects in the DECam CCDs on star and galaxy profiles are expected to cause mean shears in the $e_1$ direction that are of the order of a few times $10^{-4}$ (cf. Table 1 of \citealt{Gruen15}).  The PSF correlations in Figure~\ref{fig:tests:psfe} are also expected to contribute a similar order of magnitude, but our model of the PSF model errors does not entirely describe this mean shear \citep{shearcorr}. 

\subsection{Catalogue Flags}
\label{sec:cats:flags}

Each catalogue uses its own flagging scheme to determine which
galaxies can safely be used in science applications.

\imshape\ uses a similar flagging scheme as in \citetalias{jarvis16}, based on a small number of 
 ``error flags'' that remove extreme objects, and a larger number of 
 ``info flags'' that remove the tails of histograms in various quantities.  They are combined into a single {\sc flags\_select=0} value in our final catalogues.
The flags are applied when computing the calibration scheme, so they 
should always be used identically in precision applications, by requiring:
\begin{equation}
 \textsc{flags\_select}=0.
\end{equation}

The flag values are described in Appendix \ref{sec:im3shape:flags}.  The main changes we have made since \citetalias{jarvis16} are 
reducing our minimum \snr~ from 15 to 12, and our minimum $\rgp$ from $1.15$ to $1.13$, reflecting our improved calibration 
simulations for small faint objects.

The \metacal\ catalogue can be adapted to new data cuts, as described below in \S \ref{sec:apply:mcal}. As a default cut, which is incorporated into the {\sc flags\_select} column, we use and recommend:
\begin{eqnarray}
    \snr > 10 \nonumber \\
    T/\Tpsf > 0.5 \nonumber \\
\end{eqnarray}

\begin{figure}
\includegraphics[width=\columnwidth]{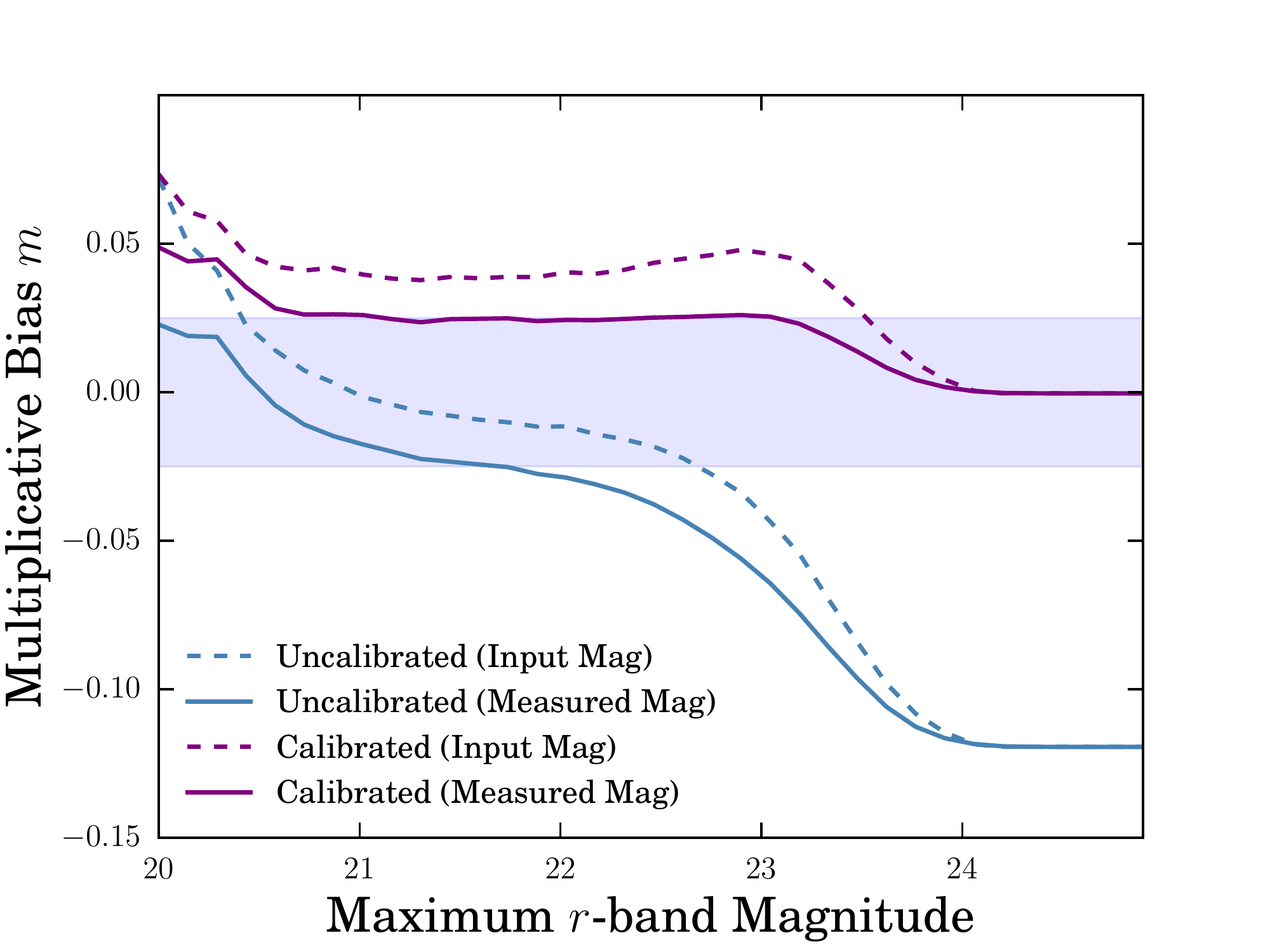}
\caption{Multiplicative bias for \imshape, measured from the \hoopoe~image simulation. 
Solid lines show the measured bias after imposing a maximum $r$-band magnitude,
using the measured values from the \sex~run on the simulation.
Dashed lines show the same, but defining the cut using the input magnitudes. 
Purple curves use the fiducial calibration scheme described in \S\ref{sec:sims}, and blue curves are uncalibrated. 
The shaded region shows the 2.5\% range that is our final \imshape~ calibration uncertainty.
This illustrates the danger of selection biases when cutting on any observable which correlates with ellipticity, as magnitude  does.}
\label{fig:sims:m-vs-mag}
\end{figure}

\subsection{Applying the \imshape~ Calibration}
\label{sec:apply:im3shape}

The \imshape~ calibration yields $m$ and $c$ values for each object, but because they include corrections for selection
biases these values are only correct when applied to the specific default \imshape~cuts. Further cuts can induce biases 
due to noise that correlates between ellipticity and other quantities.  We have verified in \S \ref{sec:im3shape:tomo} that
the specific split into tomographic bins used in concurrent DES papers does not induce a significant bias, but this cannot
be assumed for any other binnings. An example of a cut that does induce significant bias is shown in Figure~\ref{fig:sims:m-vs-mag},
which illustrates that imposing an upper magnitude limit can induce biases of $2-6\%$, depending on the limit.

The \imshape~calibration is applied in the same manner as it was in \citetalias{jarvis16}. The estimator for the mean shear on an ensemble of galaxies is:

\begin{equation}
\langle \gamma_{a} \rangle = \frac{\sum_i{w_i (e_{a,i} - c_{a,i})}}{\sum_i{w_i(1+m_i)}}
\end{equation}
where $a=1,2$ and $i$ sums over all objects.
For a shear two-point estimator, the additive $c$ correction should first be applied, then the galaxy pairs rotated to the tangential and cross directions $e^+$ and $e^\times$, and the weights and multiplicative corrections applied to these rotated values:
\begin{equation}
\xi_{\pm} = \frac{\sum_i \sum_j w_i w_j (e^+_i e^+_j \pm e^\times_i e^\times_j)}{\sum_i \sum_j w_i w_j (1+m_i)(1+m_j)}
\end{equation}
where the sums run over $(i,j)$ pairs separated by angle $\theta$.

\subsection{Applying the \metacal\ Calibration}
\label{sec:apply:mcal}
To calibrate the \metacal\ catalogues we make use of the five different sets of measurements that the code makes on each object: on the original image, and on versions positively and negatively sheared in the $e_1$ and $e_2$ directions. We can use these measurements to calibrate bias in both the shape measurement for each object, and any selection biases.  The metacalibration process can only calibrate selection biases when cuts are made on quantities which have been measured by the \metacal\ estimator, so that \eqn{eq:RSmean} can be used to calculate corrections. These include, but are not limited to, galaxy and PSF sizes and ellipticities, \snr, and fluxes.

As an example of the process, one should use this calculation to estimate the mean shear under some selection:
\begin{enumerate}
\item For a given selection criterion $S$, and for each shear component $\gamma_1$ and $\gamma_2$ determine three subsets of the catalogue:
\begin{description}
    \item[$S_0$] - by applying $S$ to the column measured on the original image,
    \item[$S_+$] - by applying $S$ to the column measured after positive shear in component $i$,
    \item[$S_-$] - by applying $S$ to the column measured after negative shear in component $i$.
\end{description}
\item For each pair of shear components $i,j$, compute $R_{\gamma_{ij}}$, the average of column $R_{ij}$ over galaxies in $S_0$
\item Compute $R_{s_{i}} = (\langle e_i \rangle_{S_+} - \langle e_i \rangle_{S_-}) / \Delta \gamma $, where the $e_i$ columns are the ones measured on the original image, and the averages are taken over the subsets in the subscripts. For DES Y1 we used $\Delta \gamma = 0.01$.
\item The complete response for the ensemble is $R_{ij} = R_{\gamma_{ij}} + \delta_{ij} R_{s_{i}}$.
\item The best estimate for the mean shear is $\gamma_i = R^{-1}_{ij} \langle e_j \rangle_{S_0}$
\end{enumerate}
The process for correcting a two-point estimator in the same way is described in \citetalias{SheldonHuff2017}. For convenience, the default {\sc flags\_select} column has four additional sheared versions, {\sc flags\_select\_X}, where {\sc X} is in \{1p,1m,2p,2m\}, representing the component and direction of the sheared version of the flag.

To enable bias correction of samples selected by photometric redshift, we have applied our \photoz\ estimators to the flux measured for each galaxy by \metacal\ both before and after the \metacal\ shears are applied (there are four additional catalogues, for $\pm \delta e_{1,2}$).  Given a galaxy selection, if the mean of the shears is to be used, for example in the null tests described in \S \ref{sec:tests}, the correction factors in \S\ref{sec:metacal} must be applied.  The calibration factor that must be applied when constructing two-point statistics is described in \citetalias{SheldonHuff2017}.  For higher-order statistics an equivalent calibration should be derived.  The selection biases in mean shear for the DES redshift bins range from 1.1 to 2.5\%.

Note that the correction factors applied to each \metacal\ object are large, because the model used is so simple, so neglecting them is unlikely to be a good approximation in any context.

\subsection{Number Density}
\label{sec:cats:neff}

Values of the (effective) number density and shape variance for three definitions for the two catalogues are shown in Table~\ref{tab:neff}. The \emph{raw} value is simply the total number of selected objects per unit area.

The variance $\sigma^2_\gamma$ of the estimated shear in a catalogue quantifies its overall constraining power. This quantity is generally split into $\sigma^2_\gamma \equiv \sigma_e^2 / n_\mathrm{eff}$, where $\sigma_e^2$ is a shape variance and $n_\mathrm{eff}$ a number density. Any pair of definitions of these two quantities that yield the correct $\sigma^2_\gamma$ may be used as a metric to quantify the constraining power.

The definition described in \citet{chang13} is given by:
\begin{equation}
n^\mathrm{C13}_\mathrm{eff} = \frac{1}{A}\sum{\frac{\sigma^2_{\mathrm{sh}} }  {\sigma^2_{\mathrm{sh}} + \sigma^2_{m,i} }}
\end{equation}
where $A$ is the surveyed sky area.  
For \metacal\ the measurement noise $\sigma^2_{m,i}$  is derived from the estimated measurement covariance matrix, accounting for the response term, and the intrinsic shape noise $\sigma^2_{\mathrm{sh}}$ then derived from this and the total observed variance (the denominator). For \imshape\ the shape noise is estimated from high signal-to-noise objects where measurement noise is minimal, and the measurement noise derived from this and the total variance.

The total shape variance $\sigma^2_e$ for one galaxy is the term $\sigma^2_{\mathrm{sh}} + \sigma^2_{m,i}$.

The definition in \citet{heymans2012} is useful here for comparison to other surveys:
\begin{equation}
n^\mathrm{H12}_\mathrm{eff} = \frac{1}{A}\frac{\left(\sum{w_i}\right)^2 }{\sum w_i^2}
\end{equation}
Since we use unit weights for \metacal\ this is the same as the raw value for that catalogue.



\begin{table}
\begin{center}
\begin{tabular}{|c|c|c|c|}
   \hline
   Catalogue &  Definition & Number / arcmin$^2$ & $\sigma_e$ \\
   \hline
   \metacal  & Raw & 6.38 & \\
             & Chang-13 & 5.96 & 0.27\\
             & Heymans-12 & 6.38 & 0.28 \\
   \imshape  & Raw & 4.02 & \\
             & Chang-13 & 3.16 & 0.25 \\
             & Heymans-12 & 3.72 & 0.28 \\
   \hline
\end{tabular}
\end{center}
\caption{Number density values and noise per component using various definitions as described in the text for the two catalogues.
}

\label{tab:neff}
\end{table}

\subsection{Systematic Error Budget} 
\label{sec:budget}

Additive errors from the PSF, including the $\alpha e_{\rm PSF}$ PSF leakage term, have been discussed in \S \ref{sec:tests:psf}.
In the following, we will describe the budget of multiplicative systematic errors $m$ to be used with both shape catalogues. In general, where we have an untreated systematic then we add the full width of its possible range to the prior on $m$.  Where we have a systematic that is treated but we believe the treatment to be imperfect, we add 50\% of the width to the prior.

We do not have a hard requirement on the multiplicative bias, since any uncertainty can be marginalized over at the parameter estimation stage, but at about 2\% uncertainty the associated error is comparable to the statistical uncertainty in the data.

\subsubsection{\metacal} \label{sec:budget:metacal}

The dominant contribution to the systematic calibration uncertainty of the
\metacal\ shear catalogue is the effect of overlapping objects.  

Additional multiplicative bias contributions arise from two effects related to the slight size bias of our PSF models, described in detail in \S\ref{sec:mcal:wrongpsf} and \S\ref{sec:mcal:psfmult}. These are strongly subdominant to the neighbour bias and its uncertainty, especially when added in quadrature, which is appropriate since the effects are \emph{a priori} uncorrelated. 
An overview is shown in Table~\ref{tab:m:metacal}, which results in a total Gaussian prior on the multiplicative bias with center $m=0.012$ and $1\sigma$ width 0.013. We note that all of the effects contributing to \metacal\ multiplicative bias are potentially highly correlated between source redshift bins in a tomographic analysis, a fact that needs to be accounted for (see Appendix \ref{app:m:tomography} for details).

\begin{table}
\begin{center}
\begin{tabular}{|l|l|r|r|}
   \hline
   Effect & Bin & Mean  & Gaussian $\sigma_m$ \\
   & correlation & [$10^{-2}$] & [$10^{-2}$] \\
   \hline
   Stellar contamination & yes & 0.0 & 0.2 \\
   PSF size bias & yes & 0.0 & 0.3 \\
   Neighbour bias & yes & 1.2 & 1.2 \\
   \hline
   Total & & 1.2 & 1.3 \\ 
   \hline
\end{tabular}
\caption{Multiplicative bias budget for \metacal. For the effect of contributions that are correlated between redshift bins on tomographic analyses, see Appendix \ref{app:m:tomography}.
}
\label{tab:m:metacal}
\end{center}
\end{table}

\subsubsection{\imshape}\label{sec:imbudget}

As in DES-SV (see \citealt{jarvis16}, their section 7.3.2), we calibrate the \imshape\ catalogue using image simulations. For DES Y1, however, we have developed a new independent pipeline for generating image simulations, which includes several improvements intended to mimic the properties of actual Y1 data as closely as possible (see Table~\ref{tab:sims:sv_y1_comparison} and \citealt{des_sim_2017}). Unlike SV, where the multiplicative bias uncertainty was estimated by the (dis)agreement of our two pipelines on simulations, our systematics budget for \imshape\ is now set by quantifiable residual uncertainties in the statistics and methodology of the simulation-based calibration.

A main part of this uncertainty is due to the effect of detected and undetected neighbours on multiplicative bias. Comparison of \imshape\ runs on identical sets of simulations with and without neighbouring galaxies \citep{des_sim_2017} (see their Figure 16) has shown a mean shift in calibration corresponding to $\Delta m=-0.034$ -- mean shears measured in simulations with neighbours are about 3 per cent larger than for a sample of fully isolated galaxies. While our simulation-based calibration is a bona fide correction of this effect that should capture its dominant influence on shape measurement, some aspects of the effect in real data might not be captured in the simulations. Among these are the relative alignment of physical neighbours and coherently sheared projected neighbours (both of which, however, influence the distribution of relative alignments only slightly), the influence of completely blended galaxies (which are rare in DES data), or the clustering and coherent alignment of undetected background galaxies (which are, however, altogether a subdominant contribution to neighbour-related bias in \imshape\ as shown in \citealt{des_sim_2017}). We therefore assume half of the neighbour-induced shift in our calibration as an uncertainty, giving $\sigma_m=0.017$, which is conservative given the degree of realism present in the simulations.

Additional systematic uncertainties in the simulation-based calibration are due to 
\begin{itemize}
\item assignment of cut-out sizes in the MEDS file -- While stamp size in the real data is based on measurements of a source's size and ellipticity performed on the coadd using \sex, in the bulk of the simulations the code mistakenly truncated each simulated galaxy's image at the bounds of a postage stamp of the original source whose position it was taking.
Larger or highly elliptical galaxies in our simulations are therefore often assigned smaller boxes than they would in the data. When we remove galaxies from the simulations that are in an incorrectly-sized box, the population of galaxies used in deriving the calibration significantly changes. 
We were unable to devise a cut based on the true input properties of the 
simulated galaxies that did not significantly alter the ellipticity and size distributions.
Reweighting was found to be unreliable (since the cut leaves very few large elliptical galaxies
to upweight) and not robust to binning in \snr~and \rgp.
Re-running the calibration on a small subset of the data with this problem fixed, we find a maximum change in multiplicative bias of 0.025.

We conservatively assume a top-hat prior of $|m|<0.025$ per redshift bin, corresponding to a Gaussian $\sigma_{m}=0.014$. While this is a non-negligible contribution to our overall error budget, rerunning the full simulation with box sizes assigned according to properties measured in the stack, as is done in the data, would require a large computational overhead and represents a non-trivial restructuring of the simulation pipeline that we defer to future work.

\item removal of bad objects from the COSMOS galaxy sample -- We have manually identified galaxies among the COSMOS library that show issues potentially affecting multiplicative bias calibration (see Appendix~\ref{app:eyeball}). The change in calibration when removing flagged galaxies is at most 0.009 among the top-hat redshift bins. Despite these efforts, the choice of which galaxies to remove remains somewhat subjective, and the change in the galaxy properties of the sample that ends up being used in the simulations could cause a small systematic difference of our calibration sample from the galaxies present in the real data. We therefore assume half of the observed shift, or 0.005, as a systematic uncertainty.

\item variation of morphology as a function of redshift -- Our calibration is described by a function of signal-to-noise ratio and size, which are the two most important parameters affecting noise and selection biases, and performed separately for galaxies better fit by bulge- and disc-type S\'ersic profiles. Noise bias does, however, depend on additional galaxy properties whose distributions at given signal-to-noise ratio and size vary as a function of redshift. When we apply the calibration derived from the full galaxy sample to a redshift bin in our simulations, we therefore find deviations from zero bias, which are at most at the level of 0.01. These residual biases are robust to all of the choices which enter the calibration scheme  (interpolation, binning etc). Since lensing analyses virtually always employ some implicit or explicit redshift-dependent re-weighting of sources, we assume an additional systematic uncertainty of this size in each redshift bin.
\end{itemize}
Summing in quadrature, these effects amount to a Gaussian systematic uncertainty of $\sigma_m=0.018$.

The volume of our simulations is large but finite, leading to a statistical uncertainty on mean $m$ of $\sigma_m=0.002$. 

In addition, \imshape\ suffers biases from the mean size residual in our PSF models. To assess the impact of error in the interpolated PSF kernel at source positions, we run \imshape~on the high \snr~simulations described in \S\ref{sec:mcal:psfmult}. These images consisted of analytic profiles under constant shear $\mathbf{g}=(0.01,0.00)$, and convolved with highly elliptical Moffat PSFs. Using these simple simulations we compute a single-number mean bias $m$.
Given the lack of variance in $g$, and the low noise in these images the statistical error on these measurements can safely assumed to be negligible at $\mathcal{O} ( 10^{-4} )$. Comparing the results from the reference simulations (no PSF size bias) with a set of images with a mean dilation 
$\left <\Delta \Tpsf/T \right > \sim 8.3 \times 10^{-4}$ 
we find a change in the mean multiplicative bias of $\Delta m = 0.006$.
While these simulations likely capture the dominant part of the effect, realistic galaxy morphology might change the result at a second order level. Adopting a conservative approach, we scale this observed change by a factor of 1.5 before incorporating it into our $m$ prior. 
After conversion to Gaussian width, maintaining variance, the total impact is $\sigma _m = 0.005$. Note that we find no change in additive biases between the two simulations.

Contamination of our \imshape\ source sample with point sources is a negligible effect at the strict cuts we have applied to the catalogue, which we include in the error budget at an estimated level below one per-mille.

Adding these effects in quadrature, as shown in an overview in \S \ref{tab:m:im3shape}, we arrive at a total Gaussian prior on the multiplicative bias with center $m=0.0$ and $1\sigma$ width 0.025. Some of the effects contributing to the multiplicative bias are correlated between source redshift bins and estimated in a way that requires us to account for this fact in a tomographic analysis (see Appendix \ref{app:m:tomography} for details).

\begin{table}
\begin{center}
\begin{tabular}{|l|l|r|r|}
   \hline
   Effect & Bin & Mean  & Gaussian $\sigma$ \\
   & correlation & [$10^{-2}$] & [$10^{-2}$] \\
   \hline
   Stellar contamination & yes & 0.0 & 0.1 \\
   PSF size bias & yes & 0.0 & 0.4 \\
   Neighbour bias & yes & 0.0 & 1.7 \\
   Calibration statistical & yes & 0.0 & 0.2 \\
   Calibration systematic & no & 0.0 & 1.8 \\
   \hline
   Total & & 0.0 & 2.5 \\ 
   \hline
\end{tabular}
\caption{Multiplicative bias budget for \imshape. The calibration systematic error includes the effects of cut-out size, removal of bad objects from the COSMOS sample, and variation of morphology besides size as a function of redshift. For the effect of contributions that are correlated between redshift bins on tomographic analyses, see Appendix \ref{app:m:tomography}.}
\label{tab:m:im3shape}
\end{center}
\end{table}

\section{Summary and Discussion}
\label{sec:conclusions}

We have presented two independent catalogues of shape measurements of galaxies imaged in Year One of the Dark Energy Survey, covering 1500
square degrees of the Southern sky and containing 34.8 million (for \metacal) and 21.9 million (for \imshape) objects. 
They have passed a battery of tests that demonstrate that, when appropriately used with calibration and error
models, they are suitable for weak lensing science.   In companion papers we also demonstrate that these catalogues lead
to consistent cosmological constraints: in \citet{shearcorr} we study constraints from cosmic shear, in \citet{gglpaper}
we examine galaxy-galaxy lensing, and in \citet{keypaper} we study both in conjunction with galaxy density correlation 
functions.

This work is the first application of the metacalibration method to real data, and demonstrates its significant power in
the face of noise and model biases, and especially for its approach to dealing with the pernicious issue of selection
biases.  This work also makes use of the most sophisticated image simulations currently used for
lensing noise and model bias calibration, which account for a wide range of systematic effects that would otherwise
produce a significantly biased \imshape\ catalogue. We emphasize the importance of carefully ensuring that simulations match the
data in as many ways as possible, including PSF patterns, masks, weights, and processing selections.  

Since the analysis of our science verification SV data in \citet{jarvis16} we have made the following improvements
to our shape pipelines, in addition to the improvements in data reduction described in \citet{y1gold}:
\begin{itemize}
    \item Implemented the metacalibration technique, to incorporate internal calibration of measurement and selection biases, into the \metacal\ pipeline.
    \item Included neighbours, sub-detection objects, stars, masks, realistic PSFs, and multiple exposures, in our calibration simulations for the \imshape\ pipeline.
    \item Explored the effects of blending on our results.
    \item Identified that PSF-associated errors arise almost solely from mis-estimation of the PSF itself.
    \item Enumerated a full list of error sources contributing to our final uncertainty.
\end{itemize}
As in the SV analysis, having two 
independent methods for shear estimation has provided us with significantly greater confidence in the robustness of the catalogue 
calibrations.

Like all weak lensing catalogues, the DES Y1 results come with an uncertainty on overall calibration in the form of a
multiplicative bias $m$.  Correctly and conservatively determining priors on this quantity is a vital part of
characterising a lensing catalogue, and in this case we obtain $\sigma_m \sim 1.2 \cdot 10^{-2}$ for \metacal~ and
$\sigma_m \sim 2.5 \cdot 10^{-2}$ for \imshape.  These values are small enough that this systematic is subdominant in
cosmic shear cosmology parameter estimation. An additional correction is required due partly to mis-estimation of the PSF,
which leads to correlation of the inferred shear with the PSF shape and a correctable residual additive bias in the catalogues.

The data presented here comprise only 20\% of the full Dark Energy Survey, and work to analyze the subsequent two
years of data has already begun.  To fully exploit that upcoming opportunity, the methods described here must be refined
and improved in a number of major ways.  We plan to further extend our calibration simulations to more precisely mimic
the processes applied to real data.  We will continue to improve and adapt our shape measurement algorithms, including 
incorporating new methods like BFD and applying novel calibration techniques like metacalibration to existing methods like \imshape\ using multiple
bands. We are also in the process of implementing a new PSF measurement pipeline to reduce the significant PSF model residuals found in our Y1 catalogues. 

The catalogues presented in this paper will be made publicly available following publication, at the URL \url{https://des.ncsa.illinois.edu/releases}.


\section*{Acknowledgements}

We thank Rachel Mandelbaum and the GREAT3 team for providing the COSMOS galaxy images used for the im3shape calibration. 
We are also grateful to the eyeballing volunteers, among them Mandeep Gill, Annalisa Mana, Ben Mawdsley, Tom McClintock, Alessandro Nastasi, and Corvin Stern, for their help with validating the COSMOS galaxy images.

Support for DG was provided by NASA through the Einstein Fellowship Program,
grant PF5-160138 awarded by the Chandra X-ray Center, which is
operated by the Smithsonian Astrophysical Observatory for NASA under
contract NAS8-03060.  ES was supported by DOE grant DE-AC02-98CH10886.
M. Jarvis, B. Jain, and G. Bernstein are partially supported by the US Department of Energy
grant DE-SC0007901 and funds from the University of Pennsylvania.

Based in part on observations at Cerro Tololo Inter-American Observatory,
National Optical Astronomy Observatory, which is operated by the Association of
Universities for Research in Astronomy (AURA) under a cooperative agreement with the National
Science Foundation.

Funding for the DES Projects has been provided by the U.S. Department of Energy, the U.S. National Science Foundation, the Ministry of Science and Education of Spain, 
the Science and Technology Facilities Council of the United Kingdom, the Higher Education Funding Council for England, the National Center for Supercomputing 
Applications at the University of Illinois at Urbana-Champaign, the Kavli Institute of Cosmological Physics at the University of Chicago, 
the Center for Cosmology and Astro-Particle Physics at the Ohio State University,
the Mitchell Institute for Fundamental Physics and Astronomy at Texas A\&M University, Financiadora de Estudos e Projetos, 
Funda{\c c}{\~a}o Carlos Chagas Filho de Amparo {\`a} Pesquisa do Estado do Rio de Janeiro, Conselho Nacional de Desenvolvimento Cient{\'i}fico e Tecnol{\'o}gico and 
the Minist{\'e}rio da Ci{\^e}ncia, Tecnologia e Inova{\c c}{\~a}o, the Deutsche Forschungsgemeinschaft and the Collaborating Institutions in the Dark Energy Survey. 

The Collaborating Institutions are Argonne National Laboratory, the University of California at Santa Cruz, the University of Cambridge, Centro de Investigaciones Energ{\'e}ticas, 
Medioambientales y Tecnol{\'o}gicas-Madrid, the University of Chicago, University College London, the DES-Brazil Consortium, the University of Edinburgh, 
the Eidgen{\"o}ssische Technische Hochschule (ETH) Z{\"u}rich, 
Fermi National Accelerator Laboratory, the University of Illinois at Urbana-Champaign, the Institut de Ci{\`e}ncies de l'Espai (IEEC/CSIC), 
the Institut de F{\'i}sica d'Altes Energies, Lawrence Berkeley National Laboratory, the Ludwig-Maximilians Universit{\"a}t M{\"u}nchen and the associated Excellence Cluster Universe, 
the University of Michigan, the National Optical Astronomy Observatory, the University of Nottingham, The Ohio State University, the University of Pennsylvania, the University of Portsmouth, 
SLAC National Accelerator Laboratory, Stanford University, the University of Sussex, Texas A\&M University, and the OzDES Membership Consortium.

The DES data management system is supported by the National Science Foundation under Grant Numbers AST-1138766 and AST-1536171.
The DES participants from Spanish institutions are partially supported by MINECO under grants AYA2015-71825, ESP2015-88861, FPA2015-68048, SEV-2012-0234, SEV-2016-0597, and MDM-2015-0509, 
some of which include ERDF funds from the European Union. IFAE is partially funded by the CERCA program of the Generalitat de Catalunya.
Research leading to these results has received funding from the European Research
Council under the European Union's Seventh Framework Program (FP7/2007-2013) including ERC grant agreements 240672, 291329, and 306478.
We  acknowledge support from the Australian Research Council Centre of Excellence for All-sky Astrophysics (CAASTRO), through project number CE110001020.

This manuscript has been authored by Fermi Research Alliance, LLC under Contract No. DE-AC02-07CH11359 with the U.S. Department of Energy, Office of Science, Office of High Energy Physics. The United States Government retains and the publisher, by accepting the article for publication, acknowledges that the United States Government retains a non-exclusive, paid-up, irrevocable, world-wide license to publish or reproduce the published form of this manuscript, or allow others to do so, for United States Government purposes.

This work received funding from the European Union's Horizon 2020 research and innovation programme grant agreement 681431.

The metacalibration calculations were performed using computational resources
at SLAC National Accelerator Laboratory. We thank the SLAC computational team for their consistent support.
Besides computing resources at SLAC, this research used computing resources at the National Energy Research Scientific Computing Center, a DOE Office of Science User
Facility supported by the Office of Science of the U.S. Department of
Energy under Contract No. DE-AC02-05CH11231.
It also used resources at the Ohio Supercomputing Center.

\bibliography{literature,des_y1kp_short}

\appendix

\section{The COSMOS Eyeball Project}
\label{app:eyeball}

It was noted after they had been run that the simulations described in \S\ref{sec:sims}
contained a small number of obvious artifacts,
originating from defects in the input COSMOS profiles.
These included deblending failures, and objects with diffuse light profiles
truncated at the edges of the postage stamp.
Two such objects are shown in Figure~\ref{fig:cosmos_eyeball_example}.
To assess the level to which these objects affect shape measurements on the simulations
we initiated a small-scale crowdsourcing project within the scientific community of the 
Dark Energy Survey. 
Our specific aim here was to compile a list of COSMOS galaxies
in our input catalogues that are qualitatively ``bad",
and so should be excluded from our simulations. 

The exercise was set up as follows. Each deconvolved COSMOS galaxy was 
reconvolved with a small nominal PSF and rendered into a postage stamp image 
at HST pixel resolution with no additional noise.
The images were compiled in random order, and
via a simple web interface
users were assigned batches of $\sim100$ images.
Galaxies were assigned to the categories shown in Table~\ref{tab:eyeball_numbers}.  

\begin{figure}
\centering
\includegraphics[width=0.48\columnwidth]{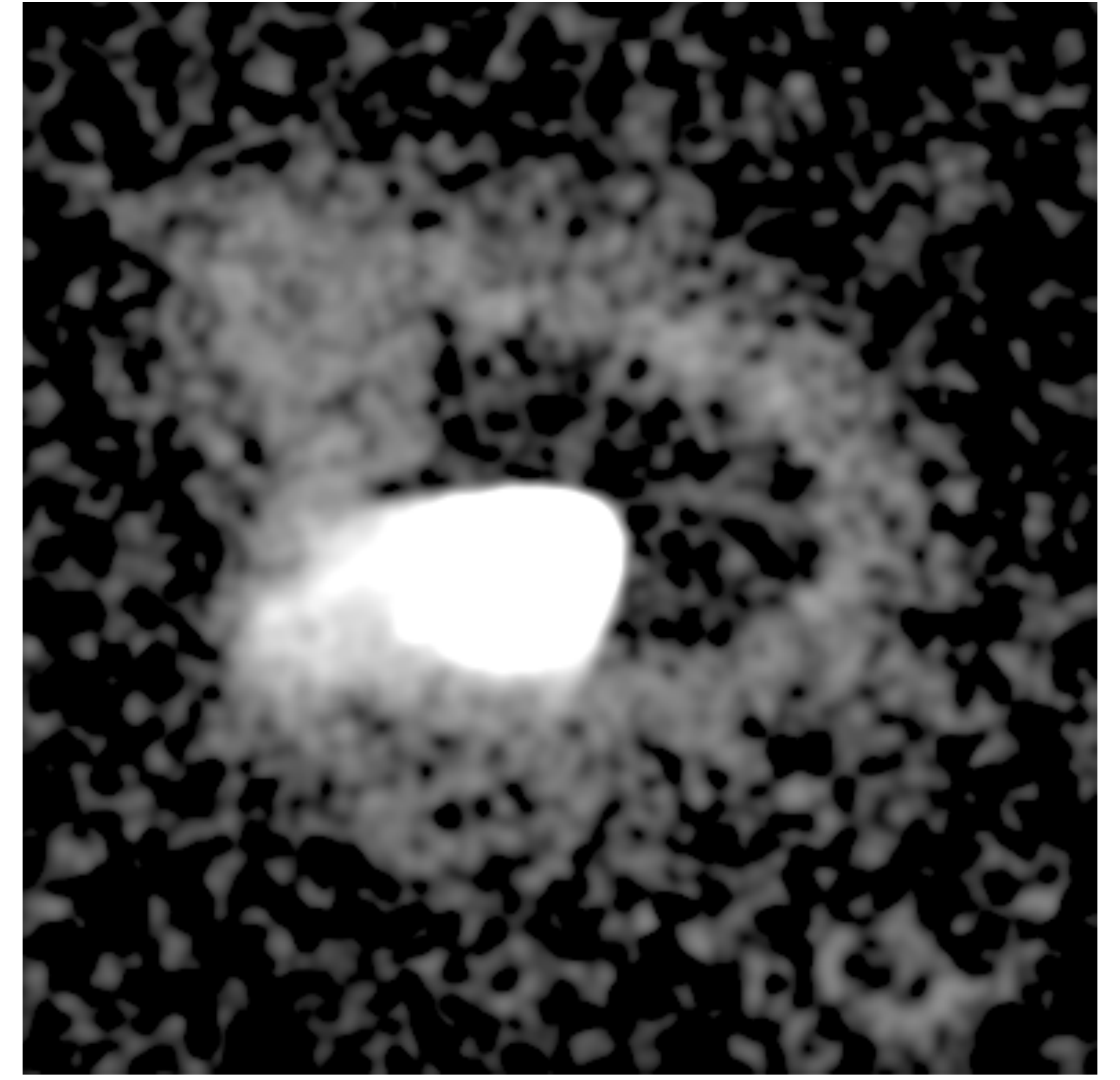}
\includegraphics[width=0.48\columnwidth]{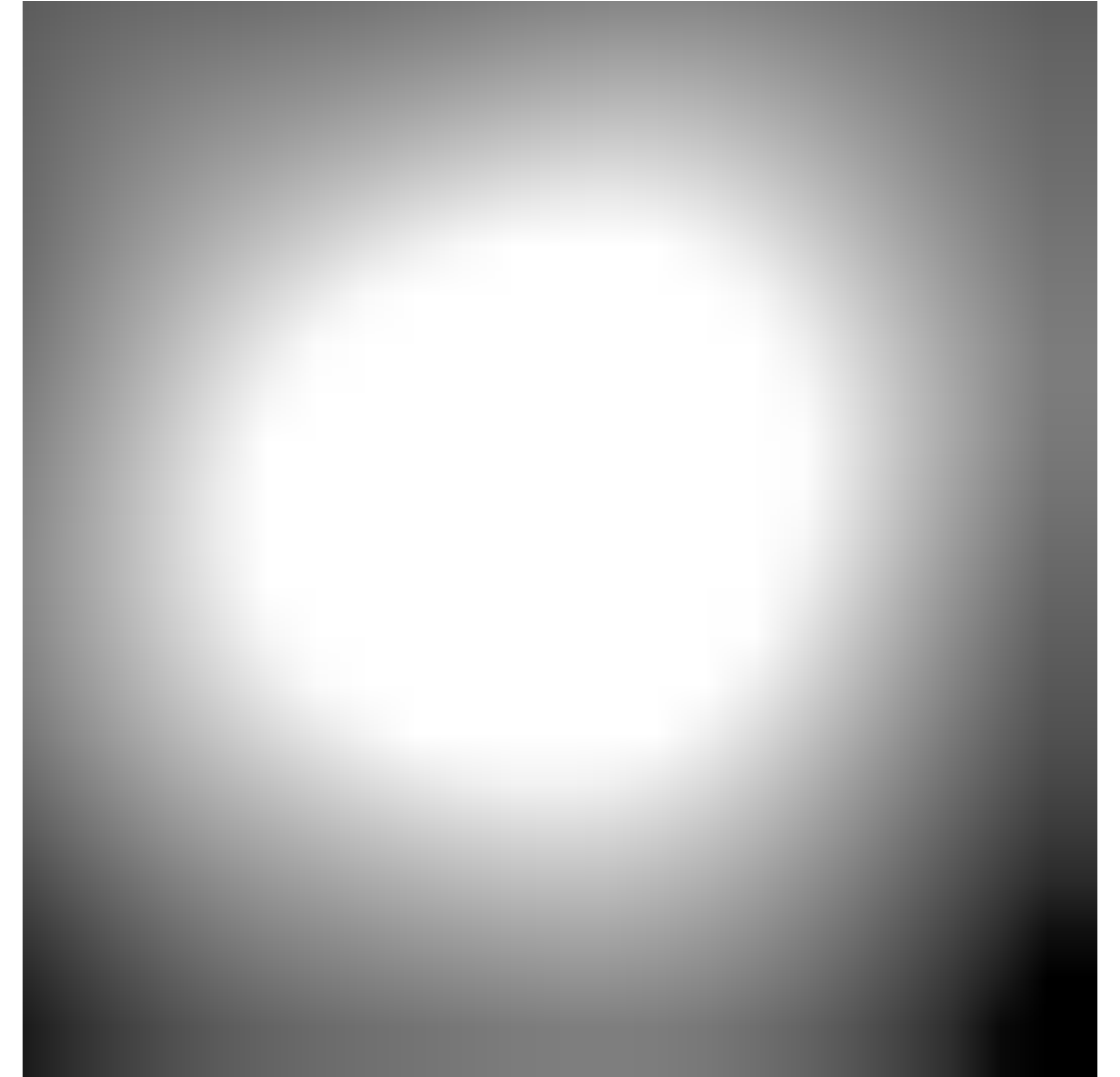}
\caption{Examples of profiles flagged as ``bad" by the  COSMOS classification exercise described in the text. The two galaxies shown here were classed as \emph{artefact} (left) and \emph{box too small} (right). For a breakdown of the number in each category see \ref{tab:eyeball_numbers}.}\label{fig:cosmos_eyeball_example}
\end{figure}

\begin{table}
\centering
\begin{tabular}{l|ccc}
   \hline
   Category                     & COSMOS          & Galaxies    & COSMOS Profiles \\
                               & Profiles        & in \hoopoe  & in \hoopoe       \\
   \hline
   Total                       &  87624          & 17.97 M     & 27612            \\
   \hline
   Good                        &  76707          & 16.93 M     & 25878            \\
   Box Too Small               &  3743           & 0.16 M      & 424              \\
   Artefact                    &  1024           & 0.35 M      & 410              \\
   Two Galaxies                &  542            & 0.40 M      & 375              \\
   Galaxy Missing              &  4212           & 0.08 M      & 354              \\
   Off Centre                  &  915            & 0.05 M      & 171              \\
   Other                       &  481            & 0.10 M      & 127              \\
   \hline
\end{tabular}
\caption{The number of input galaxies in the Y1 DES image simulations presented in Chapter 3 falling under each category in the profile inspection exercise described.
The first three columns show (left to right) the total number of COSMOS galaxies in each category from the full source catalogue from which the simulation draws profiles;
the number of simulated galaxies affected; and the corresponding number of COSMOS profiles 
(note that the second and third columns are not identical since each COSMOS profile is drawn into multiple positions).}
\label{tab:eyeball_numbers}
\end{table}

\noindent
To test the impact of the aberrant COSMOS profiles on the \imshape\
calibrations we fit for multiplicative and additive bias in the \hoopoe~dataset three times with different 
selection criteria: 
(a) \imshape~ quality cuts only; 
(b) removing any objects classed as ``bad" for any reason;  and 
(c) the same as (b), but additionally cutting any galaxies that fall within a circular aperture of 100 pixels 
around each flagged COSMOS profile. 
The results, in four DES Y1-like tomographic bins, are shown in Figure~\ref{fig:cosmos_eyeball_results}.

The straightforward cut (b) induces a shift $\Delta m$ that is comfortably within the
level of statistical error of the fit.
The second test suggests the corrupted profile may induce a small neighbour bias on
surrounding profiles, which manifests as a modulation in $m$.
It is worth pointing out that some of the categories listed in Table~\ref{tab:eyeball_numbers} may be
benign.
Off-centred galaxies and those with neighbours, for example, should
not cause a problem, since we re-run \sex~object detection and
deblending on the simulations.
Our final cut on the simulation rejects instances of COSMOS profiles
categorised under ``artefact", ``box too small", or ``galaxy missing".
We test that additionally cutting the other categories does not induce
a statistically significant change in bias.
Based on the results in Figure~\ref{fig:cosmos_eyeball_results}, we also incorporate a 
Gaussian component of width
$\sigma_m=0.005$ in the residual $m$ prior for \imshape.

\begin{figure}
\centering
\includegraphics[width=\columnwidth]{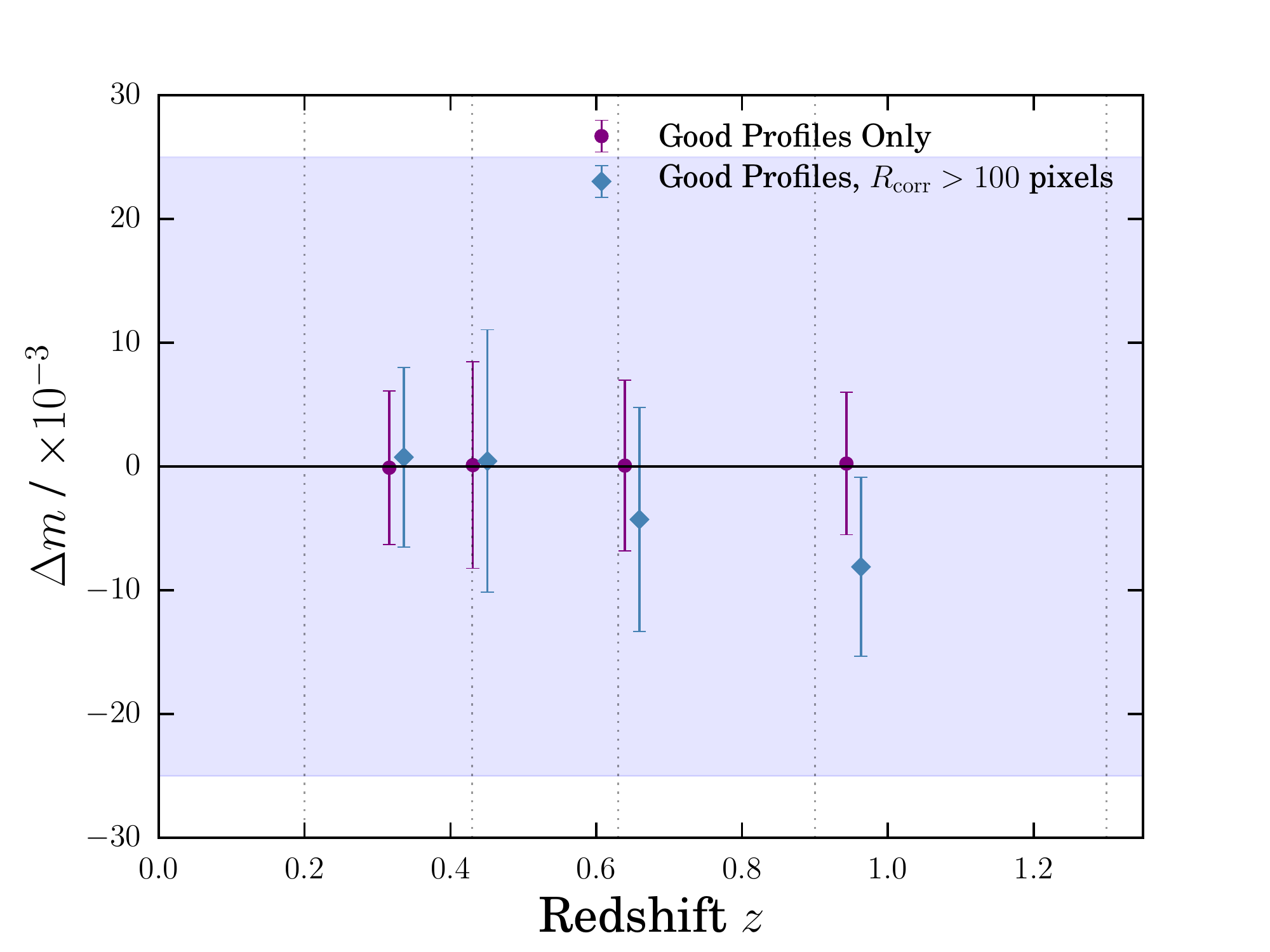}
\includegraphics[width=\columnwidth]{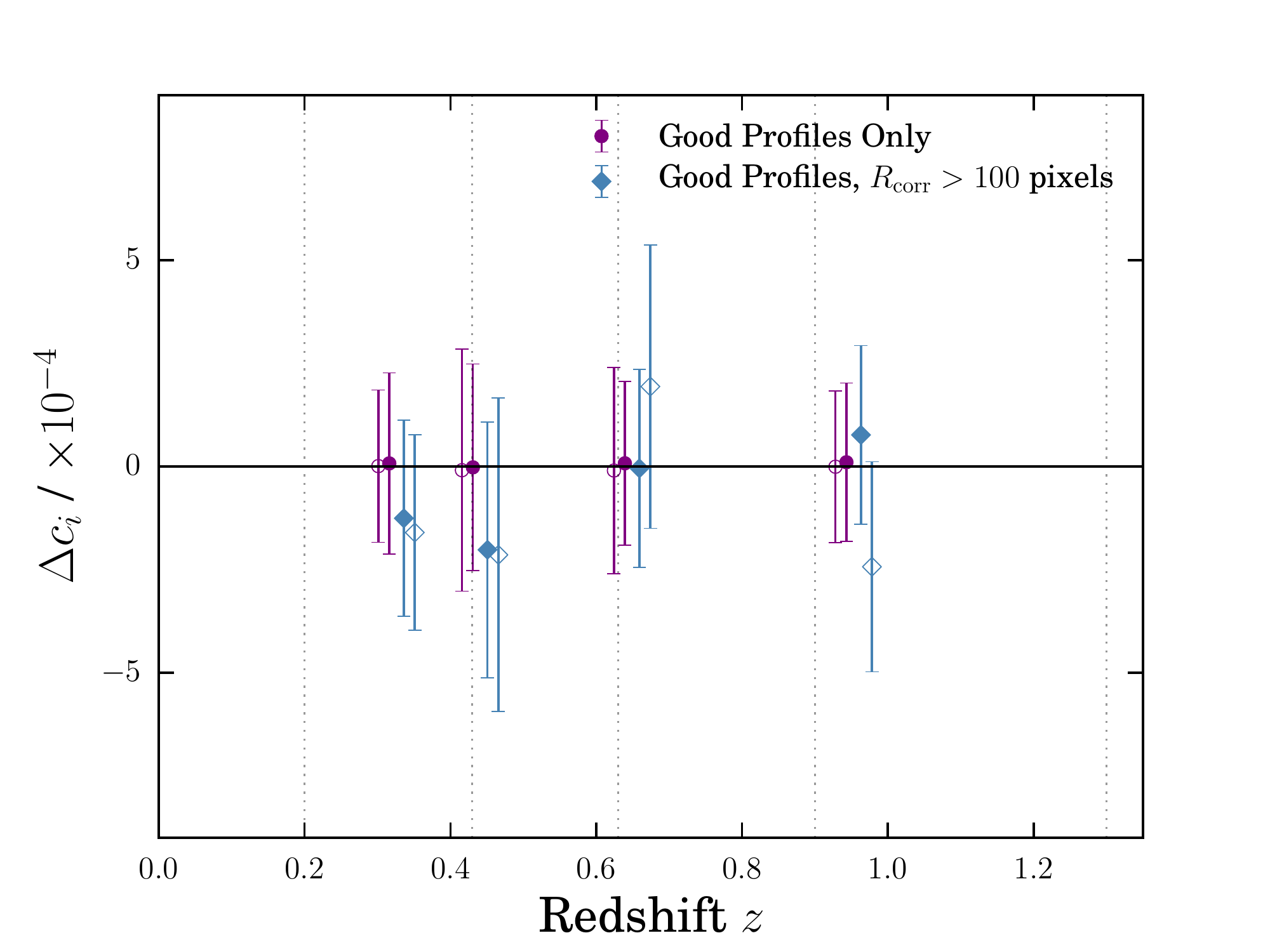}

\caption{Change in the multiplicative bias (top) and additive bias (bottom) after removal of bad COSMOS profiles, relative to the value derived using all galaxies.  In the lower panel the filled markers show the $c_1$ component and the open ones show $c_2$.}
\label{fig:cosmos_eyeball_results}
\end{figure}

\section{Sensitivity of \hoopoe~ Simulations to Observable Distributions}
\label{sec:sims:sensitivity}

Though our calibration appears to pass the internal tests presented in \S \ref{sec:cal},
it is still possible that residual biases could arise due to differences with the data
seen in \S\ref{sec:sims:comparisons}.
The most notable differences are in flux
and \rgp.
The raw distributions of \rgp\ and flux are shown by the solid lines in Figure~\ref{fig:sims:reweighted_histograms}, with the parent DES data shown by the shaded histograms.

We assess the importance of these differences by reweighting the \hoopoe~simulations to match the data.
In the case of \rgp\ we simply divide galaxies into bins of size and assign a uniform weight to each bin, such that the
simulated distribution $p($\rgp$)$ matches the data.
In the second case we carry out the same procedure for galaxy flux.
This time, however, an independent set of weights is computed for bulge and disc galaxies, 
such that they each match the corresponding sub-populations of the data. 
The reweighted distributions are shown by the dashed lines in Figure~\ref{fig:sims:reweighted_histograms}. 

As pointed out by \citet{fc16}, who carried out a similar test for KiDS, reweighting can be problematic if the quantities in question are covariant with ellipticity. In such cases reweighting to match a 1D projected distribution $p(q)$ may be inadequate to correct (or even worsen) differences in the 2D joint distribution $p(q,e)$. In each case we check both the 2D distributions (not shown here) and the 1D $p(e)$ histograms (shown in the right-hand panels of Figure~\ref{fig:sims:reweighted_histograms}). Neither reweighting operation is found to produce such spurious differences. 

Finally, galaxies are divided into four Y1-like tomographic bins, as before,
the fiducial calibration is applied, and the residual $m$ is calculated in each bin.
The results are shown in Figure~\ref{fig:sims:m_sensitivity_reweight}.
The maximum change under reweighting $\Delta m$ in both cases is $\mathcal{O} (10^{-3})$.
This is not found to have a coherent direction across $z$ bins,
and is well within both the statistical error margin (the blue shaded boxes)
and the $1\sigma$ width of our prior (the dashed horizontal lines).

\begin{figure}
\includegraphics[width=\columnwidth, height=0.6\columnwidth]{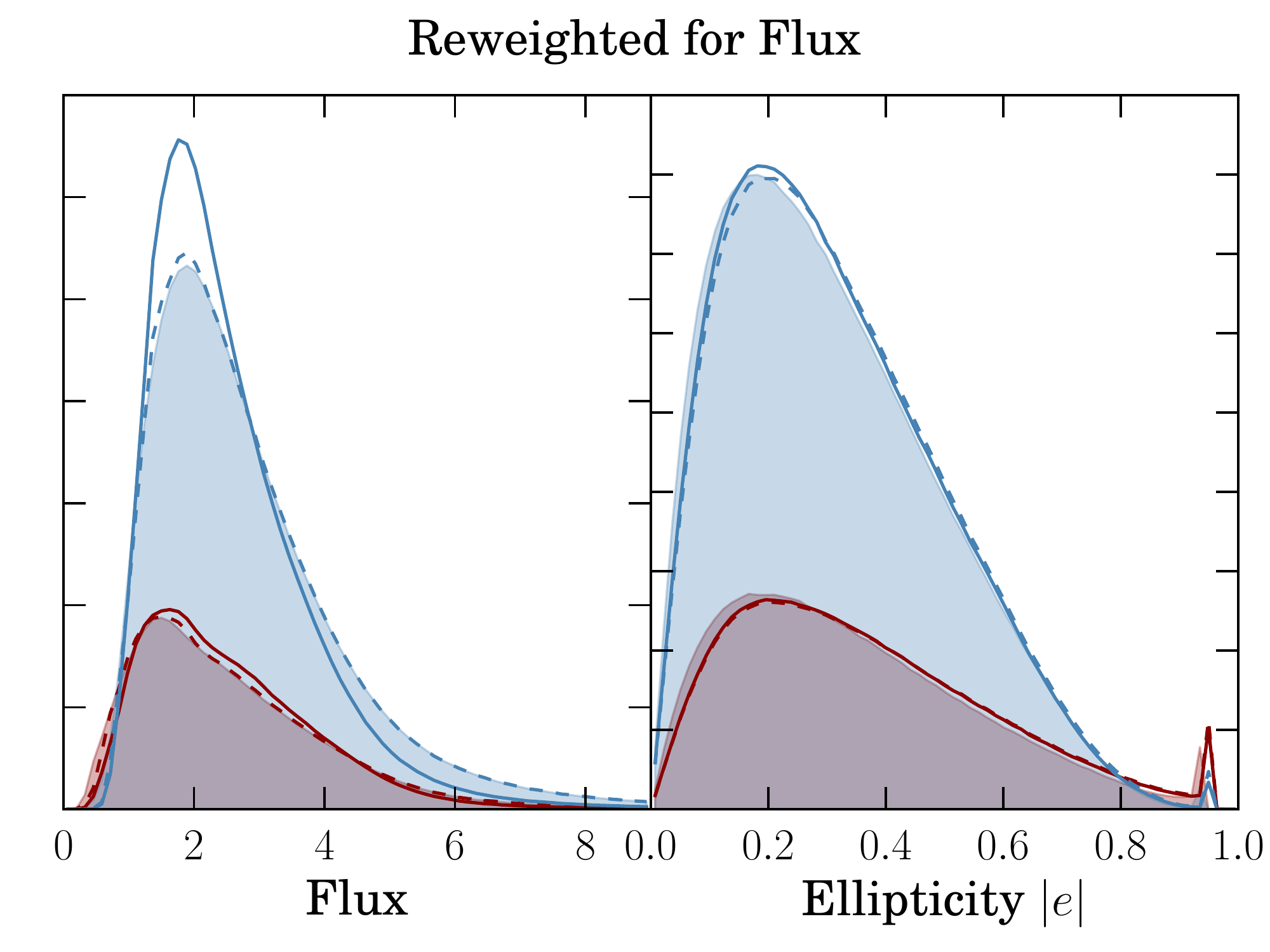}
\includegraphics[width=\columnwidth, height=0.6\columnwidth]{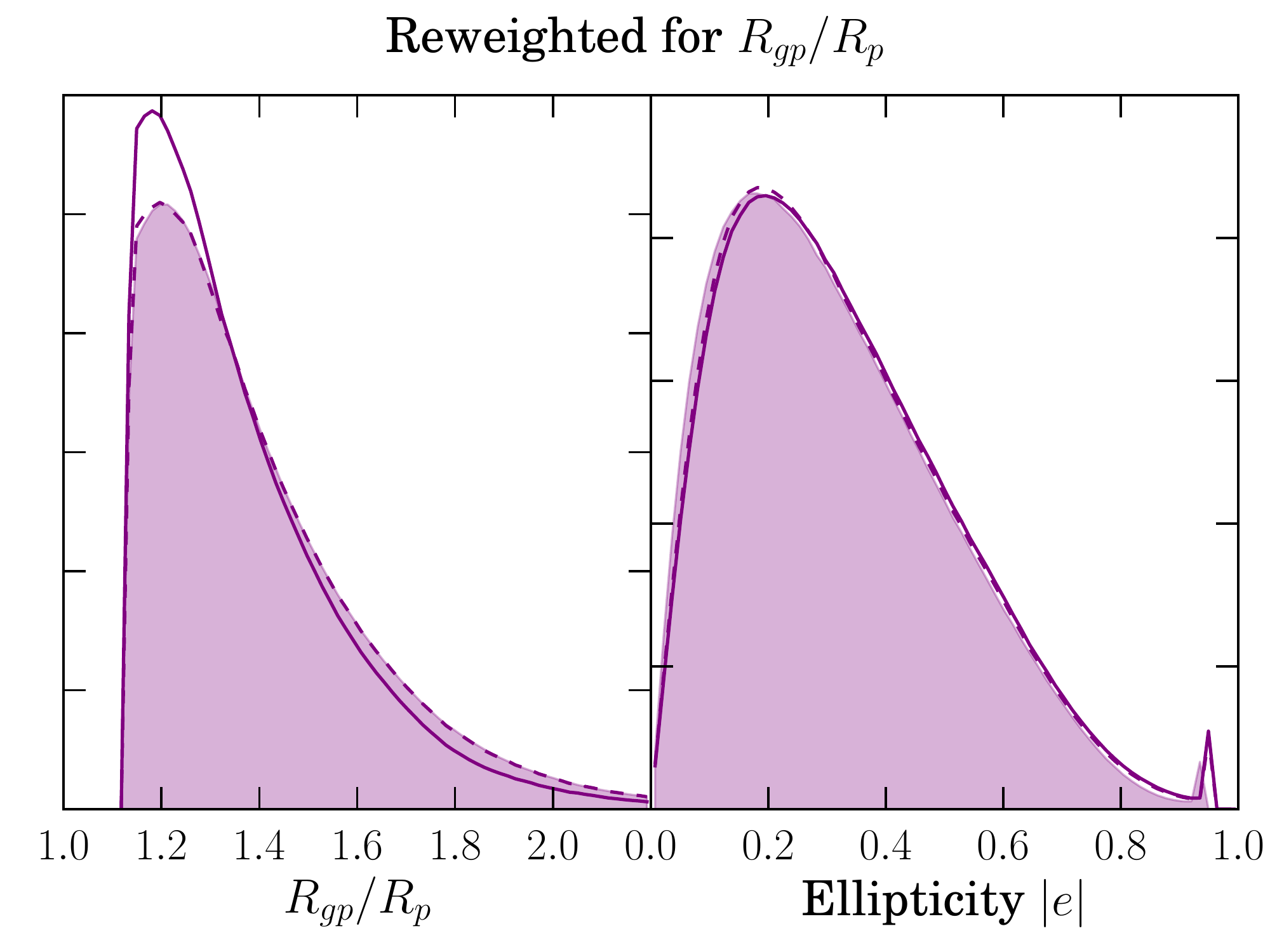}
\caption{Histograms of flux (upper left), size (lower left) and ellipticity (upper/lower right) in the data (shaded region) and simulations used for \imshape\ calibration before (solid) and after (dashed) objects are re-weighted to match
the flux (top) and size (bottom) distributions in the data.
In the upper panel we show bulge and disc galaxies separately in red and blue respectively.
}
\label{fig:sims:reweighted_histograms}
\end{figure}

\begin{figure}
\includegraphics[width=\columnwidth]{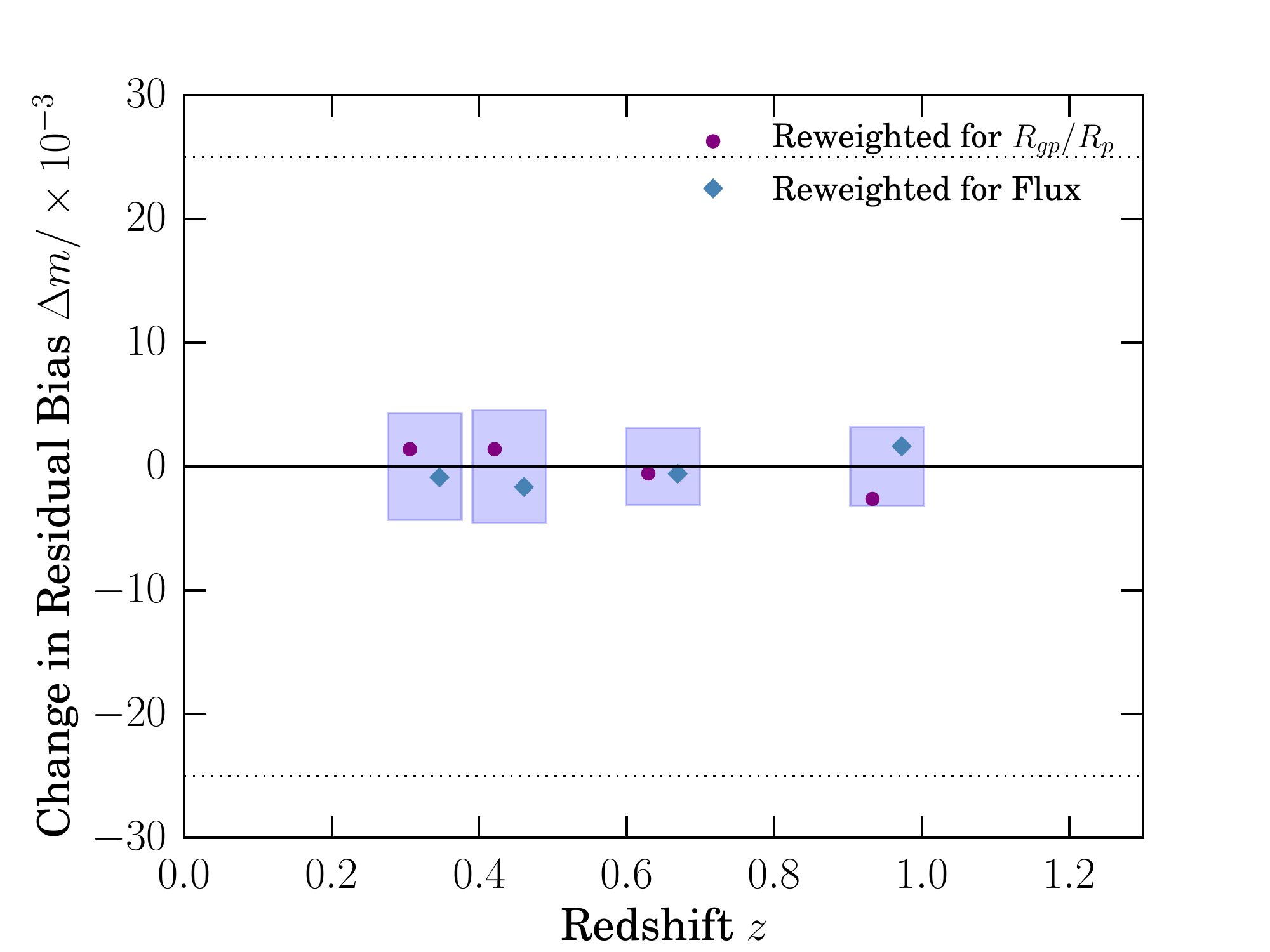}
\caption{Change in the residual \imshape\ bias resulting from reweighting the simulations prior to calibration, shown in the four tomographic bins used in the DES Y1 shear 2pt analysis. The purple circles show the result when reweighting to compensate for the excess of small galaxies shown in the centre-right panel of Figure~\ref{fig:sims:histograms}, while the blue diamonds are reweighted for bulge/disc flux. The blue bands mark the $1\sigma$ statistical error on $m$, while the horizontal dotted lines are the $\pm 1\sigma$ bounds of the $m$ prior for \imshape.
\label{fig:sims:m_sensitivity_reweight}
}
\end{figure}

\section{Validating the \hoopoe~Simulations}
\label{sec:sims:validation}

In this appendix we describe a series of exercises to test the level at which features of
our \hoopoe~simulations which are systematically different from the data
affect the multiplicative bias calibration.
Any such effects which have a non-vanishing impact must be included in our prior on residual $m$ after calibration.

The first limitation comes from the fact that a finite selection of COSMOS galaxies is used
to simulate a much larger sample of DES galaxies.
The cache of input profiles, though continuously updated is relatively small,
which results in the same COSMOS galaxies appearing repeatedly within particular regions of the simulated images.
Such effects could conceivably lead to additive or multiplicative biases,
if the frequency of repetition is sufficiently high. 
To test this we divide the \hoopoe~galaxies according to COSMOS identifier.
For each unique profile we construct a $k$-d tree data structure
on the coadd pixel grid. This is repeatedly queried to locate the nearest
instance of the same COSMOS profile.

We find a mean recurrence scale
of $\sim150$ pixels or 40.5 arcseconds,
though there is a significant asymmetry in the distribution of distances
with a heavy tail out to 1000 pixels and higher.
The fraction of galaxies with a relatively close self-neighbour is, however, also non-vanishing. 
We thus perform the following test.
\hoopoe~ galaxies are first assigned to four top-hat redshift bins, 
as described in \S \ref{sec:im3shape:tomo}.
In each bin we fit for multiplicative and additive biases 
(a) using all galaxies and 
(b) using only galaxies with no instance of the same profile within a radius of 100 pixels.
The raw number removed by the cut is relatively small,
but it could conceivably favour small round objects.
To ensure we are measuring the true impact of self-neighbours, and not a selection effect from
the cut devised to remove them, we reweight the surviving galaxies.
Weights are assigned based on \snr~and \rgp,
such that, when applied, the 2D histogram 
$p(\snr,\rgp)$ matches the data.
We find no significant change in multiplicative nor additive bias in any of the redshift bins
($\Delta m \sim 10^{-4}$, $\Delta c_{i} \sim 10^{-5}$).

A second limitation concerns the nature of the input COSMOS profiles themselves.
The simulations make use of an early release of the deep COSMOS catalogue.
Due to masking errors and deblending failures a fraction of this input catalogue is visibly defective.
We use an internal crowdsourcing exercise, the details of which can be found in Appendix \ref{app:eyeball},
to categorise the COSMOS galaxies into six groups according to their visual characteristics. 
In the final cut we remove profiles flagged as ``artefacts" or oversized relative to their boxes.
In total this removes 0.51M/18M objects from the simulated shape catalogue.
Using a similar nearest neighbour search as above, we estimate mean distance to the nearest
``bad" COSMOS profile to be $\sim 90$ pixels.
We recompute the biases $m$ and $c_i$ under three scenarios:
(a) using all galaxies,
(b) cutting COSMOS profiles classed as artefacts or oversized and
(c) the same as (b), but also cutting galaxies drawn within 100 pixels of a bad COSMOS profile.
We find the computed biases are stable to well within $1 \sigma$ in all apart from the upper redshift bin.
Here we lose the bulk of the galaxies removed by this cut,
which is perhaps unsurprising given that these objects tend to be small,
faint and thus most susceptible to deblending failures.
The change in all scenarios is at the level of the
$1 \sigma$ statistical error at $\Delta m \sim 0.005-0.0075$.
Though small, this is non-trivial and so we incorporate this uncertainty as a
systematic contribution to our $m$ prior (see \S\ref{sec:budget}). 

The use of the Y1 detection catalogue to source the positions of simulated galaxies
is intended to capture the galaxy clustering patterns across the survey.
It does have some drawbacks, chiefly that it omits undetected or strongly blended galaxies
(see \S\ref{subsubsection:faint_galaxies}). 
A second potential limitation is this: not all detections in the Y1 source catalogue correspond to real galaxies. 
Spurious detections can be produced by CCD chip edges and by image artefacts such as satellite trails and ghosts. 
These detections are removed prior to shape measurement and do not feature in the final {\sc Gold} catalogues,
but the raw detection catalogues, which are used as inputs to our simulations, 
do not provide sufficient information to distinguish real from false detections during runtime. 

We tried a simple detection algorithm to flag these features, using boxcar averaged source densities, 
but this was not found to reliably detect diagonal or curved streaks.
The \hoopoe~images consequentially include infrequent but visually striking lines of COSMOS galaxies in these locations.
To quantify the impact, we implemented a second crowdsourcing exercise,
analogous to the one described in Appendix \ref{app:eyeball}.
We first ran the boxcar detection algorithm on the simulated coadd images,
and created visual bookmarks for these detections.
Participants were then asked to inspect approximately half of the simulated tiles,
each of which was split into $5\times5$ square patches.
Patches in which the detection positions exhibited visible structure were flagged for removal.
As before we then divide \hoopoe~galaxies into DES-like redshift bins and recompute $m$ and $c_i$,
first including the flagged regions and then excising them.
Using all galaxies (no redshift binning) we find a shift $\Delta m = 3.7 \times 10^{-5}$,
which is equivalent to less than $2\%$ of the $1\sigma$ statistical uncertainty on $m$.
In four redshift bins, and again reweighting to ensure the $p(\snr,\rgp)$ distribution
still matches the data, we measure 
$\textbf{m}=(-0.0969, -0.1583 , -0.1697 , -0.2160)$ 
with the spurious detection lines cut and 
$\textbf{m}=(-0.0973, -0.1581 , -0.1691, -0.2160)$
when they are included.
That is, the cut alters $m$ by at most $\Delta m=0.0007$. 
Since any systematic shift is subdominant to statistical uncertainty,
we do not consider spurious detections
further as a source of systematic calibration error.

\section{Multiplicative biases in tomographic measurements}
\label{app:m:tomography}

The uncertainties $\sigma_m$ on multiplicative bias $m$ of the \metacal\ and \imshape\ catalogues given in \S\ref{sec:budget} are valid for our overall source sample without redshift binning or weighting. The true multiplicative biases present in our catalogues likely vary as a function of redshift.  We do not have a reliable model for this variation, so instead must use different multiplicative bias parameters $m_i$ for different bins in tomographic analyses.  In this appendix we consider how $m$ values should be statistically correlated between redshift bins.  Which choice of covariance matrix is more conservative depends on the type of parameter that we are measuring (see also \citealt{photoz}, their appendix A).

Consider two hypothetical parameters to be estimated from a tomographic lensing measurement with two bins, denoted by $S$ (proportional to the sum of amplitudes in the low and high redshift bin) and $D$ (proportional to the difference of these amplitudes). Practical examples for $S$ include $S_8=\sigma_8\sqrt{\Omega_m/0.3}$, and for $D$ include photo-$z$ bias parameters of the low and high redshift bin.  The two most obvious ways to marginalize over $m$ in this scenario are:
\begin{itemize}
\item to marginalize over a single parameter $m$ with Gaussian prior $\sigma_m$ -- this is the same as using a fully correlated $m$ per bin, and is the most conservative choice possible for $S$.  For $D$, though, it underestimates the error when $m$ varies with redshift.
\item to marginalize over two parameters $m_1$ and $m_2$ with fully independent Gaussian priors of width $\sigma_m$.  This is the most conservative choice for $D$ but underestimates the systematic uncertainty of $S$ due to $m$ by a factor $\sqrt{2}$.
\end{itemize}

If we want to be conservative for both these types of parameter then we must increase $\sigma_m$.  If we have $n$ redshift bins with equal signal-to-noise, then we should use uncorrelated $m$ values with $\sigma_{m_{i}}=\sqrt{N}\sigma_m$.  The generalization of this to bins with unequal signal-to-noise ratios $\rho_i$ is to use  $\sigma_{m_i}=a\times\sigma_m$ where:
\begin{equation}
a=\sqrt{ 
\frac{\sum_{i,j} \rho_i^2 \rho_j^2}
{\sum_{i}\rho_i^4}
}\;,
\end{equation}
For cosmic shear and redMaGiC galaxy-galaxy lensing with the binning schemes similar to \citet{shearcorr} and \citet{gglpaper}, we find approximately $a=\sqrt{2.6}$ and $\sqrt{3.1}$, respectively. We take the larger value of a as the default tomographic rescaling of $\sigma_{m_i}$.

Analyses using redshift-weighted or binned versions of our shape catalogues should take this re-scaling into account. For \metacal, it applies to all the contributions to $\sigma_m$, and can be multiplied with the $\sigma_m=0.013$ width. For \imshape, some of the contributions are either anti-correlated between redshift bins or estimated based on their maximum value among a set of redshift bins, in which case the re-scaling is not necessary. The correct $\sigma_{m_i}$ is:
\begin{equation}
\sigma_{m_i}=\sqrt{0.018^2+a^2\times(0.001^2+0.004^2+0.017^2+0.002^2)} \; .
\end{equation}

\section{\imshape~ Flags}
\label{sec:im3shape:flags}

\imshape\ uses two sets of flags to remove objects.  Many of these flags will cause selection biases,
so they are also applied in the calibration simulations so that this effect will be taken into account.
These flags are described in Tables \ref{table:im3shape:errorflags} and \ref{table:im3shape:infoflags}.

In addition to these flags, the calibration process does not calibrate objects with $\snr>200$ or $\rgp>3$.
Objects outside this range have {\sc FLAGS\_SELECT} $>0$ in the catalogue.

\begin{table}
    \centering

\begin{tabular}{ll}
\hline
 Value  &   Meaning  \\
\hline
$ 2^{0} $ &  \imshape~ failed completely \\
$ 2^{1} $ &  Minimizer failed to converge \\
$ 2^{2} $ &  $e<10^{-4}$: im3shape fit fail \\
$ 2^{3} $ &  $e_1$ or $e_2$ outside $(-1,1)$ \\
$ 2^{4} $ &  Radius $> 20$ arcsec \\
$ 2^{5} $ &  $\rgp > 6$ or NaN \\
$ 2^{6} $ &  Negative or NaN \rgp \\
$ 2^{7} $ &  $\snr<1$ or NaN \\
$ 2^{8} $ &  $\chi^2$ per effective pixel $> 3$ \\
$ 2^{9} $ &  Normalized residuals $< -20$ in any pixel \\
$2^{10} $ &  Normalized residuals $> 20$ in any pixel \\
$2^{11} $ &  RA more than 10 arcsec from nominal \\
$2^{12} $ &  Dec more than 10 arcsec from nominal \\
$2^{13} $ &  Failed to measure the FWHM of psf or galaxy \\
$2^{14} $ &  r-band sextractor flag has 0x4 or above \\
\hline
    \end{tabular}
    \caption{\imshape\ error flags, for extreme objects. These are not individually propagated into released catalogues.}
    \label{table:im3shape:errorflags}
\end{table}

\begin{table}
    \centering

    \begin{tabular}{ll}
\hline
Value  &  Meaning\\
                       \hline
$  2^{0} $  &  Area masked out in the {\sc Gold} catalogue \\
$  2^{1} $  &  Region flagged in the {\sc Gold} catalogue \\
$  2^{2} $  &  {\sc Modest} classifies as star \\
$  2^{3} $  &  Mask fraction $> 0.75$ \\
$  2^{4} $  &  levmar\_like\_evals $> 10000$ \\
$  2^{5} $  &  r-band sextractor flag 0x1, (bright neigbours) \\
$  2^{6} $  &  r-band sextractor flag 0x2, (blending) \\
$  2^{7} $  &  More than 25\% of flux masked \\
$  2^{8} $  &  $S/N < 12$ \\
$  2^{9} $  &  $S/N > 10000$ \\
$ 2^{10} $  &  $\rgp < 1.13$ \\
$ 2^{11} $  &  $\rgp > 3.5$ (very large galaxy) \\
$ 2^{12} $  &  Radius $> 5$ arcsec \\
$ 2^{13} $  &  Radius $< 0.1$ arcsec \\
$ 2^{14} $  &  Centroid more than 1 arcsec from nominal \\
$ 2^{15} $  &  $\chi^2$ per effective pixel $< 0.5$ \\
$ 2^{16} $  &  $\chi^2$ per effective pixel $> 1.5$ \\
$ 2^{17} $  &  Normed residuals $< -0.2$ somewhere \\
$ 2^{18} $  &  Normed residuals $> 0.2$ somewhere \\
$ 2^{19} $  &  Very large PSF \\
$ 2^{20} $  &  Negative PSF FWHM \\
$ 2^{21} $  &  One or more error flags is set \\
    \hline
    \end{tabular}
    \caption{\imshape\ info flags, for objects with any undesirable features. These are included in the released catalogues as {\sc FLAGS}.}
    \label{table:im3shape:infoflags}
\end{table}

\section{\metacal\ Response Behaviour} \label{sec:mcal:response-example}

The \metacal\ response factors $R_\gamma$ and $R_s$ described in \S \ref{sec:metacal:overview} can 
vary with any galaxy feature, since they 
are calculated on a per-object basis.  To illustrate the general behaviour of these factors and reir 
relative importance, Figure \ref{fig:mcal:response-behaviour} shows the size of the different 
terms.  Note that the two quantities plotted are calculated slightly differently - the $R_s$ part is 
the correction for bias caused by cutting out all objects below the $x$ coordinate value, 
whereas $R_\gamma$ is the mean correction for all objects in a bin centred on the $x$ coordinate 
value. 

For the specific estimator chosen here the selection bias associated with signal-to-noise is nearly negligible, being well below 1\%, whereas the size selection bias is much larger, peaking at 4\%. Our fiducial size cut was $T/\Tpsf > 0.5$, corresponding to a 2\% correction.

\begin{figure}
\includegraphics[width=\columnwidth]{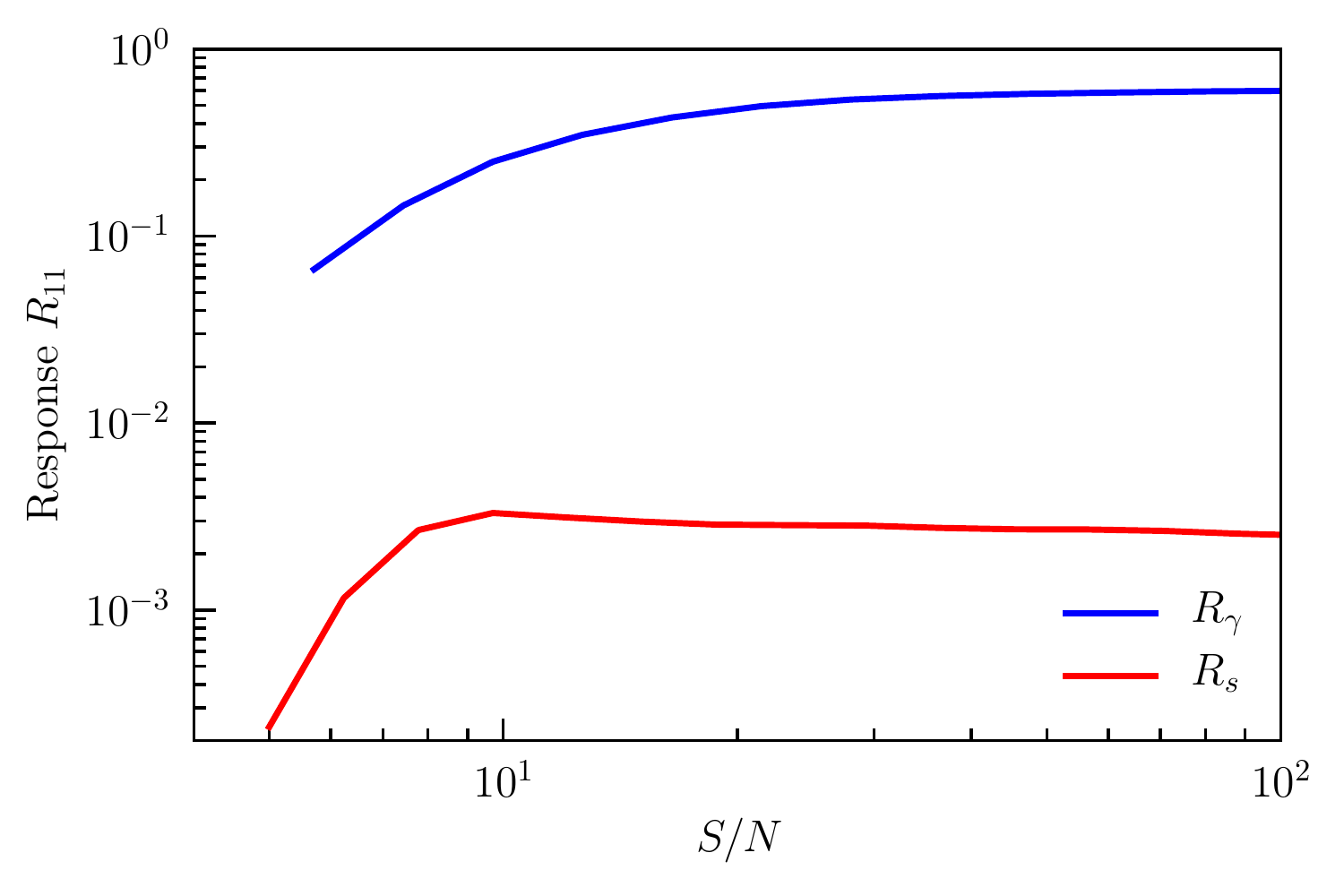}
\includegraphics[width=\columnwidth]{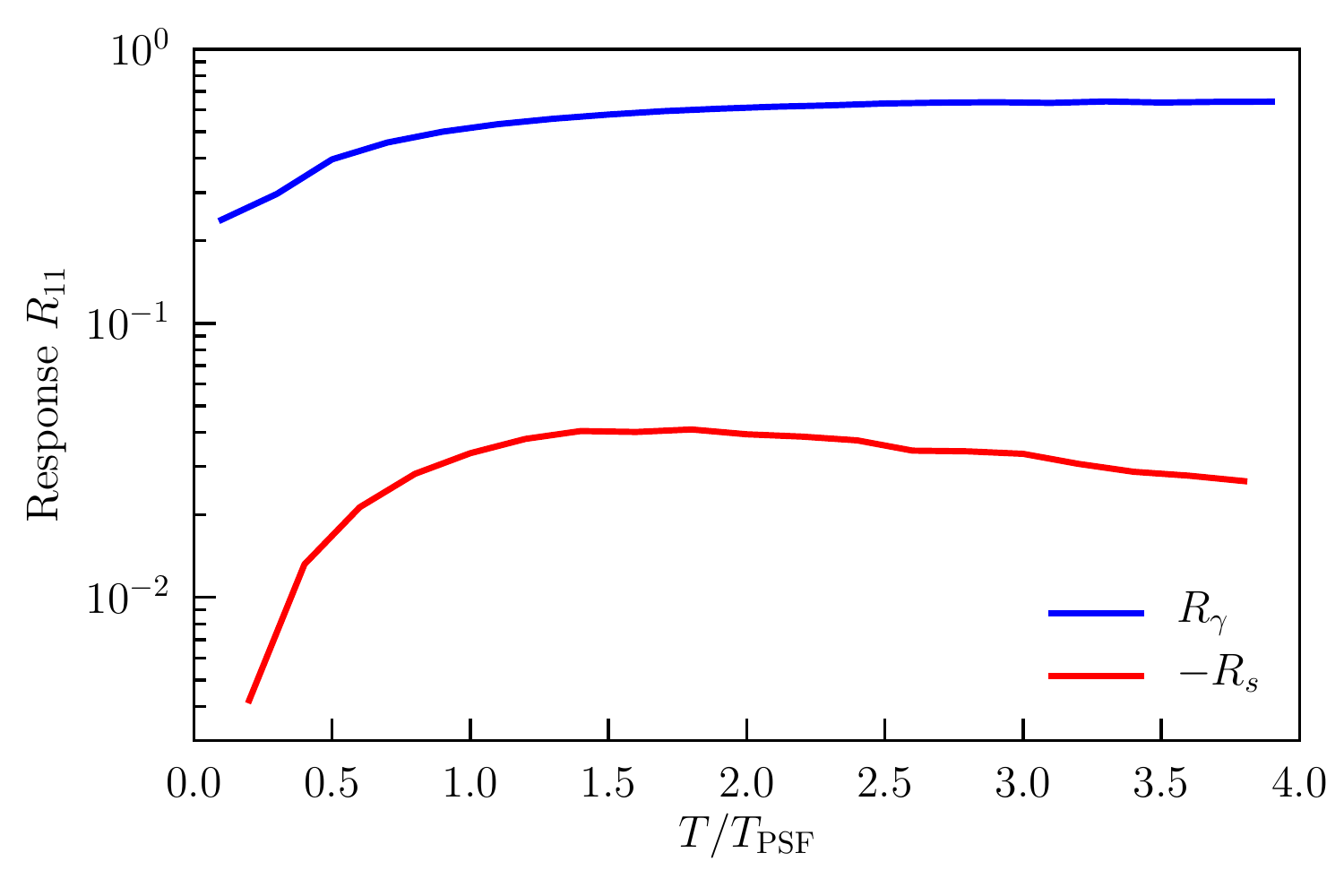}
\caption{Variation of the 1-1 components of the selection bias response correction $R_s$ and mean calibration response $R\gamma$ with signal-to-noise (top) and galaxy size relative to PSF size (bottom).}
\label{fig:mcal:response-behaviour}
\end{figure}

\section{Author Affiliations}
\label{sec:affiliations}
$^{1}$ Institute for Astronomy, University of Edinburgh, Edinburgh EH9 3HJ, UK\\
$^{2}$ Brookhaven National Laboratory, Bldg 510, Upton, NY 11973, USA\\
$^{3}$ Jodrell Bank Center for Astrophysics, School of Physics and Astronomy, University of Manchester, Oxford Road, Manchester, M13 9PL, UK\\
$^{4}$ Center for Cosmology and Astro-Particle Physics, The Ohio State University, Columbus, OH 43210, USA\\
$^{5}$ Department of Physics, The Ohio State University, Columbus, OH 43210, USA\\
$^{6}$ Department of Physics and Astronomy, University of Pennsylvania, Philadelphia, PA 19104, USA\\
$^{7}$ Kavli Institute for Particle Astrophysics \& Cosmology, P. O. Box 2450, Stanford University, Stanford, CA 94305, USA\\
$^{8}$ SLAC National Accelerator Laboratory, Menlo Park, CA 94025, USA\\
$^{9}$ Institut de F\'{\i}sica d'Altes Energies (IFAE), The Barcelona Institute of Science and Technology, Campus UAB, 08193 Bellaterra (Barcelona) Spain\\
$^{10}$ Fermi National Accelerator Laboratory, P. O. Box 500, Batavia, IL 60510, USA\\
$^{11}$ Kavli Institute for Cosmological Physics, University of Chicago, Chicago, IL 60637, USA\\
$^{12}$ Department of Astronomy, University of Illinois, 1002 W. Green Street, Urbana, IL 61801, USA\\
$^{13}$ National Center for Supercomputing Applications, 1205 West Clark St., Urbana, IL 61801, USA\\
$^{14}$ Universit\"ats-Sternwarte, Fakult\"at f\"ur Physik, Ludwig-Maximilians Universit\"at M\"unchen, Scheinerstr. 1, 81679 M\"unchen, Germany\\
$^{15}$ Jet Propulsion Laboratory, California Institute of Technology, 4800 Oak Grove Dr., Pasadena, CA 91109, USA\\
$^{16}$ Department of Physics \& Astronomy, University College London, Gower Street, London, WC1E 6BT, UK\\
$^{17}$ Department of Physics, ETH Zurich, Wolfgang-Pauli-Strasse 16, CH-8093 Zurich, Switzerland\\
$^{18}$ Centro de Investigaciones Energ\'eticas, Medioambientales y Tecnol\'ogicas (CIEMAT), Madrid, Spain\\
$^{19}$ Institute of Astronomy, University of Cambridge, Madingley Road, Cambridge CB3 0HA, UK\\
$^{20}$ Kavli Institute for Cosmology, University of Cambridge, Madingley Road, Cambridge CB3 0HA, UK\\
$^{21}$ Max Planck Institute for Extraterrestrial Physics, Giessenbachstrasse, 85748 Garching, Germany\\
$^{22}$ Cerro Tololo Inter-American Observatory, National Optical Astronomy Observatory, Casilla 603, La Serena, Chile\\
$^{23}$ Department of Physics and Electronics, Rhodes University, PO Box 94, Grahamstown, 6140, South Africa\\
$^{24}$ LSST, 933 North Cherry Avenue, Tucson, AZ 85721, USA\\
$^{25}$ CNRS, UMR 7095, Institut d'Astrophysique de Paris, F-75014, Paris, France\\
$^{26}$ Sorbonne Universit\'es, UPMC Univ Paris 06, UMR 7095, Institut d'Astrophysique de Paris, F-75014, Paris, France\\
$^{27}$ Laborat\'orio Interinstitucional de e-Astronomia - LIneA, Rua Gal. Jos\'e Cristino 77, Rio de Janeiro, RJ - 20921-400, Brazil\\
$^{28}$ Observat\'orio Nacional, Rua Gal. Jos\'e Cristino 77, Rio de Janeiro, RJ - 20921-400, Brazil\\
$^{29}$ Institute of Space Sciences, IEEC-CSIC, Campus UAB, Carrer de Can Magrans, s/n,  08193 Barcelona, Spain\\
$^{30}$ Department of Physics, IIT Hyderabad, Kandi, Telangana 502285, India\\
$^{31}$ Excellence Cluster Universe, Boltzmannstr.\ 2, 85748 Garching, Germany\\
$^{32}$ Faculty of Physics, Ludwig-Maximilians-Universit\"at, Scheinerstr. 1, 81679 Munich, Germany\\
$^{33}$ Department of Physics, California Institute of Technology, Pasadena, CA 91125, USA\\
$^{34}$ Department of Astronomy, University of Michigan, Ann Arbor, MI 48109, USA\\
$^{35}$ Department of Physics, University of Michigan, Ann Arbor, MI 48109, USA\\
$^{36}$ Instituto de Fisica Teorica UAM/CSIC, Universidad Autonoma de Madrid, 28049 Madrid, Spain\\
$^{37}$ Astronomy Department, University of Washington, Box 351580, Seattle, WA 98195, USA\\
$^{38}$ Santa Cruz Institute for Particle Physics, Santa Cruz, CA 95064, USA\\
$^{39}$ Australian Astronomical Observatory, North Ryde, NSW 2113, Australia\\
$^{40}$ Argonne National Laboratory, 9700 South Cass Avenue, Lemont, IL 60439, USA\\
$^{41}$ Departamento de F\'isica Matem\'atica, Instituto de F\'isica, Universidade de S\~ao Paulo, CP 66318, S\~ao Paulo, SP, 05314-970, Brazil\\
$^{42}$ Department of Astronomy, The Ohio State University, Columbus, OH 43210, USA\\
$^{43}$ Department of Astrophysical Sciences, Princeton University, Peyton Hall, Princeton, NJ 08544, USA\\
$^{44}$ Instituci\'o Catalana de Recerca i Estudis Avan\c{c}ats, E-08010 Barcelona, Spain\\
$^{45}$ Institute of Cosmology \& Gravitation, University of Portsmouth, Portsmouth, PO1 3FX, UK\\
$^{46}$ Lawrence Berkeley National Laboratory, 1 Cyclotron Road, Berkeley, CA 94720, USA\\
$^{47}$ Department of Physics and Astronomy, Pevensey Building, University of Sussex, Brighton, BN1 9QH, UK\\
$^{48}$ School of Physics and Astronomy, University of Southampton,  Southampton, SO17 1BJ, UK\\
$^{49}$ Instituto de F\'isica Gleb Wataghin, Universidade Estadual de Campinas, 13083-859, Campinas, SP, Brazil\\
$^{50}$ Computer Science and Mathematics Division, Oak Ridge National Laboratory, Oak Ridge, TN 37831\\
$^{51}$ Department of Physics, Stanford University, 382 Via Pueblo Mall, Stanford, CA 94305, USA\\
$\ddagger$ Einstein Fellow
\label{lastpage}

\end{document}